\documentclass[fleqn,usenatbib]{mnras}

\usepackage{newtxtext,newtxmath}
\usepackage[T1]{fontenc}

\DeclareRobustCommand{\VAN}[3]{#2}
\let\VANthebibliography\thebibliography
\def\thebibliography{\DeclareRobustCommand{\VAN}[3]{##3}\VANthebibliography}

\usepackage{graphicx}	% Including figure files
\usepackage{amsmath}	% Advanced maths commands
\usepackage{siunitx}
\usepackage{multirow}
\usepackage{enumitem}
\usepackage{booktabs}
\usepackage{pdflscape}
\usepackage{rotating}
\usepackage{afterpage}
\usepackage{longtable}
\usepackage{caption}
\newcommand\notsotiny{\@setfontsize\notsotiny{6.31415}{7.1828}}
\newcolumntype{C}[1]{>{\centering\arraybackslash}p{#1}}
\newcommand{\supplementarysection}{%
  \setcounter{figure}{0}% Reset figure counter
  \renewcommand{\thefigure}{A\arabic{figure}}
  \renewcommand{\theHfigure}{A\arabic{figure}}
}

\title[Exogeological inferences from white dwarfs]{Exogeological inferences from white dwarf pollutants: the impact of stellar physics}

\author[A. M. Buchan et al.]{Andrew M. Buchan$^{1}$\thanks{E-mail: andy.buchan@warwick.ac.uk (AMB)}, Pier-Emmanuel Tremblay$^{1}$, Antoine B\'edard$^{1}$, Evan B. Bauer$^{2,3}$ 
\newauthor{and Tim Cunningham$^{3}$}
\\
$^{1}$Department of Physics, University of Warwick, Coventry CV4 7AL, UK\\
$^{2}$Lawrence Livermore National Laboratory, Livermore, CA 94550, USA\\
$^{3}$Center for Astrophysics, Harvard \& Smithsonian, 60 Garden Street, Cambridge, MA 02138, USA
}

\date{Accepted XXX. Received YYY; in original form ZZZ}

\pubyear{\the\year{}}

\begin{document}
\label{firstpage}
\pagerange{\pageref{firstpage}--\pageref{lastpage}}
\maketitle

\begin{abstract}

Many white dwarfs have accreted material from their own planetary systems. These objects can be used to infer the composition of exoplanetary material and identify evidence for key geological processes. However, the white dwarf atmospheric physics distorts the inferred material composition away from the true composition, mainly through differential atomic diffusion of the accreted metals. Correcting for this effect is essential, but is dependent on various physical assumptions associated with the white dwarf itself. 
%It is therefore important to establish the impact of these various, and to understand which assumptions are relevant for which white dwarfs. 
%We focus on assumptions related to convective overshoot, thermohaline mixing and atmospheric boundary conditions on the atomic diffusion timescales. 
We first focus on the effect of assumptions related to convective overshoot and thermohaline mixing on the atomic diffusion timescales. 
For white dwarfs with H-dominated atmospheres between \SI{12000}{K} and \SI{18000}{K}, we find that including a complete treatment of convective overshoot decreases the inferred Fe and O abundances in accreted material. For these white dwarfs, we also find that including thermohaline mixing decreases Fe and O abundances. For He-dominated systems, the effect of convective overshoot is comparatively minor.
%We also find that the choice of atmospheric boundary conditions is highly important for white dwarfs with He-dominated atmospheres, and has a direct impact on the inferred core to mantle ratio of accreted material. 
We then explore the overall effect of other physical assumptions by comparing publicly available grids of diffusion timescales. We find that the choice of model grid can have a large impact for white dwarfs with He-dominated atmospheres, notably on the inferred core to mantle ratio of accreted material. We identify several systems for which the geological interpretation is robust against these systematics. We also present a `discrepancy metric' which can be used to estimate the potential impact of changing the stellar physics without requiring detailed modelling.

%PG\,1225$-$079 and GD\,133 show robust evidence of incomplete condensation, while WD\,J183352+321757 and HE\,0106$-$3253 additionally show evidence of accretion of core-like material.

\end{abstract}


%%%%%%%%%%%%%%%%%%%%%%%%%%%%%%%%%%%%%%%%%%%%%%%%%%

%%%%%%%%%%%%%%%%% BODY OF PAPER %%%%%%%%%%%%%%%%%%

\section{Introduction}

White dwarfs, the remnants of main sequence stars with masses below $\approx8\,$M$_{\odot}$, can be used to deduce the geology of exoplanetary systems (e.g., \citealt{Zuckerman2007,Farihi2013,Harrison2018,Putirka2021}). This is possible because many white dwarfs accrete material from reservoirs within their own planetary system \citep{Jura2003,Koester2014}. Once this material reaches the photosphere of the white dwarf, its constituent metals, here defined as elements heavier than He, can be detected from their absorption lines in the observed spectrum.

By using atmospheric models to predict synthetic spectra, the relative abundances of these metals can be measured. Given this composition and further corrections for differential atomic diffusion where the overall composition of the accreted material can be estimated (e.g., \citealt{Fontaine2015a,Fontaine2015b,Koester2020}), we may utilise geological modelling (e.g., \citealt{Buchan2021}), or comparison against known Solar System bodies (e.g., \citealt{Swan2023a}), to infer its geological history. In order to draw any geological inferences from relative metal abundances, a minimum of 2 metals must be detected. As of September 2024, there are 1739 white dwarfs with reliably detected photospheric pollution \citep{Williams2024}. The majority of polluted white dwarfs only have 1 metal detected (typically Ca), leaving 376 which are potentially amenable to geological analysis \citep{Williams2024}. The number of such systems is set to increase by an order of magnitude in the coming years. Surveys including 4MOST, WEAVE-WD, DESI and SDSS will between them provide medium-resolution spectroscopy for upwards of 150\,000 white dwarfs.

Geological analysis of material accreted by white dwarfs suffers from a major caveat: the material which we measure in the photosphere does not necessarily have the same composition as what was actually accreted. This is because all metals sink through the atmosphere or interior of the white dwarf, and different metals sink on different timescales \citep{Paquette1986a,Paquette1986b,Dupuis1992,Koester2009}. All else being equal, elements which sink on shorter timescales will appear relatively depleted, a phenomenon called `differential sinking'. Any attempt to model geology based on white dwarf pollution must take differential sinking into account.

Calculating these timescales requires detailed modelling of the atmosphere and envelope of the white dwarf. This includes the following four key transport processes and their uncertainties (see \citealt{Bedard2024} for a recent review).

\textit{Atomic diffusion.} This refers to the collective effect of microscopic diffusion processes (gravitational settling, thermal diffusion, chemical diffusion, and radiative levitation), which is a function of atomic mass and charge and is thus different for each element. In white dwarfs, the dominant contribution is provided by gravitational settling \citep{Paquette1986b}. Modelling this process requires two key pieces of input physics: the ionisation state and diffusion coefficient of each element. In most existing calculations, ionisation states are obtained from a simple Thomas--Fermi or Hummer--Mihalas equation of state, while diffusion coefficients are based on a collision integral formalism using a screened Coulomb interaction potential \citep{Paquette1986a,Stanton2016}. Recent improvements based on average-atom models and effective potential theory have shown that these traditional assumptions break down in cool He-dominated envelopes, leading to large systematics in diffusion timescales \citep{Heinonen2020}. However, these improved calculations were limited to only two elements (Si and Ca) due to computational costs, and therefore current complete grids of diffusion timescales still rely on the older framework \citep{Fontaine2015a,Bauer2019,Koester2020}.

\textit{Convection.} As white dwarfs cool, convection develops in their outer envelope due to the recombination of the main chemical constituent. The convection zone appears when the effective temperature, $T_{\rm eff}$, drops below about \SI{18000}{K} in H-dominated white dwarfs \citep{Cunningham2019}, and about \SI{50000}{K} in He-dominated white dwarfs \citep{Cukanovaite2021}. Because convection generates efficient chemical mixing, differential sinking can only operate below the convective region, where the diffusion timescales are longer than at the surface. In most studies, the depth of the convection zone has been predicted using 1D envelope models and the mixing-length theory \citep{BohmVitense1958}. This procedure depends on many physical ingredients and assumptions, notably the H/He radiative opacity and equation of state, the atmospheric boundary condition, and the assumed mixing-length parameter \citep{Tassoul1990,Fontaine2001}. In the last decade, 3D hydrodynamical simulations have produced a new state-of-the-art treatment of white dwarf convection \citep{Tremblay2013,Kupka2018,Cukanovaite2019,Cunningham2019,Cunningham2021}. These simulations crucially remove the need for the mixing-length parameter to describe the convective efficiency, but still rely on the same input microphysics.

\textit{Convective overshoot.} This refers to the process by which convective cells can enter into underlying, formally stable, layers of the white dwarf \citep{Freytag1996,Cunningham2019,Kupka2018}. Convective overshoot effectively extends the depth of the uniformly mixed region, and forces differential sinking to take place in deeper layers, also increasing the inferred rate of accretion. In 1D models, the extent of convective overshoot is essentially a free parameter, usually expressed as some fraction of the pressure scale height \citep{Koester2020,Bedard2022_STELUM}. In 3D models, this feature is instead predicted from first principles, thus providing a much greater constraining power. However, 3D overshoot studies have so far only been performed for warm ($11\,000\,\mathrm{K} \lesssim T_{\rm eff} \lesssim 18\,000\,\mathrm{K}$) H-dominated white dwarfs \citep{Cunningham2019}.

\textit{Thermohaline mixing.} This is an instability which may be triggered by the inverted composition gradient caused by the accretion of heavy elements onto the H/He envelope \citep{Deal2013,Wachlin2017,Cresswell2025}. If such an instability is active, it would lead to extremely rapid sedimentation of metals, dramatically increasing the inferred rate of accretion. A secondary effect, more relevant to this work, is that all metals would sink on similar timescales, effectively rendering differential sinking inactive \citep{Bauer2018}. This may, in turn, lead to a change in the interpretation of the composition of the accreted material. Thermohaline mixing is expected to be important in H-dominated white dwarfs, but not in their He-dominated counterparts \citep{Deal2013,Bauer2019}.

Differing treatment of these processes can influence the inferred metal sinking timescales of polluted white dwarfs, and could therefore influence the inferred geological history of accreted material. In this work, we investigate the circumstances under which different assumptions and techniques can result in different relative metal sinking timescales, and how this affects the geological interpretation.

This paper is structured as follows. In Sections~\ref{sec:geo_intro} and \ref{sec:phases} we introduce physical concepts which are key for all the test cases we consider. In Section \ref{sec:methods} we introduce the diffusion timescale grids and physical models used to describe the variables in each test case. In Sections~\ref{sec:overshoot} and \ref{sec:thermohaline}, we address the uncertainties due to convective overshoot and thermohaline mixing, respectively. In Section~\ref{sec:public_grids}, we estimate the overall effect of other physical assumptions by comparing available grids of sinking timescales. In each of Sections~\ref{sec:overshoot}-\ref{sec:public_grids},  we start by identifying white dwarfs which are most likely to be affected, then assess the impact of changing the underlying physics, and finally discuss the results and broader implications. Section~\ref{sec:robust} highlights systems for which there is no such impact, and so have robust geological interpretations. We summarise our work in Section~\ref{sec:conclusions}.

\section{Differentiation and incomplete condensation}
\label{sec:geo_intro}

In this work, the geological processes which we focus on are \mbox{core--mantle} differentiation and incomplete condensation. These are the two most important processes in setting the Earth's bulk composition and first order structure \citep{McDonough2003,WANG2019}. 

Incomplete condensation refers to a process that depends on the local temperature of a nascent body, which in turn is set by its distance from its host star. If this temperature is sufficiently high, as in the case of Earth, volatile elements cannot (completely) condense into the body and the resulting composition is depleted in volatile metals, including O, Na, C and N (e.g., \citealt{WANG2019}). Bulk-Earth-like devolatilisation has been detected in white dwarf pollutants (e.g., \citealt{Xu2014,Harrison2021,OBrien2025}). More extreme heating can cause more moderately volatile elements to become depleted, or equivalently for refractory elements such as Ca, Ti and Al to become enriched, resulting in compositions analogous to thermally processed meteorite populations in the Solar System (e.g., \citealt{Xu2013}).

Core--mantle differentiation refers to the process by which a body such as Earth segregates into an iron-rich core and a silicate mantle. This does not affect the bulk composition of the body. However, if the body is subsequently fragmented in some way (the details of which we are agnostic towards), those fragments may, in general, be enriched in core- or mantle-like material. Accretion of a core-rich fragment would result in pollution high in siderophile elements such as Fe and Ni (e.g., \citealt{Wilson2015}). Accretion of a mantle-rich fragment is detectable by enrichment in lithophile elements such as Mg and Ca, relative to the siderophile elements (e.g., \citealt{Zuckerman2011}). 

\section{Phases of accretion}
\label{sec:phases}

The correction which is required in order to account for differential sinking depends on the phase of accretion. We denote the duration of an accretion event as $t_{\rm event}$, and assume that mass accretes, starting at time $t=0$, at a constant rate during this event before abruptly stopping, as in \citet{Koester2009}. Denoting the average metal sinking timescale as $\overline{\tau}$, we can identify three distinct possible phases of accretion that result as $t$ increases.

For $t \ll \overline{\tau}$, the photospheric metal abundances rapidly increase, and their relative abundances closely match the composition of the accreted material (assuming that the atmosphere was pristine prior to accretion). This is the `build-up' or `increasing' phase. For $\overline{\tau} \ll t < t_{\rm event}$, it can be shown that the photospheric abundances of all metals plateau, and the relative abundance of a given metal (compared to the accreted material) is weighted by its sinking timescale. This is known as the `steady state' phase. For $t > t_{\rm event}$, it can be shown that a given photospheric metal abundance decreases exponentially, at a rate which depends on the sinking timescale of that metal. This is the `declining' or `decreasing' phase. In this phase, the abundance ratio of any pair of metals diverges exponentially towards 0 or infinity. We can anticipate that for systems in the declining phase of the accretion, the impact of any change in the adopted sinking timescales will be greatest. If $\overline{\tau} \gtrsim  t_{\rm event}$, the system does not reach steady state, but passes straight from build-up to declining phase. For further details, see \citet{Koester2009}.

In a more realistic treatment, the accretion rate may exponentially decay over time \citep{Jura2009}. In such a model, there is no steady state phase, although analogues of the build-up and declining phases still exist. This can affect the resulting geological interpretation \citep{OBrien2025}. Consideration of this effect is beyond the scope of this work.

\section{Methods}
\label{sec:methods}

Our strategy is to model the metal abundances at white dwarfs using a Bayesian framework, identifying evidence for differentiation and/or incomplete condensation. We model each system twice, using different sinking timescales (or a different treatment of thermohaline mixing), according to the specific test being conducted. We aim to identify white dwarfs for which the geological interpretation is affected.

%To carry out these tests, we implement multiple prescriptions of sinking timescales, and a treatment of thermohaline mixing. 
This section describes our Bayesian framework, our timescale grids and our implementation of thermohaline mixing. We also define 5 tests which make use of these different physical treatments, and describe our selection of samples of white dwarfs where these different assumptions are most likely to affect the geological interpretation of their pollution.

\subsection{Bayesian modelling}

We use the Bayesian framework implemented in \texttt{PyllutedWD} \citep{Harrison2018,Buchan2021}. This framework fits a number of geological models to the metal abundances inferred for a particular white dwarf and identifies the best one (i.e., that with highest Bayesian evidence\footnote{The best model is not necessarily the one which can fit the data best. The Bayesian evidence is a trade-off between goodness-of-fit and simplicity, i.e., fewer model parameters, and correspondingly fewer geological processes.}). The Bayesian framework can identify evidence for core--mantle differentiation and/or incomplete condensation and quantify the sigma significance of that evidence.

We have updated the framework such that the absolute abundance of photospheric metals is now set by the mass of the accreted body. The prior distribution corresponds to a collisional cascade, such as might be generated when rocky bodies encounter the Roche limit of a white dwarf \citep{Wyatt2014,KenyonBromley2017}. For an ideal collisional cascade, the number density of bodies in mass range $m$ to $m + \textrm{d}m$ is proportional to $m^{-11/6}\textrm{d}m$ \citep{Dohnanyi1969}. This distribution penalises high masses (and, by extension, high accretion rates). We set the upper and lower mass limits to $6.0\times10^{24}$\,kg (approximately the mass of Earth) and $10^{13.5}$\,kg respectively. These limits are arbitrary, but cover a broad enough range for the purposes of this work. Outside of this modification, and the changes described in Sections~\ref{sec:sinking_models} and \ref{sec:thermohaline_intro}, the framework is not significantly changed from the version described in \citet{Buchan2021}. 

\subsection{Sinking timescale grids}
\label{sec:sinking_models}

We consider various grids of sinking timescales based on two white dwarf modelling codes: the Koester envelope code \citep{Koester2020} and the STELUM evolution code in its static mode \citep{Bedard2022_STELUM}. In the latter case, the stellar structure is computed with STELUM itself, but the diffusion timescales are then computed with separate routines\footnote{The practical reason is that the current version of STELUM includes detailed equations of state only for H, He, C, and O.}, originally based on the work of \citet{Paquette1986a,Paquette1986b} with updates described in \citet{Fontaine2015a} and \citet{Heinonen2020}. We still refer to these grids as STELUM grids for conciseness. We briefly describe the similarities and differences between the Koester and STELUM frameworks which are relevant to this paper.

For the H/He envelope structure calculations, both codes rely on the equation of state of \citet{Saumon1995} and the OPAL radiative opacities of \citet{Iglesias1996}. For the outer boundary condition, the Koester code uses non-grey atmospheres \citep{Koester2010}; for He-rich models, where traces elements can dominate the overall opacity, the atmospheres optionally include a fixed H abundance and various metal abundances. The STELUM models used in this work adopt a more approximate treatment using grey atmospheres and, for He-rich models, OPAL envelope opacities with a fixed metallicity to roughly account for the effect of trace elements. Improvements to this aspect of the code are currently being made and will be reported in a future publication. Both the Koester and STELUM codes describe convection following the ML2 version of the mixing-length theory \citep{Tassoul1990}, but assume different mixing-length parameter values of 0.8 and 1.0, respectively. For the diffusion calculations, both frameworks rely on diffusion coefficients obtained from the screened Coulomb potential formalism \citep{Paquette1986a,Stanton2016}. However, they use different metal ionisation models, i.e. Thomas--Fermi for the Koester code \citep{Feynman1949} and Hummer--Mihalas for the STELUM framework \citep{Hummer1988}.

Of particular interest for this paper is the treatment of convective overshoot. In the Koester code, overshoot is included by simply increasing the size of the convectively mixed region by an arbitrary amount, expressed in terms of the pressure scale height. STELUM also has this option but generally adopts a more physically motivated description, based on the expectation that the velocity of the convective flows decreases exponentially with depth \citep{Freytag1996, Cunningham2019}. The mixing coefficient $D$ due to overshoot at radial depth $r$ can then be parametrised as:
\begin{equation}
D(r) = D_0 \exp \left( \frac{-2 |r-r_0|}{f H_0} \right),
\label{eq:overshoot}
\end{equation}
where $r_0$, $H_0$, and $D_0$ are the radial depth, pressure scale height, and mixing coefficient at the base of the convection zone, respectively, and $f$ is the free parameter controlling the extent of overshoot. The boundary of the overshoot region is then taken as the depth where the mixing coefficient becomes smaller than the atomic diffusion coefficient.

In this work, we use six grids of sinking timescales based on either the Koester or STELUM framework and various treatments of convective overshoot. These grids, and their abbreviated names used in the rest of the paper, are as follows.

\begin{table*}
    \centering
    \caption{Summary of the diffusion timescale grids used in this work. Atm refers to the dominant chemical element in the atmosphere of the white dwarf. The final row shows the approximate grid range for the thermohaline dilution factor $\chi$ described in Section~\ref{sec:thermohaline_intro}. Here, the boundaries are approximate because the MESA models are not defined by $T_{\rm eff}$/$\log(g)$.}
    \begin{tabular}{clccccccc}
              \toprule
         Grid&   Code&Atm&  Min $T_{\rm eff}$ /[K]&  Max $T_{\rm eff}$ /[K]&  Min $\log(g)$& Max $\log(g)$&  ML2/$\alpha$&  Overshoot\\
                  &   && & & & &  convection&  parametrisation\\
         \midrule
         \multirow{2}{*}{SN} &  \multirow{2}{*}{STELUM} &H&  5000&  30\,000&  7.5&  9.0&  1.0 &  None\\
        &  &He& 7000& 30\,000& 7.5& 9.0& 1.0 & None\\
         \multirow{2}{*}{SF}&   \multirow{2}{*}{STELUM}&H&  5000&  30\,000&  7.5&  9.0&  1.0 &  Fixed\\
 &  &He& 7000& 30\,000& 7.5& 9.0& 1.0 & Fixed\\
         \multirow{2}{*}{SV}&   \multirow{2}{*}{STELUM}&H&  5000&  30\,000&  7.5&  9.0&  1.0 &  Variable\\
 &  &He& 7000& 30\,000& 7.5& 9.0& 1.0 & Variable\\
         \multirow{2}{*}{S3D}&   \multirow{2}{*}{STELUM}&H&  11\,000&  20\,000&  7.5&  9.0&  1.0 &  3D\\
 &  &He& -& -& -& -& -& -\\
         \multirow{2}{*}{KN}&   \multirow{2}{*}{Koester}&H&  5000&  20\,000&  7.5&  8.5&  0.8 &  None
\\
 &  &He& 3000& 15\,000& 7.5& 8.5& 0.8 & None
\\
         \multirow{2}{*}{KF}&   \multirow{2}{*}{Koester}&H&  5000&  20\,000&  7.5&  8.5& 0.8 &  Fixed
\\
 &  &He& 3000& 15\,000& 7.5& 8.5& 0.8 & Fixed\\
 Thermohaline dilution factor $\chi$ & MESA &H& $\approx$6000& $\approx$20\,500& $\approx$7.5& $\approx$8.5& - & -\\
 \bottomrule
 %MWDD& H& 5000& 30\,000& 7.5& 9.0& & None&(3)\\
 %& He& 7000& 30\,000& 7.5& 9.0& & None&(3)\\
    \end{tabular}
    \label{tab:grid_summary}
\end{table*}

\begin{itemize}[leftmargin=*]
    \item Koester, without overshoot (KN) 
    \item Koester, with fixed overshoot of one pressure scale height (KF) 
    \item STELUM, without overshoot (SN) 
    \item STELUM, with fixed overshoot of one pressure scale height (SF) 
    \item STELUM, with overshoot predicted from equation\,(\ref{eq:overshoot}) with variable $f$ following \citet{Bedard2023} (SV) 
    \item STELUM, with overshoot directly calibrated from the 3D simulations of \citet{Cunningham2019} (S3D)
\end{itemize}

The KN and KF timescales are directly sourced from the tables presented in \citet{Koester2020} and publicly available online\footnote{\url{https://www1.astrophysik.uni-kiel.de/~koester/astrophysics/astrophysics.html}, accessed 2023 October 05}. For He-dominated white dwarfs, we use the `DBAZ' grids, which rely on metal-polluted atmosphere models and thus have the Ca abundance as an additional parameter ($-15 \leq \log \mathrm{Ca/He} \leq -7$, with other elements scaled to Ca assuming bulk Earth composition). KN ignores overshoot, while KF includes overshoot down to a fixed depth of one pressure scale height. This specific depth reflects the most conservative prediction available from \citet{Cunningham2019} across the simulated $T_{\rm eff}$ range.

We calculated the STELUM SN and SF grids as counterparts to the Koester KN and KF grids. We also calculated two additional STELUM grids using more realistic treatments of overshoot. In the SV grid, we use equation\,(\ref{eq:overshoot}) with the $f$ parameter varying according to the depth of the formal convection zone, as suggested in \citet{Bedard2023}. More specifically, let $q_0$ denote the fractional mass depth of the convection zone ($q \equiv 1-m_r/M_*$, where $m_r$ is the mass within radius $r$ and $M_*$ is the total mass). For H-rich models, we assume $f = 0.80$ if $\log q_0 \leq -16$, $f = 0.05$ if $\log q_0 \geq -10$, and a linear interpolation in between. For He-rich models, we assume $f = 0.80$ if $\log q_0 \leq -12$, $f = 0.05$ if $\log q_0 \geq -6$, and a linear interpolation in between. These parametrisations are such that the overshoot region is relatively large at high $T_{\rm eff}$ when the convection zone is shallow, and becomes smaller as $T_{\rm eff}$ decreases and the convection zone deepens (see Figure\,1 of \citealt{Bedard2023}). Our assumptions are designed to roughly match both 3D simulations of shallow convection \citep{Cunningham2019} and constraints on deep convection obtained from 1D modelling of C dredge-up in cool He-rich white dwarfs \citep{Bedard2022_DQ}. Finally, in the S3D grid, the depth of the overshoot region is directly taken from the 3D predictions of \citet{Cunningham2019}. This approach is the most reliable but is currently limited to warm ($11\,000\,\mathrm{K} \lesssim T_{\rm eff} \lesssim 18\,000\,\mathrm{K}$) H-dominated white dwarfs, hence the need for the other grids. Key information about our six timescale grids, including the range of temperature and surface gravity covered, is summarised in Table\,\ref{tab:grid_summary}.

We note that the physical assumptions of the SN grid are very similar to those underlying the sinking timescales that have been publicly available so far on the Montreal White Dwarf Database (MWDD; \citealt{MWDD})\footnote{\url{https://www.montrealwhitedwarfdatabase.org/evolution.html}}. Because the latter are based on an older version of STELUM \citep{Fontaine2015a,Fontaine2015b}, the two sets of timescales are not strictly identical, but relative timescales typically agree within a few percent. The largest differences are seen for He-dominated atmospheres due to the MWDD assuming a mixing-length parameter of 1.25 (rather than our value of 1), while for H-dominated atmospheres the differences are purely numerical and thus much smaller. Therefore, our comments on the SN grid in this work are also applicable to the MWDD timescales\footnote{This refers to the MWDD timescales prior to this work, as we plan to implement our new timescales on the MWDD in the near future; see data availability statement.}.

\subsubsection{Simultaneous use of the S3D and SV grids}
\label{sec:S3DSV}

The temperature range of the S3D grid is relatively narrow ($11\,000\,\mathrm{K} \leq T_{\rm eff} \leq 20\,000\,\mathrm{K}$) compared to the other STELUM grids for H-dominated white dwarfs ($5\,000\,\mathrm{K} \leq T_{\rm eff} \leq 30\,000\,\mathrm{K}$). For the purposes of exploring the parameter space fully, when using the S3D grid we fill out the full temperature range by using values from the SV grid when $5\,000\,\mathrm{K} \leq T_{\rm eff} \leq 11\,000\,\mathrm{K}$ or $20\,000\,\mathrm{K} \leq T_{\rm eff} \leq 30\,000\,\mathrm{K}$. This represents the current fullest possible treatment of overshoot for H-dominated systems.

Throughout the rest of this work, unless otherwise noted, the S3D grid should be taken to mean the extended version of the S3D grid which includes values from the SV grid.

\subsection{Thermohaline mixing}
\label{sec:thermohaline_intro}

We implement a simplified treatment of thermohaline mixing for H-dominated white dwarfs in our forward model following \citet{Bauer2018,Bauer2019}. For a given accretion rate $\dot M$, effective temperature $T_{\rm eff}$ and surface gravity $\log(g)$, we interpolate on a grid of MESA model outputs\footnote{Available at \url{https://github.com/evbauer/DA_Pollution_Tables}} which quantify the strength of thermohaline mixing in terms of a factor which we denote $\chi$. This factor is defined by the equation
\begin{equation}
    \chi = \log_{10}\left(\frac{S^{\rm +}_{\rm Ca}}{S^{\rm -}_{\rm Ca}}\right),
\end{equation} where $S_{\rm Ca}$ is the surface abundance of Ca and is calculated both with and without thermohaline mixing (labelled $+$ and $-$ respectively). Since thermohaline mixing causes more rapid transport of metals away from the surface, we generally have $S^{\rm +}_{\rm Ca}<S^{\rm -}_{\rm Ca}$ and therefore $\chi < 0$. Because of this behaviour, $\chi$ can be thought of as a `dilution factor' for the accreted mass present in the photosphere.

When thermohaline mixing is switched on in our code, we scale all metal abundances by a factor of $\chi$. Additionally, if $\chi < \log_{10}\left(0.5\right)$, we consider that thermohaline mixing is strong enough to effectively disable differential sinking \citep{Bauer2018}. In this case, the relative photospheric metal abundances are fixed to equal those of the pollutant. This differs from the default, non-thermohaline, treatment because it removes the weighting associated with individual metal sinking timescales. 

If the system has reached the declining phase of accretion, such that the time of observation $t$ is greater than the time when accretion ends $t_{\rm event}$, the above adjustments are made at time $t_{\rm event}$. We then divide the metal abundances by a correction factor equal to $\left(t - t_{\rm event}\right)/\tau_{X}$ for each element X where $\tau_{\rm X}$ is its sinking timescale. This ensures that elements decay away in the declining phase as expected following the end of the accretion event, using the adjusted abundances as a starting point.

We implement additional logic to handle parameter values beyond the edges of the grid outlined in Table~\ref{tab:grid_summary}. Extremely high accretion rates and high temperatures are snapped to the nearest grid edge. For low accretion rates, we set $\chi = 0$ at $\dot M = 1 \textrm{g}\,\textrm{s}^{-1}$ and linearly interpolate between this boundary condition and the lowest  $\dot M$ value in our grid. This forces thermohaline mixing to become inactive for sufficiently low $\dot M$, even for values of temperature and surface gravity which always have thermohaline mixing active within the grid boundaries. For very low temperatures, we return $\chi = 0$; i.e., we assume thermohaline mixing is never active when $T_{\rm eff}\lesssim6000\,\textrm{K}$. This boundary is set by the edge of the MESA grid, although we expect thermohaline mixing to be less relevant at low temperatures in any case since very high accretion rates are required to trigger it. We have no particular logic for extreme values of $\log(g)$, but the grid range is broad in this dimension (covering roughly 7.5 to 8.5 in cgs units) so we simply restrict our attention to white dwarfs falling into this range.

\subsection{Test cases}
\label{sec:test_definitions}

We define five test cases which form the core of this work, numbered from Test~1 to Test~5, which are summarised in Table~\ref{tab:criteria}. Tests~1 and 2 investigate convective overshoot, Test~3 investigates thermohaline mixing, and Tests~4 and 5 relate to publicly available timescale grids.

The tests are primarily defined by a pair of timescale grids $a$ and $b$, which specify the grids to be used in our Bayesian modelling. The exception is Test~3, which investigates the effect of switching on thermohaline mixing when using the full treatment of 3D convective overshoot in H-dominated atmospheres. In this test, we keep the timescale grid fixed (so there is no grid $b$), but alter whether or not thermohaline mixing is active.
%We do not consider thermohaline mixing for white dwarfs with He-dominated atmospheres because for such white dwarfs, its effect is predicted to be weak \citep{Deal2013,Bauer2019}.

\begin{table*}
    \centering
\caption{Summary of tests and white dwarf sample selection criteria. For Test~1, Grid $b$ should be taken to mean the unexpanded version of the S3D grid (see Section~\ref{sec:S3DSV}).}
\label{tab:criteria}
    \begin{tabular}{ccccccc} 
    \toprule
          Test number&Test type&Atmosphere&  Grid $a$&  Grid $b$&  Min. elements& Min. metric\\
          \midrule
          1&Overshoot&  H&  SN&  S3D &  2& 1\\ 
          2&Overshoot&  He&  SN&  SV&  2& 1\\
          3&Thermohaline&  H&  S3D&  -&  2& 1\\
          4&Public grids&  H&  SN&  KN&  2& 1\\ 
          5&Public grids&  He&  SN&  KN&  4& 2\\
          \bottomrule
    \end{tabular}

\end{table*}

\subsection{Sample selection}
\label{sec:sample_selection}

We select a manageable sample of white dwarfs to model for each test scenario. We select all our samples from the Planetary Enriched White Dwarf Database (PEWDD\footnote{\url{https://github.com/jamietwilliams/PEWDD}, commit 3b905e0, accessed 2024 December 12}, \citealt{Williams2024}), using a number of criteria to identify appropriate white dwarfs which are most likely to be affected in each test case.

Firstly, we consider H- and He-dominated white dwarfs separately, so we filter by the appropriate atmospheric type. Secondly, we filter out any white dwarfs which fall outside the joint $T_{\rm eff}$ and $\log(g)$ range covered by both sinking timescale grids under consideration (i.e., we do not perform any timescale extrapolation). Thirdly, we impose a minimum number of `modellable elements'. This refers to the number of the following elements which were detected, and which have valid abundances and uncertainties: Al, Ti, Ca, Ni, Fe, Cr, Mg, Si, Na, O, C, N. These are the elements which can be modelled by \texttt{PyllutedWD}. Finally, for a given white dwarf, and a given pair of sinking timescale grids $a$ and $b$, we calculate $M_{a,b}$, a metric measuring the likely impact of swapping $a$ for $b$. For details, see the Appendix. We impose a minimum value of this metric. Table~\ref{tab:criteria} includes the full set of criteria used to select each sample. The resulting samples may be found in Table~\ref{tab:sample}.

For Test~1 only, the second grid ($b$) used in sample selection is the unmodified S3D grid, without the expanded temperature range (see Section~\ref{sec:S3DSV}). This is because, in this test, we wish to identify white dwarfs which can be modelled using the full 3D treatment of convective overshoot.

For Test~3, there is no grid $b$, so the selection criteria require some modification. This test concerns thermohaline mixing, the effect of which is to effectively disable differential sinking. We capture this effect by replacing $b$ with a dummy grid in which sinking timescales are the same for all elements. We also impose the constraint that \SI{10000}{K}$ \le \textrm{T}_{\rm eff} \le $\SI{20502}{K}. The upper limit is the highest temperature covered by the thermohaline mixing grid (although the shape of this grid is irregular). The lower limit approximates the temperature at which the impact of thermohaline mixing is expected to become small.

For Test~5, we increased the minimum required number of modellable elements and the minimum metric score in order to keep the sample size practical.

\subsubsection{White dwarf atmospheric parameters}

By default, we adopt the values of $T_{\rm eff}$ and $\log(g)$ as given in PEWDD from different literature sources, data and fitting techniques. However, in many cases, the values of these parameters are not based on state-of-the-art space-based photometry and astrometry \citep{Gaia2023}, and several entries assume $\log(g)=8.0$. Therefore, we quote an additional set of parameters in Table~\ref{tab:sample}, using the `mixed' atmosphere model values from \citet{GentileFusillo2021} for He-dominated white dwarfs, and `pure-H' values for H-dominated white dwarfs. If the $T_{\rm eff}$ values differ by more than 2\%, or the $\log(g)$ values differ by more than 0.04 dex, adopted as characteristic uncertainties of current white dwarf model atmospheres and photometric flux calibration \citep{Elms2024}, we run our model using both sets of parameter values, and comment on any impact where relevant. Atmospheric metal abundance estimates depend on the values of $T_{\rm eff}$ and $\log(g)$, so in general the \textit{Gaia} atmospheric parameters will not be fully consistent with the metal abundances.

\subsubsection{Corrections and assumptions for specific white dwarfs}

GD133 has no formal error on Mg, but \citet{Xu2014} adopt a value of 0.2 dex in their analysis, so we do likewise. \citet{Klein2011} give no error on the Ni abundance for HS2253+8023, so we assume a nominal value of 0.2 dex. At the time of writing, PEWDD contains a typographical error inherited from \citet{Steele2021}, in which an upper bound on the Ni abundance of WD\;1124$-$293 is listed as an abundance with no error - we correct for this issue here.

\afterpage{
\newgeometry{bottom=1.2cm,top=1.2cm,left=1.4cm,right=1.4cm}
\onecolumn
\setlength{\LTcapwidth}{1.2\textwidth}
\setlength{\tabcolsep}{3.5pt}
\footnotesize
\begin{landscape}
%\begin{longtable}{C{3.3cm}C{1.0cm}C{6.3cm}{0.5cm}C{1cm}C{1cm}C{1cm}C{1cm}C{1cm}C{1cm}C{1cm}C{1cm}C{0.5cm}C{0.5cm}C{0.5cm}}
\begin{longtable}{cC{1cm}ccC{1cm}C{1cm}C{1cm}C{1cm}C{1cm}C{1cm}C{1cm}C{1cm}C{0.5cm}C{1cm}c}
\caption{\label{tab:sample}Summary of the 5 samples of white dwarfs selected from PEWDD using the criteria outlined in Table~\ref{tab:criteria}. ME is the number of modellable elements. MS is the metric score. Sources: (1) \citet{Rogers2024}, (2) \citet{Xu2014}, (3) \citet{Xu2019}, (4) \citet{Limbach2024}, (5) \citet{Klein2011}, (6) \citet{Hollands2022}, (7) \citet{Hoskin2020}, (8) \citet{Farihi2011}, (9) \citet{Izquierdo2020}, (10) \citet{Blouin2018}, (11) \citet{Melis2017}, (12) \citet{Gaensicke2012}, (13) \citet{Kawka2012}, (14) \citet{Zuckerman2011}, (15) \citet{OBrien2023}, (16) \citet{Kawka2019}, (17) \citet{Steele2021}, (18) \citet{Swan2019}, (19) \citet{VennesKawka2013}, (20) \citet{Tremblay2020}, (21) \citet{Klein2021}, (22) \citet{GentileFusillo2017}, (23) \citet{Raddi2015}, (24) \citet{Swan2023a}, (25) \citet{Xu2017}, (26) \citet{Koester2000}, (27) \citet{BadenasAgusti2024}, (28) \citet{Doyle2023}.}\\
\toprule
           Name&  PEWDD ID&  PEWDD Identifier&  Atm&  PEWDD Teff /[K]&  PEWDD Teff err /[K]&   PEWDD log(g)& PEWDD log(g) err& Gaia Teff /[K]& Gaia Teff err /[K]& Gaia log(g)& Gaia log(g) err& ME&MS&  Source\\
\bottomrule
\endfirsthead
\toprule
           Name&  PEWDD ID&  PEWDD Identifier&  Atm&  PEWDD Teff /[K]&  PEWDD Teff err /[K]&   PEWDD log(g)& PEWDD log(g) err& Gaia Teff /[K]& Gaia Teff err /[K]& Gaia log(g)& Gaia log(g) err& ME&MS&  Source\\
\bottomrule
\endhead
\midrule
\multicolumn{15}{c}{Test~1 (7 entries, 5 unique white dwarfs)}\\
                          \midrule
         Gaia J0611$-$6931&  263&  Gaia J0611$-$6931 (Phot; Opt) Rogers 2023&  H&  16\,530&  561&   7.81& 0.03& 16\,147& 550& 7.86& 0.05& 4&1.5&  (1)\\
         Gaia J0611$-$6931&  264&  Gaia J0611$-$6931 (Phot; UV) Rogers 2023&  H&  16\,530&  561&   7.81& 0.03& 16\,147& 550& 7.86& 0.05& 6&1.0&  (1)\\
         GD133&  1037&  GD133 Xu 2014&  H&  12\,600&  200&   8.10& 0.10& 12\,157& 153& 8.03& 0.02& 4&1.3&  (2)\\
         Gaia J0611$-$6931&  1216&  Gaia J0611$-$6931 (Spec; Opt) Rogers 2023&  H&  17\,749&  248&   8.14& 0.04& 16\,147& 550& 7.86& 0.05& 4&1.1&  (1)\\
         GD 56&  1792&  GD 56 Xu 2019&  H&  15\,270&  300&   8.09& 0.10& 14\,904& 303& 8.00& 0.03& 4&1.8&  (3)\\
         HE 0106$-$3253&  1793&  HE 0106$-$3253 Xu 2019&  H&  17\,350&  200&   8.12& 0.10& 16\,053& 313& 8.02& 0.03& 4&1.9&  (3)\\
         WD 0310$-$688&  3543&  WD 0310$-$688 Limbach 2024&  H&  15\,865&  263&   8.08& 0.02& 15\,865& 263& 8.08& 0.02& 3&1.8&  (4)\\
         &  &  &  &  &  &   & & & & & & &&  \\
                          \midrule
         \multicolumn{15}{c}{Test~2 (12 entries, 9 unique white dwarfs)}\\
                          \midrule
         PG1225$-$079&  160&  PG1225$-$079 Model 1 Klein 2011&  He&  10\,500&  &   7.70& & 11\,248& 141& 8.13& 0.02& 6&1.1&  (5)\\
         Gaia J0644$-$0352&  583&  Gaia J0644$-$0352 (Phot; Opt) Rogers 2023&  He&  17\,000&  327&   7.98& 0.02& 17\,587& 761& 8.01& 0.07& 8&1.3&  (1)\\
         SDSS J0956+5912&  649&  SDSS J0956+5912 (No Overshoot) Hollands 2022&  He&  8\,100&  100&   8.02& 0.04& 9\,053& 458& 8.06& 0.14& 8&1.5&  (6)\\
         SDSS J0956+5912&  650&  SDSS J0956+5912 (Overshoot) Hollands 2022&  He&  8\,100&  100&   8.02& 0.04& 9\,053& 458& 8.06& 0.14& 8&1.5&  (6)\\
         WD J2047$-$1259&  809&  WD J2047$-$1259 Hoskin 2020&  He&  17\,970&  170&   8.04& 0.05& 17\,802& 660& 8.12& 0.06& 7&2.0&  (7)\\
         GD 61&  1274&  GD 61 Farihi 2011&  He&  17\,820&  &   8.20& & 16\,726& 418& 8.10& 0.04& 5&1.0&  (8)\\
         GD 424&  1389&  GD 424 (WHT) Izquierdo 2021&  He&  16\,560&  75&   8.25& 0.02& 16\,326& 525& 8.21& 0.05& 4&2.8&  (9)\\
         GD 424&  1390&  GD 424 (Keck) Izquierdo 2021&  He&  16\,560&  75&   8.25& 0.02& 16\,326& 525& 8.21& 0.05& 9&1.3&  (9)\\
         HS2253+8023&  1461&  HS2253+8023 Model 2 Klein 2011&  He&  14\,400&  &   8.40& & 13\,587& 238& 8.16& 0.03& 8&1.5&  (5)\\
         HS2253+8023&  1543&  HS2253+8023 Model 3 Klein 2011&  He&  14\,800&  &   8.70& & 13\,587& 238& 8.16& 0.03& 7&2.7&  (5)\\
         Ross 640&  1832&  Ross 640 Blouin 2018&  He&  8\,070&  140&   7.92& 0.01& 8\,964& 78& 8.08& 0.02& 4&1.0&  (10)\\
         WD 1232+563&  3519&  WD 1232+563 Xu 2019&  He&  11\,787&  423&   8.30& 0.06& 12\,110& 725& 8.05& 0.13& 7&1.8&  (3)\\
         &  &  &  &  &  &   & & & & & & &&  \\
                          \midrule
         \multicolumn{15}{c}{Test~3 (13 entries, 8 unique white dwarfs)}\\
                          \midrule
         Gaia J0611$-$6931&  263&  Gaia J0611$-$6931 (Phot; Opt) Rogers 2023&  H&  16\,530&  561&   7.81& 0.03& 16\,147& 550& 7.86& 0.05& 4&1.2&  (1)\\
         Gaia J0611$-$6931&  264&  Gaia J0611$-$6931 (Phot; UV) Rogers 2023&  H&  16\,530&  561&   7.81& 0.03& 16\,147& 550& 7.86& 0.05& 6&2.5&  (1)\\
         SDSS J1043+0855&  877&  SDSS J1043+0855 Melis \& Dufour 2017&  H&  18\,330&  &   8.05& & 16\,067& 997& 7.96& 0.10& 8&1.0&  (11)\\
         GD133&  1037&  GD133 Xu 2014&  H&  12\,600&  200&   8.10& 0.10& 12\,157& 153& 8.03& 0.02& 4&1.4&  (2)\\
         Gaia J0510+2315&  1214&  Gaia J0510+2315 (Phot Opt) Rogers 2023&  H&  20\,130&  145&   8.13& 0.02& 21\,255& 457& 8.18& 0.03& 4&2.0&  (1)\\
         Gaia J0510+2315&  1215&  Gaia J0510+2315 (Phot UV) Rogers 2023&  H&  20\,130&  145&   8.13& 0.02& 21\,255& 457& 8.18& 0.03& 3&1.5&  (1)\\
         Gaia J0611$-$6931&  1216&  Gaia J0611$-$6931 (Spec; Opt) Rogers 2023&  H&  17\,749&  248&   8.14& 0.04& 16\,147& 550& 7.86& 0.05& 4&1.1&  (1)\\
         Gaia J0611$-$6931&  1217&  Gaia J0611$-$6931 (Spec; UV) Rogers 2023&  H&  17\,749&  248&   8.14& 0.04& 16\,147& 550& 7.86& 0.05& 6&2.6&  (1)\\
         G29-38&  1468&  G29-38 Xu 2014&  H&  11\,820&  100&   8.40& 0.10& 11\,528& 206& 8.04& 0.03& 8&1.6&  (2)\\
         PG 1015+161&  1676&  PG 1015+161 Gansicke 2012&  H&  19\,200&  180&   8.22& 0.06& 18\,377& 473& 7.97& 0.04& 5&1.1&  (12)\\
         GD 56&  1792&  GD 56 Xu 2019&  H&  15\,270&  300&   8.09& 0.10& 14\,904& 303& 8.00& 0.03& 4&1.1&  (3)\\
         HE 0106$-$3253&  1793&  HE 0106$-$3253 Xu 2019&  H&  17\,350&  200&   8.12& 0.10& 16\,053& 313& 8.02& 0.03& 4&3.6&  (3)\\
         PG 1015+161&  1794&  PG 1015+161 Xu 2019&  H&  20\,420&  350&   8.11& 0.10& 18\,377& 473& 7.97& 0.04& 4&1.3&  (3)\\
         &  &  &  &  &  &   & & & & & & &&  \\
                          \midrule
         \multicolumn{15}{c}{Test~4 (17 entries, 15 unique white dwarfs)}\\
                          \midrule
         NLTT 1675&  22&  NLTT 1675 Kawka 2012&  H&  6\,020&  50&   8.04& 0.07& 6\,022& 91& 8.05& 0.05& 4&1.3&  (13)\\
         NLTT 43806&  27&  NLTT 43806 Zuckerman 2011&  H&  5\,900&  &   8.00& & 5\,808& 51& 8.17& 0.03& 9&2.3&  (14)\\
         WD J1935$-$3252&  30&  WD J1935$-$3252 O'Brien 2023&  H&  5\,430&  10&   8.00& 0.01& 5\,305& 47& 7.97& 0.04& 4&6.8&  (15)\\
         NLTT 7547&  34&  NLTT 7547 Spectral Kawka 2019&  H&  5\,198&  54&   7.89& 0.04& 5\,311& 100& 7.97& 0.07& 5&2.7&  (16)\\
         NLTT 7547&  35&  NLTT 7547 (Balmer) Kawka 2019&  H&  5\,460&  80&   8.04& 0.18& 5\,311& 100& 7.97& 0.07& 5&1.4&  (16)\\
         NLTT 7547&  36&  NLTT 7547 Balmer Kawka 2019&  H&  5\,460&  80&   8.04& 0.18& 5\,311& 100& 7.97& 0.07& 5&1.4&  (16)\\
         WD 1124$-$293&  522&  WD 1124$-$293 Steele 2021&  H&  9\,367&  &   7.99& & 9\,401& 82& 7.99& 0.02& 3&1.1&  (17)\\
         WD2115$-$560&  546&  WD2115$-$560 Swan 2019&  H&  9\,600&  340&   7.97& 0.08& 9\,528& 76& 7.96& 0.02& 5&1.1&  (18)\\
         NLTT 25792&  810&  NLTT 25792 Vennes \& Kawka 2013&  H&  7\,903&  16&   8.04& 0.03& 7\,837& 112& 8.10& 0.04& 4&1.8&  (19)\\
         WD2157$-$574&  889&  WD2157$-$574 Swan 2019&  H&  7\,010&  250&   8.06& 0.08& 7\,114& 55& 8.10& 0.02& 6&1.6&  (18)\\
         GD133&  1037&  GD133 Xu 2014&  H&  12\,600&  200&   8.10& 0.10& 12\,157& 153& 8.03& 0.02& 4&1.0&  (2)\\
         WD0354+463&  1618&  WD0354+463 Vennes \& Kawka 2013&  H&  8\,240&  120&   7.96& 0.10& & & & & 4&2.6&  (19)\\
         WDJ113444.64+610826.68&  1783&  WDJ113444.64+610826.68 Tremblay 2020&  H&  7\,590&  40&   7.96& 0.02& 7\,502& 59& 7.97& 0.02& 2&1.6&  (20)\\
         WD1455+298&  1784&  WD1455+298 Vennes \& Kawka 2013&  H&  7\,383&  19&   7.97& 0.03& 7\,342& 53& 7.97& 0.02& 3&2.1&  (19)\\
         G74-7&  1785&  G74-7 Vennes \& Kawka 2013&  H&  7\,306&  22&   8.06& 0.02& 7\,196& 48& 7.93& 0.02& 4&3.1&  (19)\\
         WD1257+278&  1786&  WD1257+278 Vennes \& Kawka 2013&  H&  8\,609&  20&   8.24& 0.02& 8\,586& 71& 8.04& 0.02& 4&1.5&  (19)\\
         G 166-58&  1790&  G 166-58 Xu 2019&  H&  7\,390&  200&   7.99& 0.10& 7\,342& 53& 7.97& 0.02& 4&2.6&  (3)\\
         &  &  &  &  &  &   & & & & & & &&  \\
                          \midrule
         \multicolumn{15}{c}{Test~5 (25 entries, 19 unique white dwarfs)}\\
                          \midrule
         PG1225$-$079&  160&  PG1225$-$079 Model 1 Klein 2011&  He&  10\,500&  &   7.70& & 11\,248& 141& 8.13& 0.02& 6&2.8&  (5)\\
         GALEXJ2339&  367&  GALEXJ2339 Klein 2021&  He&  13\,735&  500&   7.93& 0.09& 14\,796& 366& 8.08& 0.05& 7&7.3&  (21)\\
         GD 17&  540&  GD 17 Gentile Fusillo 2017&  He&  8\,300&  500&   8.00& & 9\,150& 158& 8.11& 0.04& 5&5.3&  (22)\\
         PG1225$-$079&  582&  PG1225$-$079 Model 2 Klein 2011&  He&  10\,800&  &   8.00& & 11\,248& 141& 8.13& 0.02& 6&2.7&  (5)\\
         GD 16&  594&  GD 16 Gentile Fusillo 2017&  He&  11\,000&  500&   8.00& & 11\,207& 166& 8.12& 0.03& 4&3.0&  (22)\\
         SDSS J0956+5912&  649&  SDSS J0956+5912 (No Overshoot) Hollands 2022&  He&  8\,100&  100&   8.02& 0.04& 9\,053& 458& 8.06& 0.14& 8&9.1&  (6)\\
         SDSS J0956+5912&  650&  SDSS J0956+5912 (Overshoot) Hollands 2022&  He&  8\,100&  100&   8.02& 0.04& 9\,053& 458& 8.06& 0.14& 8&9.1&  (6)\\
         SDSS J124231.07+522626.6&  652&  SDSS J124231.07+522626.6 Raddi 2015&  He&  13\,000&  300&   8.00& & 12\,571& 611& 8.08& 0.11& 8&3.3&  (23)\\
         WD1350$-$162&  724&  WD1350$-$162 Swan 2019&  He&  11\,640&  290&   8.02& 0.05& 11\,892& 560& 7.99& 0.11& 8&2.5&  (18)\\
         SDSS J1038$-$0036&  781&  SDSS J1038$-$0036 Hollands 2022&  He&  7\,560&  150&   8.06& 0.04& 8\,088& 180& 8.05& 0.06& 7&3.4&  (6)\\
         WD2216$-$657&  785&  WD2216$-$657 Swan 2019&  He&  9\,190&  210&   8.05& 0.05& 9\,821& 101& 8.07& 0.02& 4&4.2&  (18)\\
         J0939&  854&  J0939 Swan 2023&  He&  9\,020&  200&   8.07& 0.05& 9\,649& 153& 8.09& 0.04& 6&2.2&  (24)\\
         J1227&  855&  J1227 Swan 2023&  He&  7\,420&  80&   8.07& 0.04& 7\,947& 295& 8.07& 0.11& 6&2.4&  (24)\\
         HS2253+8023&  989&  HS2253+8023 Model 1 Klein 2011&  He&  14\,000&  &   8.10& & 13\,587& 238& 8.16& 0.03& 8&2.5&  (5)\\
         J0956&  1061&  J0956 Swan 2023&  He&  8\,720&  100&   8.13& 0.05& 9\,053& 458& 8.06& 0.14& 9&3.6&  (24)\\
         PG1225$-$079&  1404&  PG1225$-$079 Model 3 Klein 2011&  He&  11\,100&  &   8.30& & 11\,248& 141& 8.13& 0.02& 6&3.2&  (5)\\
         HS2253+8023&  1461&  HS2253+8023 Model 2 Klein 2011&  He&  14\,400&  &   8.40& & 13\,587& 238& 8.16& 0.03& 8&2.6&  (5)\\
         WD 1425+540&  1591&  WD 1425+540 (Model 1) Xu 2017&  He&  14\,490&  &   7.95& & 14\,549& 233& 8.07& 0.03& 8&3.1&  (25)\\
         WD 1425+540&  1592&  WD 1425+540 (Model 2) Xu 2017&  He&  14\,490&  &   7.95& & 14\,549& 233& 8.07& 0.03& 8&3.6&  (25)\\
         L 745-46A&  1606&  L 745-46A Koester \& Wolff 2000&  He&  7\,500&  200&   8.00& 0.20& 7\,650& 53& 7.98& 0.02& 4&3.5&  (26)\\
         WD 1232+563&  1828&  WD 1232+563 Badenas-Agusti 2024&  He&  11\,778&  86&   8.24& 0.06& 12\,110& 725& 8.05& 0.13& 6&2.6&  (27)\\
         Ross 640&  1832&  Ross 640 Blouin 2018&  He&  8\,070&  140&   7.92& 0.01& 8\,964& 78& 8.08& 0.02& 4&7.9&  (10)\\
         Gaia J0218+3625&  2315&  Gaia J0218+3625 Doyle 2023&  He&  14\,700&  500&   7.86& 0.05& 15\,848& 652& 7.94& 0.08& 9&2.1&  (28)\\
         WDJ183352.68+321757.25&  3508&  WDJ183352.68+321757.25 Tremblay 2020&  He&  7\,650&  60&   8.05& 0.02& 7\,743& 156& 7.93& 0.05& 5&11.3&  (20)\\
         WD 1232+563&  3519&  WD 1232+563 Xu 2019&  He&  11\,787&  423&   8.30& 0.06& 12\,110& 725& 8.05& 0.13& 7&3.9&  (3)\\
\end{longtable}
\end{landscape}
\normalsize
\restoregeometry
\twocolumn
}

\section{Convective Overshoot}
\label{sec:overshoot}

This section describes the results of Tests~1 and 2. These tests use only timescales calculated with STELUM, but with different treatments of convective overshoot. We consider H- and He-dominated white dwarfs separately.

\subsection{H-dominated white dwarfs (Test~1)}

\begin{figure}
    \centering
    \includegraphics[width=\columnwidth, keepaspectratio]{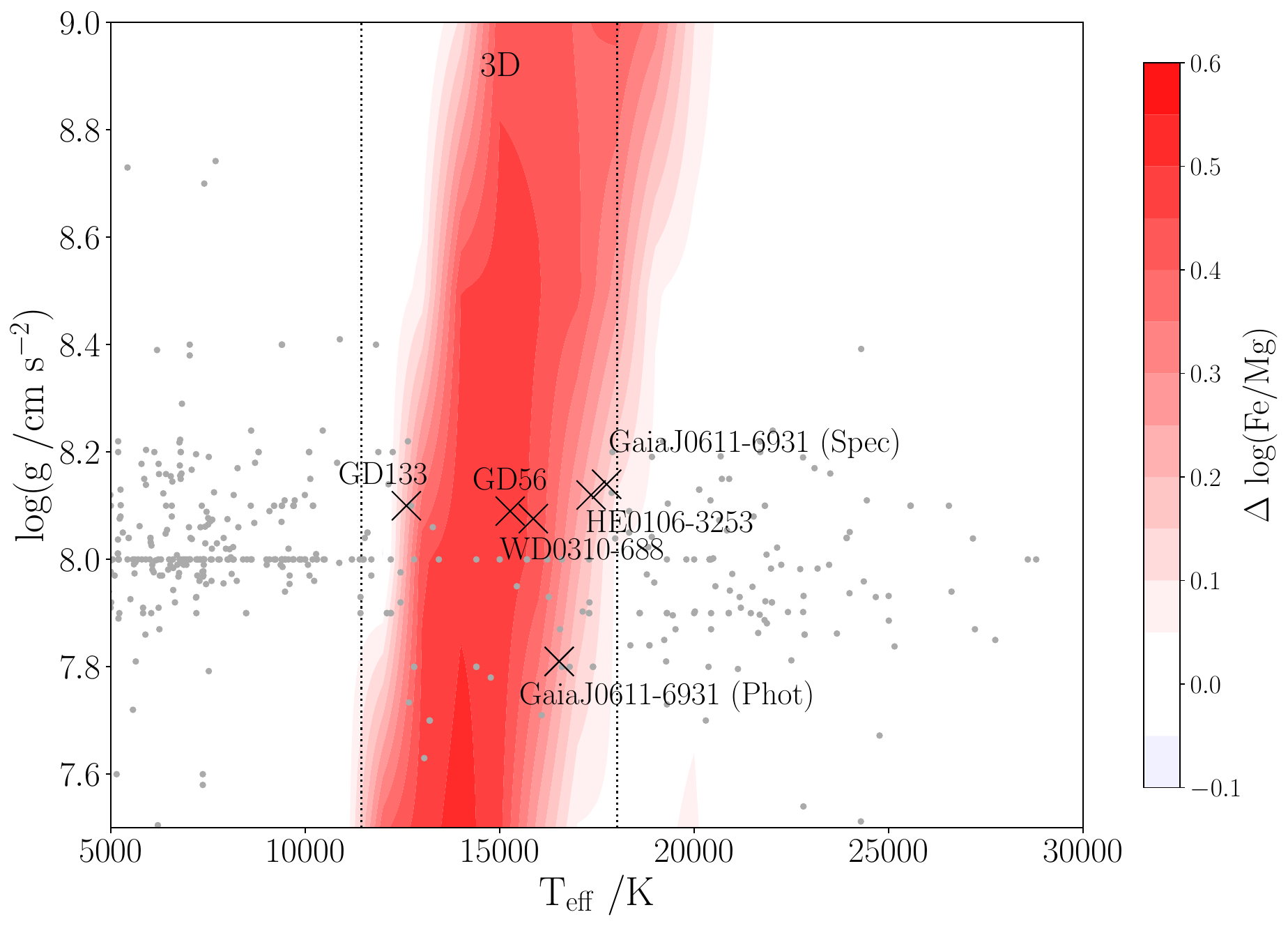}
    \caption{The change in relative sinking timescales for H-dominated white dwarfs when including 3D convective overshoot, compared to no overshoot, as a function of effective temperature ($x$ axis) and surface gravity ($y$ axis). The colour scale shows the change in the Fe/Mg sinking timescale ratio, on a log scale (base 10). Under the assumption of steady state accretion, the change in this ratio equals the change in the inferred Fe/Mg abundance ratio in the pollutant. The ratio is therefore significant when its absolute value is greater than the approximate uncertainty of metal abundances, typically 0.1 to 0.2 dex. The crosses indicate our sample of H-dominated white dwarfs which were modelled individually with these different treatments of overshoot (Test~1). The grey dots show all other H-dominated white dwarfs in PEWDD. Points may overlap when an individual white dwarf has multiple entries in PEWDD. For example, the Gaia\,J0611$-$6931 (Phot) point corresponds to two entries (both listed explicitly in Table~\ref{tab:sample}). The vertical dashed lines indicate the temperature range where convective overshoot has been explored with 3D simulations (as in the S3D grid). Outside of this range, we use the SV grid.}
    \label{fig:da_overshoot_timescales}
\end{figure}

Test~1 investigates the effect of including a full 3D treatment of convective overshoot, compared to no convective overshoot at all. We select a sample of white dwarfs to model with and without overshoot following the criteria outlined in Section~\ref{sec:sample_selection}. There are 112 entries in PEWDD for H-dominated white dwarfs with at least 2 modellable elements (defined in Section~\ref{sec:sample_selection}). Our selection criteria narrow this down to a sample of 7 entries, corresponding to 5 individual white dwarfs (note that white dwarfs may have multiple entries in PEWDD due to analysis with different data and/or methodology). These 7 entries, by construction, should be the most likely to be affected by the change in (relative) metal sinking times associated with switching on convective overshoot. However, it is not guaranteed that the geological interpretation will actually be affected: these are simply the strongest candidates according to our metric.

By construction, the sample falls within a fairly well-defined temperature range between about \SI{12000}{K} and \SI{18000}{K}, which is the range covered by 3D simulations \citep{Cunningham2019}. Within this range, we predict a significant discrepancy in relative sinking timescales, as shown in Figure~\ref{fig:da_overshoot_timescales}. This figure only shows the change in the Fe/Mg sinking timescale ratio, but all ratios show qualitatively similar behaviour (see Appendix).

Of the 7 entries in our sample, 2 are qualitatively affected by the inclusion of overshoot: GD\,56 and WD\,0310$-$688. However, the statistical significance of the effect is low in both cases.

\begin{figure}
    \centering
    \includegraphics[width=\columnwidth, keepaspectratio]{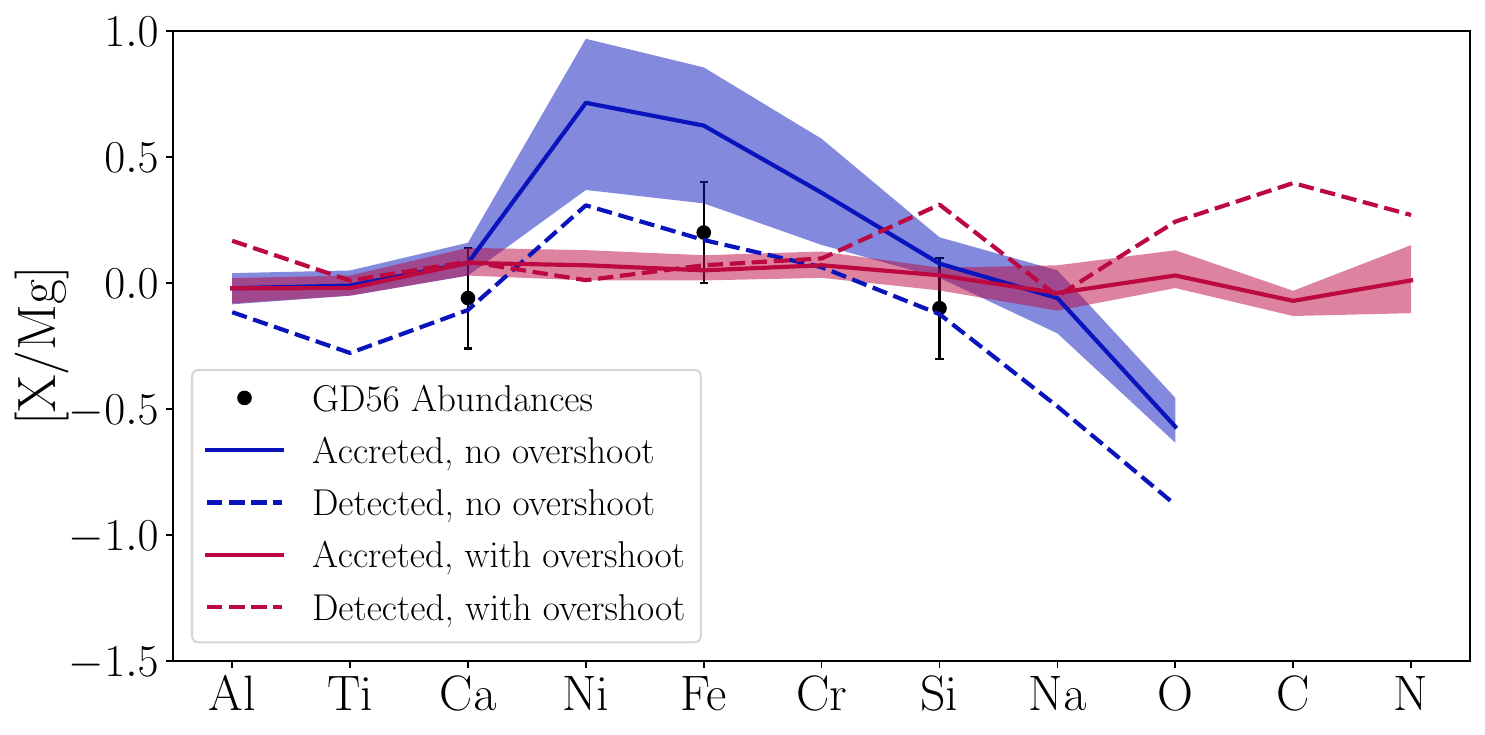}
    \caption{The geological interpretation of the metals accreted by GD\,56 changes when convective overshoot is taken into account. The black data points, with 1 sigma error bars, show the abundances of metals in the photosphere of GD\,56 on a log scale relative to Mg, and scaled to solar. The solid blue line shows the inferred composition of the accreted material when modelling the data without convective overshoot. The shaded blue region is a 1 sigma confidence interval. Our Bayesian framework finds (weak) evidence for core-mantle differentiation, with the best model invoking a core-rich composition. This composition should not be directly compared to the data points, because differential sinking causes the accreted and detected compositions to differ. The dashed blue line accounts for this correction, and may be compared with the data. For visual clarity, the 1 sigma confidence interval is omitted in this case. The red lines show the equivalent results when including convective overshoot using results from detailed 3D simulations. When taking this effect into account, the best model does not find any evidence of core-mantle differentiation. The compositions without overshoot (blue lines) predict negligible quantities of C and N, so these are not shown.}
    \label{fig:GD56}
\end{figure}

GD\,56 serves as a useful case study. We use abundance data from \citet{Xu2019}, who did not find evidence of any particular geological process for this white dwarf. Our results support this non-detection, but show that it could have been affected by overshoot. We find that, without overshoot, GD\,56 is best explained as a core-rich fragment ($41^{+14}_{-13}\%$ by moles\footnote{Here, core-rich should be understood as relative to the parent body, whose molar core fraction is estimated to be 12\%.}) of material from a body which experienced incomplete condensation at $T=358^{+474}_{-209}$\;K. In contrast, when overshoot is taken into account, the best model is primitive material: there is no evidence for incomplete condensation, or for selective accretion of core-rich material. The difference between these 2 solutions is shown in Figure~\ref{fig:GD56}. The red and blue dashed lines show the (median) inferred composition of the accreted material with and without overshoot respectively. With no overshoot, a noticeable enrichment in siderophile elements (those found in the core, such as Fe, Ni and Cr) is inferred, relative to solar abundances. The enrichment is absent when including overshoot. This can be understood as a consequence of Fe sinking more slowly (relative to Mg) in this case (see Figure~\ref{fig:da_overshoot_timescales}), such that a high initial abundance of Fe is no longer needed in order to explain its observed abundance.

The statistical significance of differentiation and incomplete condensation, when excluding overshoot, is relatively low ($1.4\sigma$ in both cases). In other words, without including a complete treatment of overshoot, one might be led to conclude that there is evidence of core-mantle differentiation and incomplete condensation, but such evidence could be identified as tentative.

% Notes from Gaia rerun:
% Nothing worth noting! Can ignore for this section

%GaiaJ0611-6931(Phot;Opt)
%Essentially no change
%Metric: 1.5

%GaiaJ0611-6931(Phot;UV)
%Essentially no change - lnZ slightly better for subseq. overshoot
%Metric: 1.0

%GD133
%No real change - heating favoured throughout (high ish Ca, low ish O)
%Metric: 1.3

%Gaia J0611-6931 (Spec; Opt)
%Essentially no change
%Metric: 1.1

%GD56:
%Metric: 1.8
%Good case study! bn invokes differentiation + heating, the others are all primitive. Evidence not especailly strong in all cases. But for subseq. oershoot, gets more confident the material is primitive. Also gets less confident in SS accretion (still most likely, but for vo and 3o, P(bu) is ~0.3). For bn, P(o excess) is 0.1, jumps to 0.4 for the others. That might be the Mg/Si talking - tSi/tMg skyrockets for vo and 3o, making it really hard to fit the (low) Si/Mg. The differentiation stuff can be explained by reference to Fe/Mg - tFe actually becomes greater than tMg for 3o, so the fragments Fe/Mg has to start below the observed value rather than above. This eliminates the apparent evidence for a core rich fragment that you would infer without overshoot!
%   |  tFe/tMg | sigma diff/Bayes factor for !bn
   
%bn |  0.35   | 1.36

%bo |  0.44   | 0.72

%vo |  0.98   | 0.36

%3o |  1.05   | 0.33

%HE 0106-3253
%Metric: 1.9
%An interesting system - it strongly wants heating and differentiation! But adding overshoot didn't make any substantial change.

\begin{figure}
    \centering
    \includegraphics[width=\columnwidth, keepaspectratio]{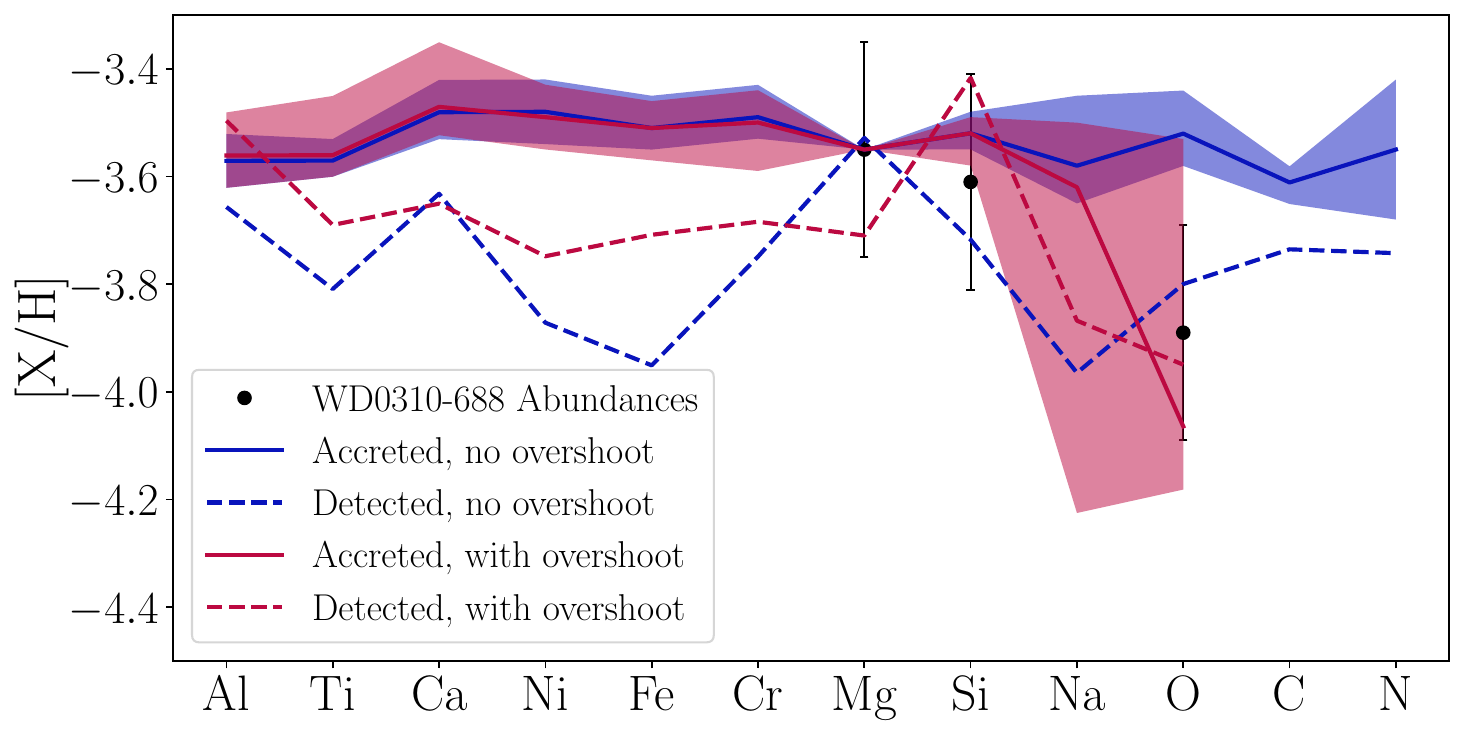}
    \caption{The geological interpretation of the metals accreted by WD\,0310$-$688 changes when convective overshoot is taken into account (red). The interpretation of the plot features is similar to Figure~\ref{fig:GD56}, but the abundances are scaled to the dominant photospheric element, in this case H, rather than Mg. We choose this scaling in cases where one or more of the fits does not replicate the Mg abundance particularly well, such that scaling to Mg results in a non-representative illustration of the overall quality of that fit. For the accreted compositions, the vertical position is arbitrary, so we fix the Mg abundance to match the detected abundance (and therefore there is no uncertainty on the Mg abundance for the accreted compositions). Our Bayesian framework finds (weak) evidence for incomplete condensation only when including convective overshoot.}
    \label{fig:WD0310-688}
\end{figure}

WD\,0310$-$688 is also a useful case study. This system exhibits an infrared excess, possibly due to an orbiting exoplanet, and is also polluted by metals \citep{Limbach2024}. The metal pollution is consistent with accretion of bulk-Earth-like material. Similar to GD\,56, our results support this conclusion, and show that it is dependent on convective overshoot, albeit weakly. The key element is ratio is O/Mg which is slightly depleted (relative to solar abundances). Without overshoot, this depletion can be explained as the effect of differential sinking: we have $\tau_{\rm O}/\tau_{\rm Mg} = 0.5$, such that O sinks out of the photosphere faster than Mg. However, when including overshoot, $\tau_{\rm O}/\tau_{\rm Mg} = 1.6$, and the best model now invokes oxygen depletion in the accreted material. The framework finds evidence (to $1.5\sigma$ significance) of incomplete condensation of volatiles in this case, comparable to bulk-Earth-like material rather than primitive material. Figure~\ref{fig:WD0310-688} illustrates the difference between the accreted material in the two cases, showing an inferred depletion of oxygen in the overshoot case.

%[Need to investigate how close the comparison to Earth is. there is still a probable oxygen excess even in that case, is that compatible?] Answer: Pretty damn close! O is a bit higher (56% compared to 48%) - otherwise v v similar with only 1 or 2 elements a bit off (Na) For the record  - there is still an oxygen excess in this case! That's how much oxygen is floating around I guess

Including convective overshoot causes the sinking timescales of O and Fe to increase by a greater factor than for the lithophile elements Ca and Mg (see Appendix). The inferred abundances of O and Fe in the accreted material therefore decrease. This results in a bias towards identification of mantle-rich and/or volatile-poor material. Similarly, identification of core-rich or volatile-rich material is hindered\footnote{The sinking timescales of Si, which typically behaves as a lithophile, are increased by a factor slightly greater than that of Fe. But this effect is small, and Si can also exhibit siderophilic behaviour \citep{Buchan2021}.}. We have identified 2 systems for which this effect is relevant. In these cases, the effect is of low statistical significance, but does nevertheless change the geological interpretation.

%[The oxygen excess is robust(ish). Probability is 72\% for 3o, 93\% for bn]

% - can we rule out mixing length as the source?

%WD0310-688:
%Metric: 1.8
%vo and 3o want some heating, bn and bo don't. It's somewhat marginal. Looks like the driving force is O - it sinks relatively much slower for vo/3o, so must be depleted by heating in order for differential sinking to bring it back up to its detected level. Si then kinda just has to take the hit, but it's OK with that! This probably means that we no longer have an oxygen excess for vo and 3o, but the EO stats don't seem to be working so will need to check...At least it consistently wants SS, probably because the absolute timescales are always low.

\subsection{He-dominated white dwarfs (Test~2)}
\label{sec:Test2}

There are 316 entries in PEWDD with a He-dominated atmosphere and at least 2 modellable elements, which our selection criteria narrow down to a sample of 12 (covering 9 unique white dwarfs). For most entries in our sample (9/12), there was no change (or only extremely marginal change) when including our best prescription of convective overshoot. However, this count includes two entries, HS\,2253+8023 Model 2 and 3, which the Bayesian framework was unable to fit regardless of the treatment of overshoot. This is due to the high Ca abundance in these cases, particularly as compared to Ti and Si, relative to the small errors quoted for these metals. Our count also includes 2 entries for SDSS\,J0956+5912 from \citet{Hollands2022} which are identical for our purposes. The difference between these entries is whether convective overshoot was included in their metal accretion rate estimates, but those estimates are not used in this work, since we effectively recalculate them ourselves. \citet{Hollands2022} found no significant effect on the inferred composition of accreted material.

The 3 systems for which we nominally find an effect are Ross\,640, WD\,1232+563 and GD\,424 (when using Keck data). For WD\,1232+563 \citep{Xu2019}, our forward model strongly favours a particular value of the initial nebular composition, and so our results are specific to that value and do not sample a range as would normally be the case. While this is not necessarily unphysical, we conservatively exclude this system from the following discussion. The other two systems also come with caveats associated with the values of $T_{\rm eff}$ and $\log(g)$. 

Ross\,640 \citep{Blouin2018} is similar to GD 56: differentiation is invoked only when ignoring convective overshoot (to 1.4$\sigma$). This is because Fe sinks faster (relative to other metals) when including overshoot, such that the Fe depletion can be explained as the consequence of differential sinking rather than accretion of mantle-like material (with molar core fraction $4\pm2\%$). But this discrepancy is only present when using the values of $T_{\rm eff}$ and $\log(g)$ found by \citet{Blouin2018} (as included in PEWDD). The equivalent values from \citet{GentileFusillo2021} using \textit{Gaia} data are significantly higher (see Table~\ref{tab:sample}), leading to different metal sinking timescales. When using these timescales, the discrepancy in interpretation is not present, leaving it ambiguous as to whether the treatment of convective overshoot is relevant for this system. However, the potential discrepancy is only minor in any case (1.4$\sigma$).

\begin{figure}
    \centering
    \includegraphics[width=\columnwidth, keepaspectratio]{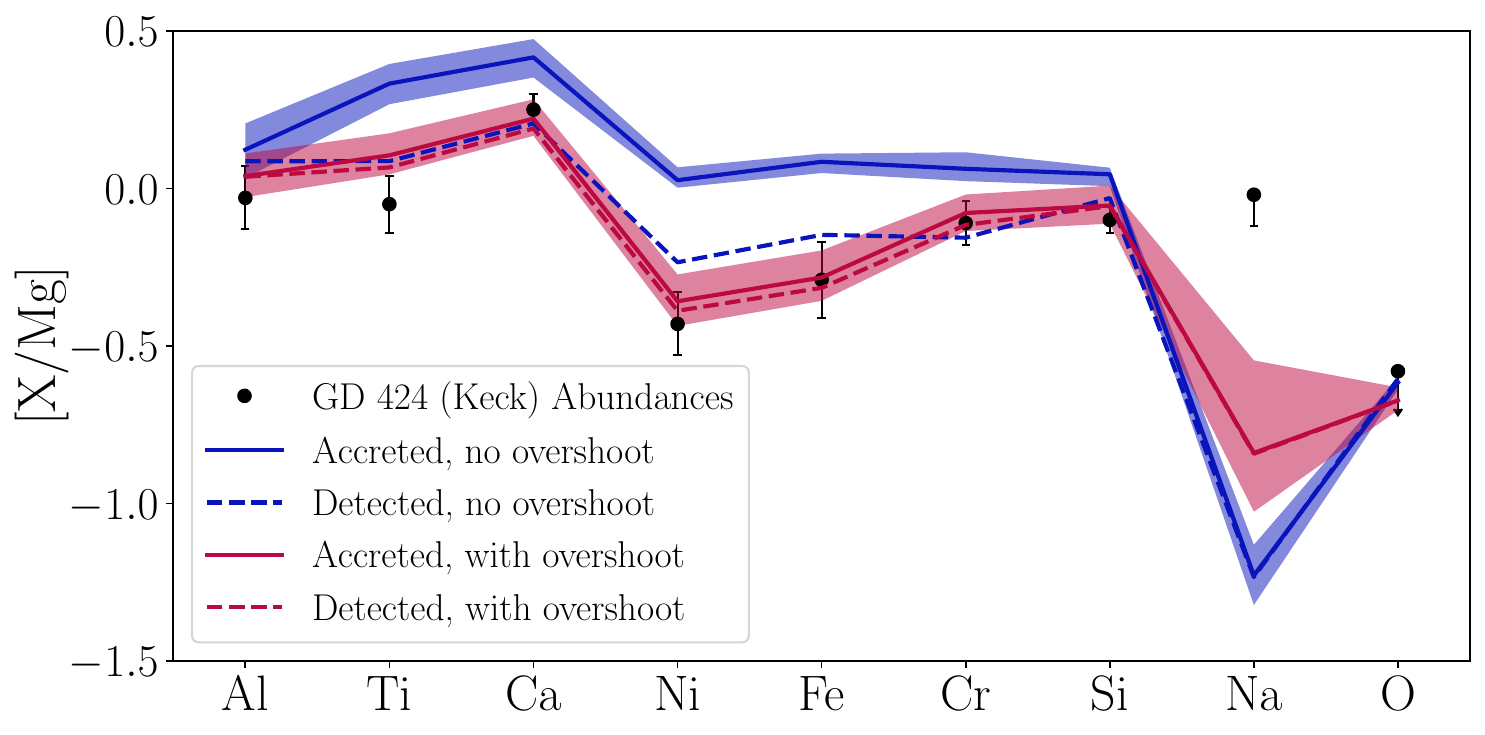}
    \caption{The geological interpretation of the metals accreted by GD\,424 changes depending on whether convective overshoot is neglected (blue) or included (red). The plot features are similar to Figure~\ref{fig:GD56}.}
    \label{fig:GD424}
\end{figure}

The results for GD\,424, using data from the High Resolution Echelle Spectrometer (HIRES) at the Keck I telescope  \citep{Izquierdo2020}, are illustrated in Figure~\ref{fig:GD424}. Differentiation is invoked only when including convective overshoot (to 2.6$\sigma$). Including overshoot does not strongly affect the sinking timescales of Ni or Fe relative to Mg in this case. With or without overshoot, the framework finds that accretion has reached the declining phase. However, with overshoot included, the best model finds that the system is not as deep into the declining phase of accretion, and that therefore the Ni and Fe depletion cannot be attributed to differential sinking. This result is driven by the Al abundance. With convective overshoot included, Al sinks on a (relatively) long timescale. If accretion had proceeded too deep into the declining phase, the predicted Al abundance, relative to other metals, would be too high. The best model now invokes accretion of a mantle-rich fragment (with molar core fraction $6\pm2\%$). Another entry for GD\,424, corresponding to data from the Intermediate dispersion Spectrograph and Imaging System (ISIS) at the William Herschel Telescope (WHT), was found to show no change in interpretation when including overshoot. For this entry, there are no Ni, Fe or Cr abundances recorded, so it is not possible to infer the presence of differentiation. However, in all four cases (Keck and WHT data, with and without overshoot), the best model invokes incomplete condensation.

As with Ross\,640, using the \textit{Gaia} values for $T_{\rm eff}$ and $\log(g)$ removes any discrepancy in geological interpretation: with these parameters, differentiation and incomplete condensation are invoked regardless of whether overshoot is considered. The parameter values taken from \citet{Izquierdo2020} and \citet{GentileFusillo2021} are sufficiently close that they can be considered consistent within the systematic errors. This suggests that the geological modelling for this system is highly sensitive to the values of these parameters.

\citet{Izquierdo2020} favour a solution for this system in which the accreted material is similar to CI chondrites or bulk Earth, and accretion is ongoing. Core--mantle differentiation was not invoked. The declining phase was ruled out because it implies a high abundance of siderophile elements (although we note that this tension might potentially be eased by invoking differentiation followed by preferential accretion of mantle-like material). We also note that \citet{Izquierdo2020} used timescales based on \citet{Koester2020}. One of the consequences is that, relative to timescales calculated using STELUM, high siderophile element abundances are more likely to be inferred. This will be discussed in Section~\ref{sec:public_grids}.

%For the record: their timescales did not have overshoot, and they used the Keck abundances

%[Now need to compare with the original conclusion, which ruled out decling phase. In particular, the Fe/Si got really high for their dec solution, why?- note that they also had Mn to work with!] [This also seems like a system where a graph might be useful!] GAIA: Now, there's no discrepancy! Both want diff and heating. Seems to back up idea that Al is driving this, weirdly!

\subsection{Overview}
\label{sec:overshoot_overview}

So far, we have focussed on individual white dwarfs. We now consider the population of white dwarfs more broadly, and ask which regions of parameter space, as defined by $T_{\rm eff}$ and $\log(g)$, are most likely to host white dwarfs affected by the treatment of convective overshoot. To do this, we calculate the mean value of our discrepancy metric across parameter space. The metric is a function of $\beta$, the `depth' of a system into its declining phase of accretion (see Appendix), and the abundances of the modellable elements. 
%Both of these are randomly (and independently) sampled.

We sample metal abundances from 50\,000 synthetic polluted white dwarfs generated from Monte Carlo-style simulations similar to those described in \citet{Buchan2024}. We do not describe this in detail, because the only practical output is a list of which elements are `detected' for each white dwarf. White dwarfs with fewer than 2 detected elements are ignored.

% For reference, if details are needed, I assumed 0.2 dex errors, Teff/logg based on MWDD, no differentiation, uniform distribution of log dist from -1 to 1 AU, collisional cascade mass distribution

For each synthetic white dwarf, we calculate the discrepancy metric 50 times, using different values for $\beta$. We sample $\beta$ randomly from an exponentially decaying distribution with a decay parameter equal to the \textit{a priori} probability of observation in the declining phase, $p$ (see Equation~\ref{eq:apriori_p}). This parametrisation reproduces important, physically motivated, features. For white dwarfs with very short sinking timescales (e.g., warm and H-dominated), $p$ is close to zero. Therefore, the sampled values of $\beta$ are also close to zero and the declining phase contribution to the discrepancy metric is negligible, as expected. For white dwarfs with long sinking timescales (e.g., cool and He-dominated), $p$ is close to 1, and the resulting distribution of $\beta$ approximates the inferred distribution of $\beta$ for cool DZs (see Figure 4.7 in \citealt{BuchanThesis}). Between these extremes, the distribution of $\beta$ varies smoothly.

\begin{figure}
    \centering
    \includegraphics[width=\columnwidth, keepaspectratio]{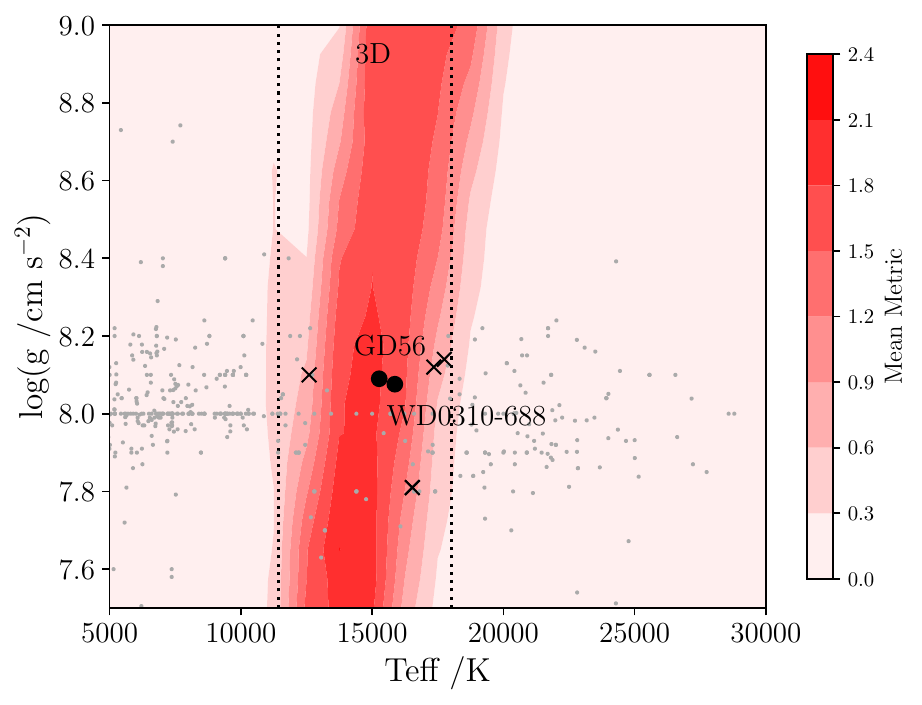}
    \caption{The susceptibility of H-dominated white dwarfs to a change in geological interpretation when including a full treatment of convective overshoot, compared to no overshoot, as a function of effective temperature and surface gravity. We show this in terms of the mean value of our discrepancy metric (see Section~\ref{sec:overshoot_overview}) for randomly generated synthetic white dwarfs. The metric does not take into account the specific abundances (and errors) of detected metals, which is important for predicting whether any particular polluted white dwarf will be susceptible. The metric is inversely proportional to the uncertainty on abundances - here we assume 0.2 dex errors. The sample used for Test~1 is shown with black crosses. The black circles show the subset of those systems for which the geological interpretation does actually change when using a full 3D parametrisation of convective overshoot. The grey dots show all other H-dominated white dwarfs in PEWDD.}
    \label{fig:DMP_DA_OVERSHOOT}
\end{figure}

Figure~\ref{fig:DMP_DA_OVERSHOOT} shows that this approach reproduces the sensitive regime between about \SI{12000}{K} and \SI{18000}{K} that was visible in Figure~\ref{fig:da_overshoot_timescales}. This is expected, because in this case the sensitive regime is in roughly the same location for all possible pairs of detected metals that the discrepancy metric could sample (this is not the case generally). Almost any H-dominated white dwarf in the sensitive regime could be affected, depending on its exact metal abundances. This is roughly 12\% of all H-dominated metal polluted white dwarfs (based on PEWDD).

%Other cases are more complicated because the sinking timescale ratios of different element pairs do not necessarily have similar behaviour as a function of $T_{\rm eff}$ and $\log(g)$ [include figure?]. The discrepancy metric is a function of all the modellable elements.

\begin{figure}
    \centering
    \includegraphics[width=\columnwidth, keepaspectratio]{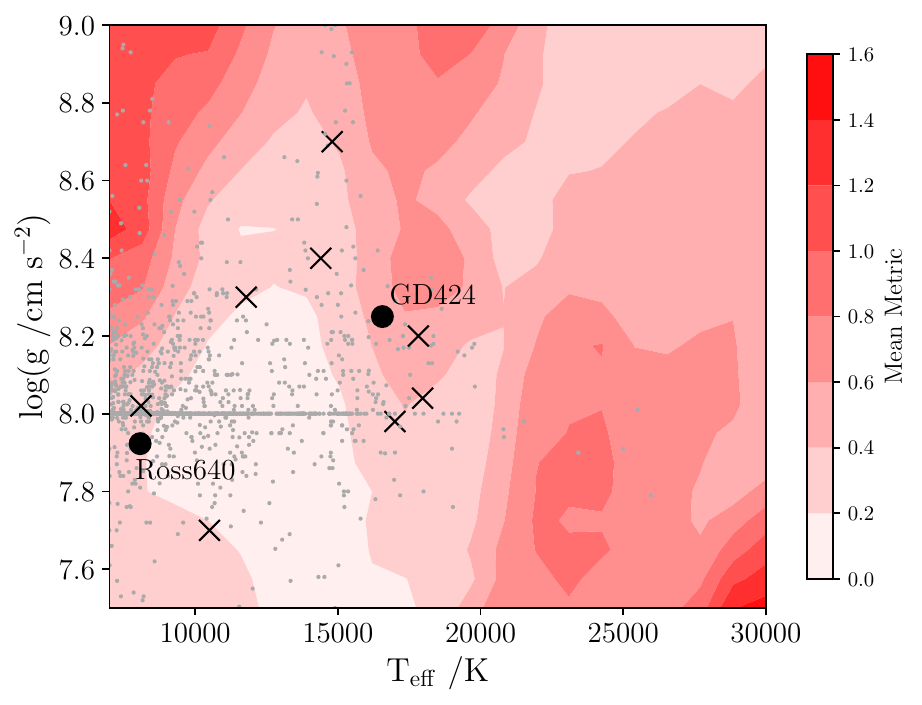}
    \caption{The susceptibility of He-dominated white dwarfs to a change in geological interpretation when including a detailed treatment of convective overshoot compared to no overshoot. The meaning of the colour bar and label types are similar to those in Figure~\ref{fig:DMP_DA_OVERSHOOT}, but the black markers show the sample for Test~2, and grey dots now correspond to He-dominated white dwarfs.}
    \label{fig:DMP_DB_OVERSHOOT}
\end{figure}

Figure~\ref{fig:DMP_DB_OVERSHOOT} shows the equivalent plot for He-dominated white dwarfs. The severely affected regions of parameter space are sparsely populated. This suggests that for most He-dominated white dwarfs, the impact of neglecting convective overshoot is likely to be minor. However, in some cases there may still be a significant impact, depending on the details of the white dwarf. In particular, white dwarfs with particularly small errors on their abundances are more likely to be impacted. Notably, all but one of the entries in our sample had at least 3 elements with particularly small errors ($\le0.1$dex). 
%It should also be noted that Figure~\ref{fig:DMP_DB_OVERSHOOT} is ultimately based on extrapolation of 1D simulations, but to fully address this question it is essential to use 3D simulations. Full 3D hydrodynamical simulations have yet to be carried out for He-dominated atmospheres.
It should be noted that Figure~\ref{fig:DMP_DB_OVERSHOOT} is based on a 1D parametrisation of overshoot, which has not been confirmed with full 3D simulations.

%, and conversely the region of least severity largely coincides with the location of most He-dominated white dwarfs. This does not mean that there are no affected white dwarfs, because there are a large number and each has a non-zero probability of being affected. We were able to identify 3 affected white dwarfs. [Note that 1232+563 seems dodgy. For the other two what stands out is that, more than most, these are ones where the two grids disagree on how deep into the declining phase we are - this possibility isn't really taken into account by the metric so might explain why it looks a bit random! But then the question is: how can we identify the ones where there's going to be a disagreement on $\beta$? Something to look at - could Al/Ti/Ca be a system for this, since they behave similarly geologically?] 

\subsection{Treatment of convective overshoot}

Broadly speaking, for H-dominated atmospheres, ignoring convective overshoot will cause inferred compositions to tend towards core- and volatile-rich, under the assumption that the accreted material formed through the same geological processes which are most significant in the formation of rocky Solar System bodies. We found that this effect is active within a temperature range of roughly \SI{12000}{K} to \SI{18000}{K}. For such white dwarfs, it is likely that accretion is in the steady state phase, limiting the potential discrepancy. Nevertheless, for such white dwarfs, convective overshoot should be considered in order to draw reliable geological conclusions. Below \SI{11000}{K}, we are reliant on a 1D model of convective overshoot. 
It is possible that convective overshoot may continue to be impactful below that temperature, but further 3D simulations probing this temperature range are required to draw firm conclusions.

For He-dominated white dwarfs, Figure~\ref{fig:DMP_DB_OVERSHOOT} shows that the effect of convective overshoot itself is predicted to be relatively minor in most cases.
%(although this should be tested using 3D simulations). 
%Particularly cool white dwarfs with high surface gravity ($T_{\rm eff}\lesssim$\SI{10000}{K}, $\log(g)\gtrsim8.2$) might be systematically affected, and in these cases overshoot ought to be considered. 
The systems for which we nominally found some impact had unusually small errors on their metal abundances.
%and even so the impact came with  The systems which were affected showed a discrepancy in $\beta$, the depth into the declining phase. [But $\beta$ is ultimately determined using sinking timescales, so overshoot did matter? Perhaps it's actually to do with the absolute timescales here! Which would explain why it's not visible here - need to investigate]

\subsubsection{Fixed overshoot}

So far, we have used what we consider to be the most complete, variable parametrisations of convective overshoot. In this section, we discuss a simpler treatment, in which the extent of overshoot is fixed to one pressure scale height. We investigate whether using fixed overshoot is preferable to using no overshoot, as benchmarked against a full treatment of overshoot (meaning either the S3D or SV grid for H- or He-dominated systems, respectively).

\begin{figure}
    \centering
    \includegraphics[width=\columnwidth, keepaspectratio]{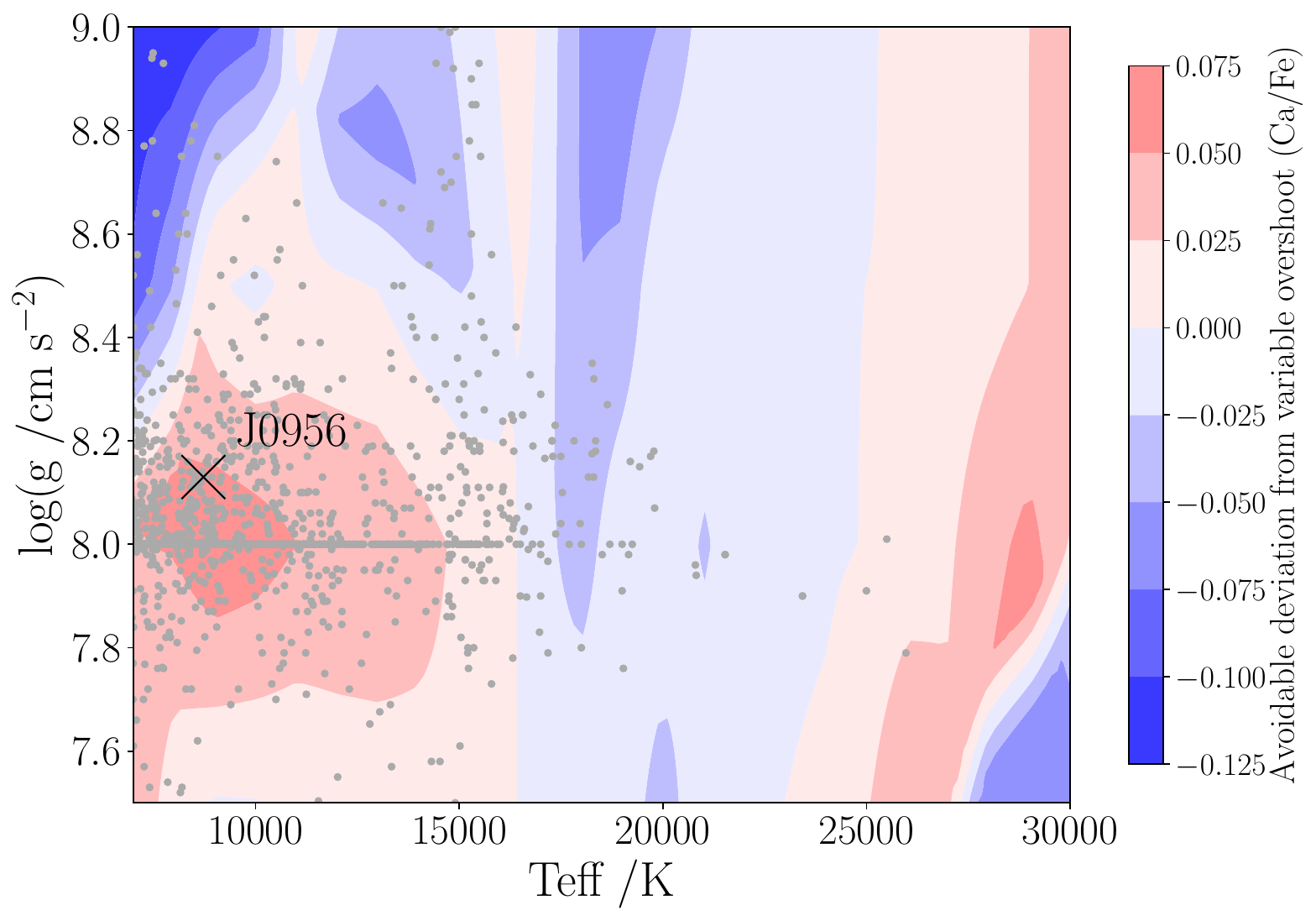}
    \caption{Plot showing which He-dominated white dwarfs (grey dots) are best modelled with no overshoot, or with fixed one pressure scale height overshoot, as benchmarked against a variable treatment of overshoot (SV grid) and using the Ca/Fe sinking timescale ratio. The red (positive) regions show parts of parameter space in which it is preferable to use no overshoot. In the blue (negative) regions, fixed overshoot is preferable. The magnitude shows the deviation (from the benchmark) that is avoided by using the preferred option (i.e., the greater the magnitude, the more strongly the relevant option should be preferred). 
    %A full treatment of overshoot is the best option, and should be used where possible.
    }
    \label{fig:FixvNo}
\end{figure}

For H-dominated systems, outside of specific combinations of temperature and metals, the SO grid reproduces relative sinking timescales from the S3D grid better than the SN grid, implying that fixed overshoot is usually preferable to no overshoot. For He-dominated systems, the situation is more complicated. We illustrate our results in Figure~\ref{fig:FixvNo}, looking at the timescale ratio of Ca/Fe (we find qualitatively similar behaviour for most combinations of Ca, Fe, O, Si and Mg). In Figure~\ref{fig:FixvNo}, the red (positive) regions show where it is preferable to use no overshoot. The blue (negative) regions are where fixed overshoot is preferable. The magnitude of the value indicates how strong the improvement is. Most He-dominated white dwarfs (grey dots) are best modelled using no overshoot, rather than fixed overshoot. Physically, this is because for white dwarfs of this approximate temperature, the extent of convective overshoot in the SV grid becomes small, such that no overshoot is a better approximation than one pressure scale height of overshoot.

\begin{figure}
    \centering
    \includegraphics[width=\columnwidth, keepaspectratio]{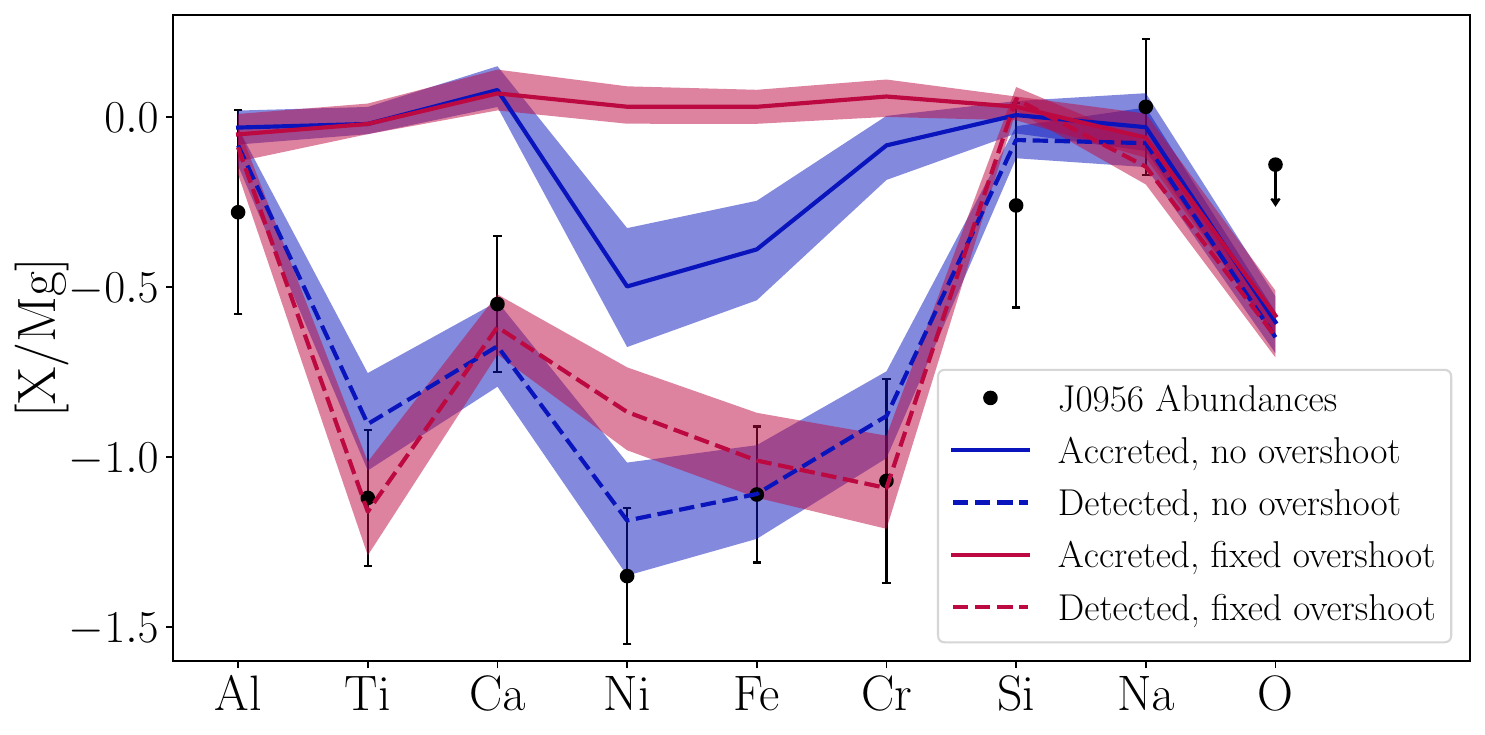}
    \caption{The geological interpretation of the metals accreted by J0956 changes when assuming no convective overshoot (blue) or overshoot fixed to one pressure scale height (red). The plot features are similar to Figure~\ref{fig:GD56}, but with the 1 sigma confidence intervals now also shown for the fits to the detected compositions. The Bayesian framework finds evidence for mantle-rich material, and therefore core-mantle differentiation, only when using no overshoot. The data point for O is an upper bound.}
    \label{fig:J0956}
\end{figure}

As an example, we consider the white dwarf J0956, using data from \citet{Swan2023a}. When modelling the system with no overshoot (SN grid), or a full treatment (SV grid) of overshoot, core-mantle differentiation is invoked to $2.4\sigma$ significance. However, when including one pressure scale height of overshoot (SF grid), no evidence of differentiation is recovered (see Figure~\ref{fig:J0956}).

\section{Thermohaline mixing}
\label{sec:thermohaline}

\begin{figure}
    \centering
    \includegraphics[width=\columnwidth, keepaspectratio]{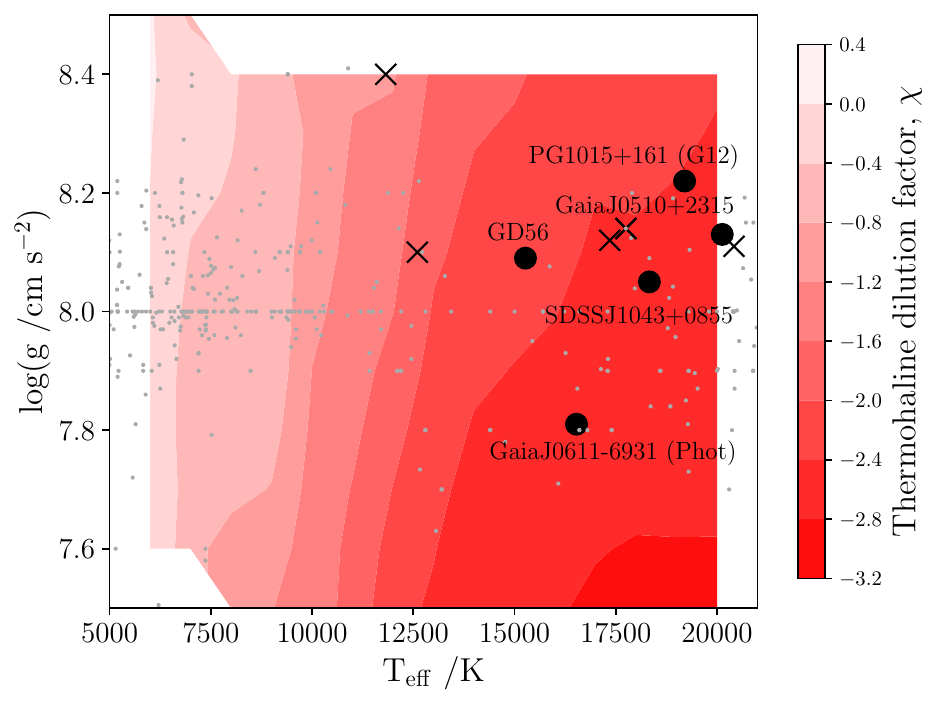}
    \caption{Strength of thermohaline mixing as a function of effective temperature and surface gravity for H-dominated white dwarfs. The strength is expressed as the dilution factor, $\chi$, which quantifies the reduction of accreted mass remaining in the photosphere, due to thermohaline mixing (see Section~\ref{sec:thermohaline_intro}). Here, we calculate $\chi$ assuming a fixed accretion rate of $10^9$ g/s. The black markers show the white dwarfs included in our sample for Test~3, with circles indicating systems for which the geological interpretation changes when including thermohaline mixing. Grey dots show all H-dominated white dwarfs in PEWDD. Black crosses show the final sample which we model. The grid is irregular due to non-linear transformations between mass and surface gravity, and temperature and cooling age.}
    \label{fig:therm_factors}
\end{figure}

This section covers the results of Test~3. This test differs from the others: rather than comparing the effect of switching between different grids of sinking timescales, we stick with one set of sinking timescales and test the impact of including thermohaline mixing on a sample of 13 entries from PEWDD with H-dominated atmospheres (corresponding to 8 white dwarfs). We select these entries using our metric, which estimates which white dwarfs are most likely to be affected by thermohaline mixing (Section~\ref{sec:sample_selection}). We illustrate the selected sample in Figure~\ref{fig:therm_factors}, which also shows the predicted strength of thermohaline mixing across $T_{\rm eff}$/$\log(g)$ space. In Section~\ref{sec:therm_results}, we discuss results for the white dwarfs highlighted in Figure~\ref{fig:therm_factors}. In Section~\ref{sec:therm_discussion}, we discuss the effect of thermohaline mixing across the parameter space more generally. We also compare accretion rates, calculated with and without thermohaline mixing, against constraints from X-ray detections for G\,29$-$38.
%With thermohaline mixing activated, differential sinking can be disabled depending on the sampled value of accretion rate and the white dwarf's temperature and surface gravity. The absolute abundances of atmospheric metals also become lower at any given time (for details, see Section~\ref{sec:thermohaline_intro}). We test the impact 

\subsection{Results for individual systems (Test~3)}
\label{sec:therm_results}

\begin{figure}
    \centering
    \includegraphics[width=\columnwidth, keepaspectratio]{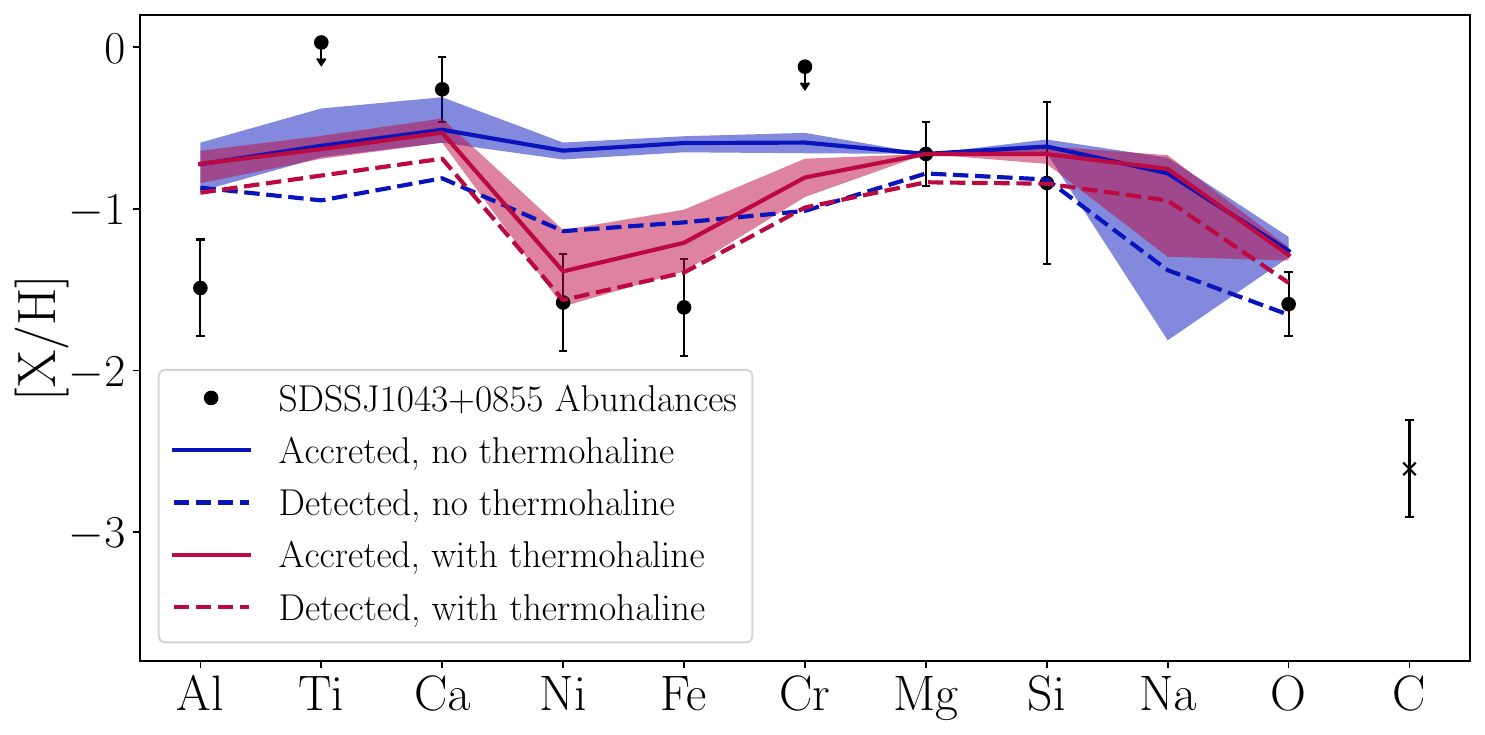}
    \caption{The geological interpretation of the metals accreted by SDSS\,J1043+0855 changes when including thermohaline mixing (red). The plot features are similar to Figure~\ref{fig:WD0310-688}. The Bayesian framework finds evidence for accretion of mantle-rich material, and therefore core-mantle differentiation, only when including thermohaline mixing. The data points for Ti and Cr are upper bounds. The C data point was excluded during fitting because our incomplete condensation model does not accurately handle sub-Solar values of C. The C depletion implies incomplete condensation, which is inferred for all compositions shown here due to the O depletion.}
    \label{fig:SDSSJ1043}
\end{figure}

Two entries, SDSS\,J1043+0855 and PG\,1015+161 (specifically the entry using data from \citealt{Gaensicke2012}), show changes in whether differentiation is invoked, depending on whether thermohaline mixing is considered. SDSS\,J1043+0855 is slightly depleted in Fe and Ni relative to Solar, but we find that this depletion can be explained solely as the result of differential sinking when thermohaline mixing is excluded. With thermohaline mixing included, differential sinking is ineffective, and the depletion is now inferred to be geological in nature (i.e., accretion of mantle-like material, to 2.4$\sigma$ significance; see Figure~\ref{fig:SDSSJ1043}). PG\,1015+161 shows the opposite behaviour (for the same underlying reason), requiring core-like material (to 1.4$\sigma$ significance), but only when thermohaline mixing is not included. In this case, the fit to the data is poor (reduced $\chi^2>2$), but this is marginal and other models (with lower Bayesian evidence) are still able to fit the data well.

The results for both of these entries are sensitive to the values of $T_{\rm eff}$ and $\log(g)$. When adopting values from \textit{Gaia} instead of spectroscopic parameters, we find that differentiation is invoked for SDSS\,J1043+0855, and is not invoked for PG\,1015+161, regardless of whether thermohaline mixing is considered. We also note that \citet{Melis2017} concluded that SDSS\,J1043+0855 has accreted material from a differentiated body, using spectroscopic parameters and excluding thermohaline mixing, but using sinking timescales from an earlier version of STELUM.

The inferred volatile content is affected for 4/11 entries, with sigma significance varying between 1.3 and 2.0. For three of these (Gaia\,J0611$-$6931 (Phot; Opt), Gaia\,J0510+2315 (Phot Opt), Gaia\, J0510+2315 (Phot UV)), O is slightly depleted. Without thermohaline mixing this is explained via differential sinking, whereas with thermohaline mixing included the accreted material is inferred to be depleted in volatiles to some extent. These results depend on the values of $T_{\rm eff}$ and $\log(g)$. \citet{Rogers2024} provide photometric and spectroscopic values, and \citet{GentileFusillo2021} provide additional values using \textit{Gaia} data. \citet{Rogers2024} also calculate metal abundances using both UV and optical data. We do not go through every combination here, but note that each of these choices can affect whether incomplete condensation is invoked. For GD\,56, the sigma significance is low (less than 1.3), and the change in result can be attributed to random chance.

Both Gaia\,J0611$-$6931 and Gaia\,J0510+2315 have a detection (or upper bound) on C from UV data, indicating sub-Solar abundances  \citep{Rogers2024}. Our modelling does not include C if it is significantly sub-Solar, due to a known issue in the condensation code. If the C abundances were taken into account, it is likely that incomplete condensation would be invoked to some extent (see \citealt{Rogers2024b}). This could potentially resolve the ambiguity noted here.

\subsection{Discussion}
\label{sec:therm_discussion}

\begin{figure}
    \centering
    \includegraphics[width=\columnwidth, keepaspectratio]{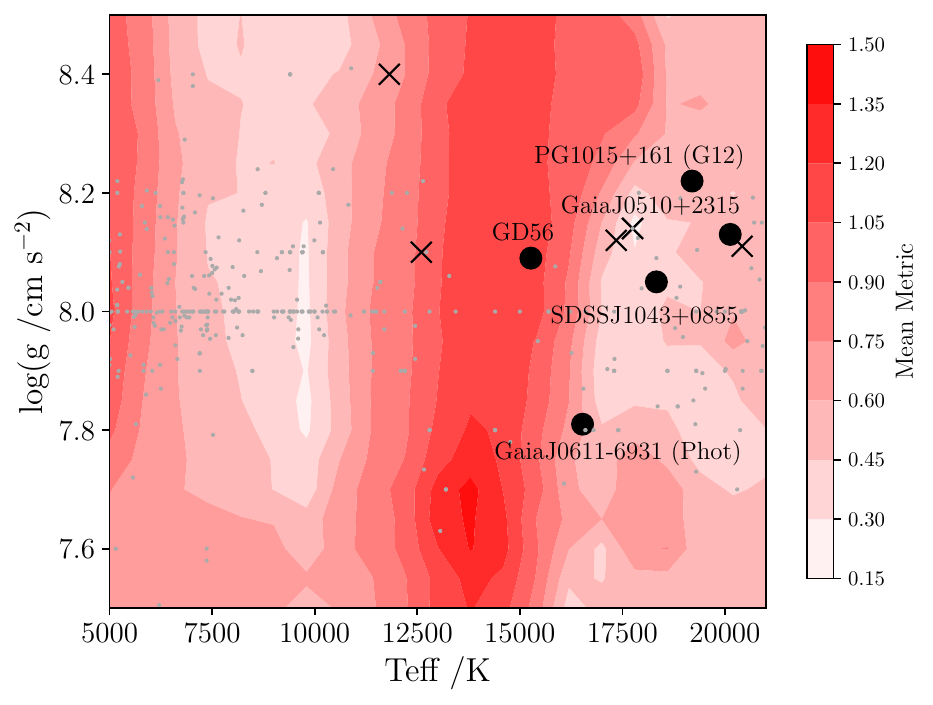}
    \caption{The susceptibility of H-dominated white dwarfs to disabling differential sinking, if thermohaline mixing is sufficiently strong. The crosses show the sample of white dwarfs used in Test~3. The black circles show the subset of those systems for which the geological interpretation changed when including thermohaline mixing. The grey dots show all other H-dominated white dwarfs in PEWDD. The metric here compares the S3D timescale grid against a dummy grid in which all sinking timescales are equal.}
    \label{fig:thermohaline_paramspace}
\end{figure}

The region of parameter space in which thermohaline mixing is most relevant depends on multiple factors. The first factor is whether thermohaline mixing is strong enough to deactivate differential sinking. Figure~\ref{fig:therm_factors} shows the predicted strength of thermohaline mixing for H-dominated white dwarfs. The strength of thermohaline mixing is shown in terms of the change in mixed mass assuming a fixed accretion rate of $10^9$g\;s$^{-1}$. Broadly speaking, thermohaline mixing is stronger for hotter white dwarfs, leading to lower mixed masses for a given accretion rate. For our modelling purposes, the consequence is that, for a given mass of atmospheric metals, higher accretion rates are inferred. This does not, in and of itself, affect geological interpretation. However, in our prescription of thermohaline mixing, differential sinking is disabled if thermohaline mixing is sufficiently strong. This can affect geological interpretation if differential sinking would cause significant changes in the detected relative metal abundances. This is the second factor in determining the relevance of thermohaline mixing, and is illustrated in Figure~\ref{fig:thermohaline_paramspace}. This figure shows the strength of differential sinking, using similar methodology to Figure~\ref{fig:DMP_DA_OVERSHOOT}, but comparing the timescales calculated using a full treatment of overshoot against a dummy grid of timescales. The dummy grid represents the case of no differential sinking, and contains the same value for every metal. Figure~\ref{fig:thermohaline_paramspace} shows that the white dwarfs which are most affected by the deactivation of differential sinking lie between approximately \SI{12000}{K} and \SI{16000}{K}. This temperature range is largely driven by the Si/Mg sinking timescale ratio, which is close to 1 for the most of the parameter space outside this temperature range but is higher within it (up to about 2). In other words, while the relevant effect of thermohaline mixing is to erase differential sinking, for Si and Mg differential sinking is relatively weak at high and low temperatures anyway. This is why the mean metric value does not continue to increase for temperatures beyond about \SI{16000}{K}, even though thermohaline mixing itself is strongest at these high temperatures.

The overall strength of thermohaline mixing can be thought of as a combination of Figures~\ref{fig:therm_factors} and \ref{fig:thermohaline_paramspace}. Figure~\ref{fig:thermohaline_paramspace} shows the potential impact of thermohaline mixing, while Figure~\ref{fig:therm_factors} shows how likely such an impact is in the first place.

%The Bayesian framework cycles through a number of models, constructed by invoking different combinations of physical processes. We focus on whether the best model invokes core-mantle differentiation and/or incomplete condensation. These are the most important processes for setting Earth's composition and first-order structure. For further details on the Bayesian framework, see \citet{Buchan2021}. The only notable change is that the variable determining the absolute quantity of detectable pollution is now the mass of the pollutant body. The associated prior distribution is assumed to match a collisional cascade. 

For the systems which might be affected by thermohaline mixing, when thermohaline mixing is ignored, Fe and O are predicted to sink relatively quickly compared to major lithophile metals. For example, for a H-dominated white dwarf with $T_{\rm eff} = $\;\SI{20000}{K} and $\log(g) = $\;8, STELUM predicts $\tau_{\rm Fe} = $\;\SI{0.006}{yr} and $\tau_{\rm O} = $\;\SI{0.007}{yr}, which is shorter than other key timescales (e.g., $\tau_{\rm Mg} = $\;\SI{0.014}{yr}, $\tau_{\rm Si} = $\;\SI{0.012}{yr}, $\tau_{\rm Ca} = $\;\SI{0.008}{yr}). Switching on thermohaline mixing erases the differences between these timescales, decreasing the inferred Fe and O content in the accreted material. Qualitatively, this results in the inference of more mantle-rich, less core-rich, less volatile-rich material.

In principle, the geology of accreted material could provide evidence either for or against it, if it were the case that one treatment consistently leads to more reasonable geological interpretations (with higher Bayesian evidence). We consistently find that the best geological model has higher Bayesian evidence, $\mathcal{Z}$, in the case where thermohaline mixing is neglected. The value of $|\Delta\ln(\mathcal{Z})|$ across all the entries we ran varies from 0.4 to 8.9, with a mean value of 4.8. In isolation, this would be equivalent to a Bayes factor of over 120. A caveat to this result is that the entries are not entirely independent of each other.

The underlying cause is that thermohaline mixing leads to inferences of higher accretion rates, and hence more massive pollutants (all else being equal). Our priors are weighted to penalise high mass pollutants, according to a collisional cascade size distribution. In other words, we assume that smaller pollutants are more plausible than larger ones \citep{Cunningham2025}, and since thermohaline mixing inherently demands larger pollutants, it is penalised. We emphasise that this result is driven by our assumptions rather than data. Nevertheless, it quantifies the intuition that the high accretion rates implied by thermohaline mixing are harder to explain dynamically. If one is confident that thermohaline mixing ought to be included, it would suggest that, broadly speaking, our assumed pollutant masses are too low, or that our assumed accretion events are too long.

% Could elaborate here - posterior odds = prior ratio x bayes factor, so unless you're 200x more confident in thermohaline mixing a priori, you should reject it. In that case (i.e., you are very confident that thermohaline mixing should be included) then it probably means the prior on accretion event duration needs to shift way down.

%Further work A more convincing argument could be made if several pollutants which can only be fitted either with or without thermohaline mixing are identified (we identified one system which could be fitted only when ignoring thermohaline mixing). Alternatively, we could model a sufficiently large population that the change in Bayesian evidence, if present, is driven by data rather than priors.

\subsubsection{Comparison against accretion rates derived from X-ray observations}

\begin{figure}
    \centering
    \includegraphics[width=\columnwidth, keepaspectratio]{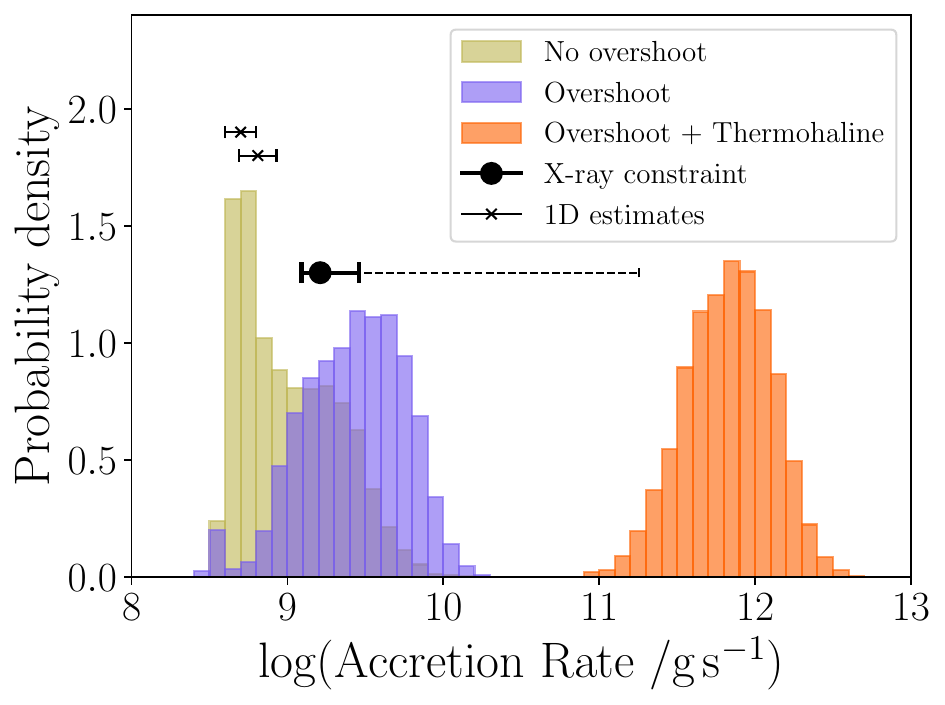}
    \caption{The accretion rate inferred for G\,29-38 by our Bayesian framework using three sets of modelling assumptions: no convective overshoot (olive), convective overshoot (purple), convective overshoot plus thermohaline mixing (orange). The distributions are generated by randomly resampling 10,000 times from the posterior distributions associated with the best model in each case. For comparison, the black circle shows the value determined from X-ray detections by \citet{Cunningham2022}, and the associated $1\sigma$ confidence interval. This value may be an underestimate, since it does not account for additional undetected emission, primarily in the extreme ultraviolet. The dashed line extending from the data point indicates the (90\% confidence) upper limit on the estimated total accretion rate. We also show, for illustrative purposes only, previously published accretion rates estimated using 1D models (upper: \citealt{Farihi2009}, lower: \citealt{Xu2014}). The vertical positions of these points are arbitrary.}
    \label{fig:G2938}
\end{figure}

The accretion rates inferred by our models may be tested against independently measured accretion rates determined via X-ray luminosity. At present, the only white dwarf with such a measurement is G\,29-38, for which the accretion rate (expressed as $\log_{10} {\rm g}\,{\rm s}^{-1}$ throughout this section) is $9.21^{+0.25}_{-0.12}$ \citep{Cunningham2022}. G\,29-38 is included in our sample, allowing for convenient comparison. We model this system with no overshoot (SN grid), with convective overshoot (S3D), and with both convective overshoot and thermohaline mixing. We adopt values of $T_{\rm eff}$ and $\log(g)$ from \textit{Gaia}. The resulting accretion rates are $9.0^{+0.4}_{-0.3}$, $9.4^{+0.3}_{-0.4}$ and $11.8^{+0.3}_{-0.3}$ respectively. The values obtained without thermohaline mixing are consistent with the X-ray measurements, given the quoted uncertainties, as illustrated in Figure~\ref{fig:G2938}. The accretion rate obtained with thermohaline mixing is nominally inconsistent with the X-ray detections (at a significance level of $6.6\sigma$). However, the X-ray constraint does not account for undetected emission, primarily in the UV. Including this additional contribution increases the inferred accretion rate, such that it may still be consistent with the estimate from thermohaline mixing (see dashed line in Figure~\ref{fig:G2938}).

% FWIW: Purple is 0.4 sigma out, olive is 0.5 sigma out

%\subsection{Reliability of metric}

%Metric does not seem to be that reliable! 2 main points - difficulty in estimating D \textit{a priori}, and the fact that the abundances themselves make a huge difference (case in point - WDJ183352.68+321757.25. Highest metric value, but interpretation as core-rich is set in stone because the Fe and Ni abundances are so high. 31.2 sigma (bn) or 31.3 sigma (kn)). The metric should be thought of as flagging up WDs which are most likely to be potentially affected by a change in sinking timescale grid, absent any knowledge of what the abundances (and errors!) themselves are, which will have a large impact.

%Also - it looks like maybe the metric does at least indicate the confidence level that could not be overthrown by a grid swap?

\section{Comparison between publicly available diffusion timescales}
\label{sec:public_grids}

This section, covering Tests~4 and 5, compares the publicly available Koester timescales grid (`KN') with a corresponding grid calculated with STELUM (`SN'). Convective overshoot is ignored in both cases in order to minimise the number of variables. The SN timescales can be considered to be equivalent to those that have been publicly available on the MWDD so far. As mentioned in Section\,\ref{sec:sinking_models}, the Koester and STELUM frameworks make a few different assumptions, notably regarding the mixing-length parameter, the atmospheric boundary condition, and the heavy-element ionisation model. The comparisons performed in this section thus provide an estimate of the overall effect that these differences can have on the geological interpretation of accreted material.

Test 4 concerns H-dominated white dwarfs, while Test~5 covers He-dominated systems. We use our metric (Section~\ref{sec:sample_selection}) to select white dwarfs most likely to be affected by the differences between the KN and SN grid in both cases. For Test~4, our sample contains 17 entries (with 15 white dwarfs). The equivalent numbers for Test~5 are 25 and 19.

\begin{figure}
    \centering
    \includegraphics[width=\columnwidth, keepaspectratio]{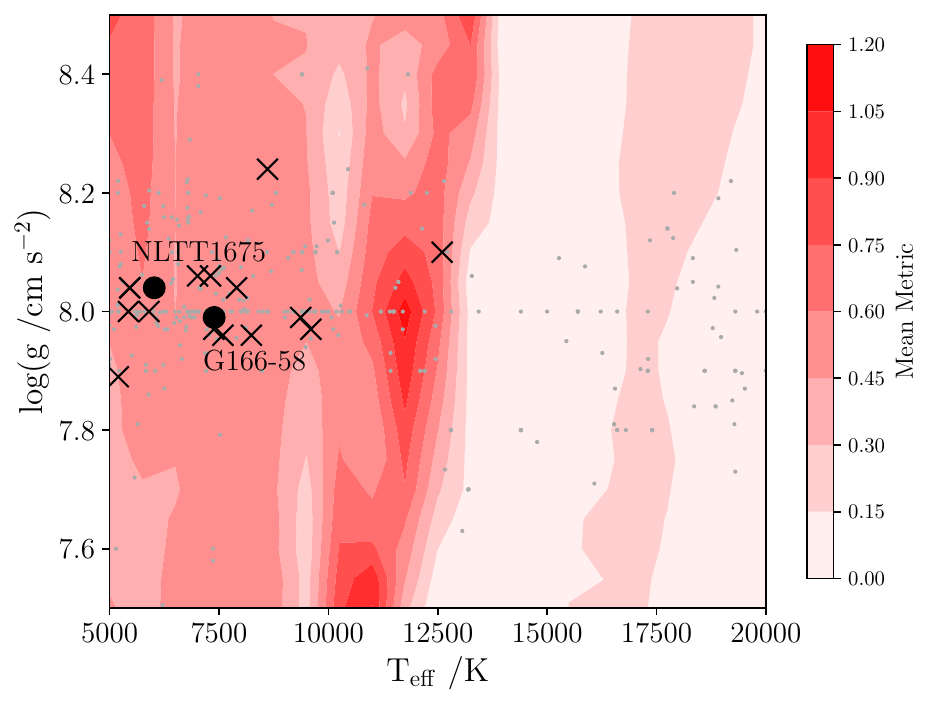}
    \caption{The susceptibility of H-dominated white dwarfs to a change in geological interpretation when switching between timescale grids SN and KN (i.e., using different modelling assumptions). The colour bar and label types are similar to those in Figure~\ref{fig:DMP_DA_OVERSHOOT}, but the black markers show the sample used for Test~4.}
    \label{fig:DMP_DA_BVK}
\end{figure}

\begin{figure}
    \centering
    \includegraphics[width=\columnwidth, keepaspectratio]{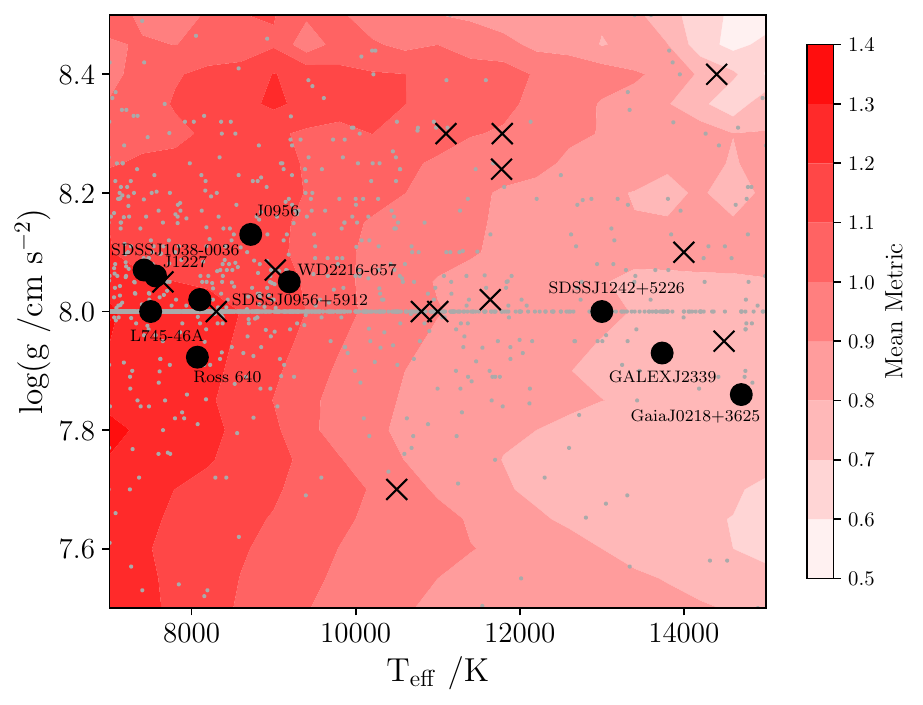}
    \caption{The susceptibility of He-dominated white dwarfs to a change in geological interpretation when switching between timescale grids SN and KN (i.e., using different modelling assumptions). The colour bar and label types are similar to those in Figure~\ref{fig:DMP_DA_OVERSHOOT}, but the black markers show the sample used for Test~5, and grey dots correspond to He-dominated white dwarfs.}
    \label{fig:DMP_DB_BVK}
\end{figure}

We illustrate the two samples in Figure~\ref{fig:DMP_DA_BVK} (H-dominated systems) and Figure~\ref{fig:DMP_DB_BVK} (He-dominated systems). These figures also repeat the methodology of Section~\ref{sec:overshoot_overview} to show the regions of parameter space in which discrepancies between the STELUM and Koester timescales are most likely to be relevant. We discuss results for selected H-dominated systems in Section~\ref{sec:bvk_h}. He-dominated systems are discussed in Section~\ref{sec:bvk_he} and 
% Since neither the KN nor SN grids represent a complete treatment of overshoot, the results in these sections are not necessarily definitive. We are primarily interested in the differences between the results obtained with each grid. 
Section~\ref{sec:bvk_overview} presents a broader overview.

%For H-dominated white dwarfs, the metals have a negligible effect on the atmospheric structure, hence boundary conditions of the envelope calculations.  The grids are based on the same set of equations for atomic diffusion, and apart from a small difference in the ML2/$\alpha$ convection parameterisation (see Table~\ref{tab:grid_summary})\footnote{We note that although ML2/$\alpha$ can be calibrated to reproduce the location of the convection zone's base in the 3D simulations, this approach neglects overshooting. As a result, no particular parameterisation is intrinsically preferred in this context.}, there is no expectation that one grid should be preferred to another. Nevertheless, these differences may propagate into the geological interpretation of accreted material

%For He-dominated objects, the Koester timescales are likely more accurate because they are calculated including the effect of trace metals, which we generally include in our timescale interpolation. The trace metals are the main electron donors, hence change the atmospheric stratification compared to the pure-He case, which has a feedback effect on the depth of the convection zone, hence on the depth where atomic diffusion occurs. This is what we call `boundary condition' effects. However, because convective overshoot is neglected, it is possible that neither grid gives a definitive result. More important in this section is whether the grids give different results from each other.

\subsection{H-dominated white dwarfs (Test~4)}
\label{sec:bvk_h}

For 2 of the 17 H-dominated entries we tested, we find that the favoured model changes when swapping between the SN and KN timescale grids. There were 3 entries which our code could not fit in any case (those associated with NLTT\,7547). There were also 2 systems for which there was, nominally, a change in the favoured model, but at least one of the model fits was unphysical (WD\,J113444.64+610826.68 and WD\,1455+298).

\begin{figure}
    \centering
    \includegraphics[width=\columnwidth, keepaspectratio]{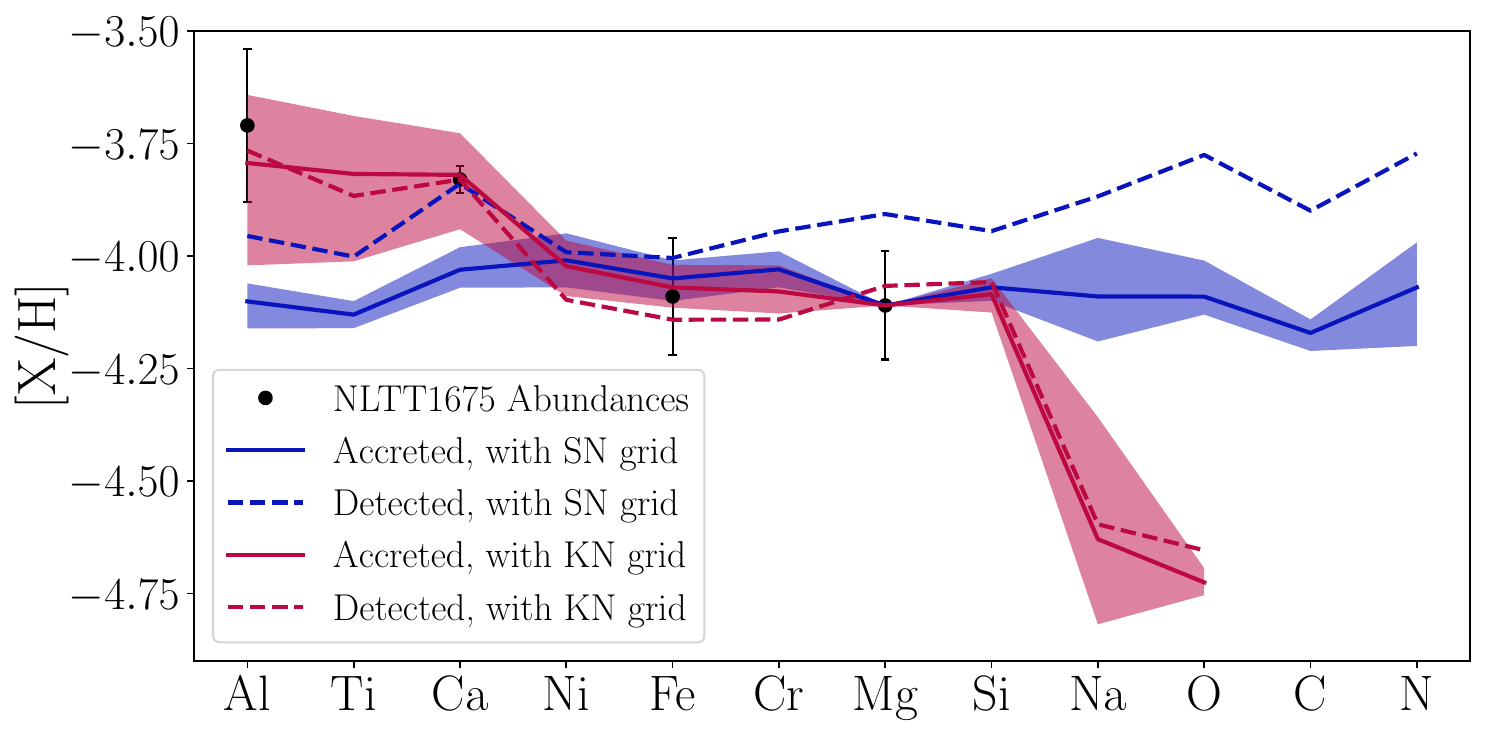}
    \caption{The geological interpretation of the metals accreted by NLTT1675 changes when switching between the STELUM (blue) and Koester (red) timescale grids, corresponding to different physical assumptions and treatments of the white dwarf physics. The plot features are similar to Figure~\ref{fig:WD0310-688}. Our Bayesian framework finds evidence of incomplete condensation only when using the Koester grid.}
    \label{fig:NLTT1675}
\end{figure}

For NLTT\,1675 \citep{Kawka2012}, incomplete condensation is invoked only when using the Koester grid (to $2.1\sigma$). The key ratio is Ca/Mg. The detected Ca/Mg is slightly elevated relative to solar, but for the Koester grid, Ca sinks relatively fast in comparison to Mg. Therefore, the best model favours a pollutant composition with elevated Ca, which is achieved by condensation at high temperature. The resulting composition is illustrated in Figure~\ref{fig:NLTT1675}. The accreted material has previously been compared with CI chondrites by \citet{Kawka2012}. Their results use timescales from \citet{Koester2009}, equivalent to an earlier version of our KN grid, and assume steady state accretion. For H-dominated white dwarfs it is often assumed that metal sinking timescales are sufficiently short, relative to the duration of an accetion event, that any other phase is highly unlikely to be observed. However, our results (with either timescale grid) favour a declining phase solution (with 94\% confidence). NLTT\,1675 is sufficiently cool that its metal sinking timescales are on the order of (at least) thousands of years. Our prior distribution for the accretion event duration goes down to 1 year, such that declining phase solutions are viable. In both cases, the median value of the accretion event duration was about 100 years (in units of log years, $2.4^{+1.8}_{-1.5}$ and $1.8^{+1.5}_{-1.2}$ for the SN and KN grids, respectively). The corresponding accretion rates (in units of log $\textrm{g}\,\textrm{s}^{-1}$) are $10.6^{+1.6}_{-1.8}$ or $9.5^{+1.2}_{-1.5}$ respectively, and total accreted masses in the range of a small moon. These accretion rates fall within the range of values inferred for other white dwarfs (see \citealt{Bauer2018}). Assuming a steady state solution, rather than declining phase, decreases the inferred Ca/Mg ratio of accreted material, bringing it closer to the chondritic value, because $\tau_{\rm Ca} < \tau_{\rm Mg}$.

% BN: Accretion timescale (log yr) is= 2.42885994383693 +1.78670819112906 -1.5337988471889, mass (log kg) is 17.5105844277604 +0.233084049484241 -0.195671030996131, acc rate (log g/s) is 10.5854708067764 +1.54799149070888 -1.77706206686753
%KN: 1.8216382068258 +1.50984366824586 -1.16168544273695 / 15.8307305668314 +0.271876914268661 -0.164819417676211 / 9.51380464188755 +1.20366176771332 -1.47181854347025

%[Note to self - the 2014 update for this one (NLTT01675 KawkaVennes2014) is  irrelevent, requotes  the same Ca abundnace]

For G\,166-58, incomplete condensation is inferred (to $1.3\sigma$ significance) only for the SN grid. This is primarily driven by the change in Ni and Fe sinking times, which are (relatively) long for the SN grid, such that additional Ni and Fe depletion due to incomplete condensation is favoured, albeit marginally.

%[Potential TODO: mention G74-7?]

\subsection{He-dominated white dwarfs (Test~5)}
\label{sec:bvk_he}

For 12/25 entries, there was no change in the overall interpretation, although this includes 3 entries for which our code could not fit the data (WD\,1350$-$162, due to high Na, and both entries for HS\,2253+8023).

There are 3 entries implying core-mantle differentiation when using the SN grid only (SDSS\,J124231.07+522626.6, SDSS\,J1038$-$0036 and J0956, to 4.7, 3.6 and 2.4$\sigma$ respectively). In all 3 cases, the accreted material is inferred to be mantle-rich. This can be understood as a consequence of the siderophile elements Ni and Fe sinking on (relatively) shorter timescales for the KN grid, such that depletions in these elements can be attributed to differential sinking. The magnitude of the timescale change is not large, but is exacerbated by the inference that all three systems are in the declining phase of accretion.

Given this behaviour, it can be expected that the opposite outcome is possible: inference of core-rich material, but for the Koester timescales only. We indeed find an example of this outcome in WD\,2216-657, to $1.8\sigma$ significance. However, in this case the interpretation is also dependent on the values of $T_{\rm eff}$ and $\log(g)$, with the inference of core-rich material occurring only with \textit{Gaia} atmospheric parameter values (and Koester timescales).

For 6 entries (GALEXJ2339, WD\,1425+540 (Model 1), L\,745-46A, WD\,1232+563 (Xu 2019), J1227 and Gaia\,J0218+3625), evidence for incomplete condensation is, nominally, ambiguous. For WD\,1425+540, the change in favoured model is marginal ($1.2\sigma$) and can be attributed to random chance, and for WD\,1232+563, our results are not trustworthy (see Section \ref{sec:Test2}).

For the remaining 4 entries, unlike the case of differentiation, there is no consistent reason for the ambiguity. For L\,745-46A, incomplete condensation is invoked with $2.7\sigma$ significance for the SN grid only. This is primarily due to the Fe sinking timescale, which is (relatively) longer when using the SN grid. This is roughly cancelled out by incomplete condensation, which slightly depletes Fe relative to the non-volatile rock forming metals. For GALEX2339 and J1227, incomplete condensation is invoked for the SN grid only (to $2.2\sigma$ and $1.7\sigma$ significance respectively) in order to match the abundance (or upper bound) of Na, a volatile metal. Na sinks (relatively) slowly for the SN grid, but not the KN grid, such that incomplete condensation is required to deplete it. Notably, GALEX2339 exhibits a high Be abundance. We do not model Be - if we did, it is unlikely that the geological processes we consider would be able to account for it. For Gaia J0218+3625 the key metal is O. For the Koester grid, O has a (relatively) long sinking timescale, such that additional heating is required to deplete it in this case only (to $2.3\sigma$ significance). In each of these four cases, except L\,745-46A, the choice of $T_{\rm eff}$ and $\log(g)$ has an impact: the ambiguity noted above is only present for either the \textit{Gaia} or PEWDD values, but not both. In other words, the presence of incomplete condensation depends not only on the timescale grid adopted, but also on the white dwarf parameter values.
% GALEX = 2.2, 1425 = 1.2, L = 2.7, 1232 1.4.
% 1227 = 1.7, 0218 = 2.3 (with KN!)

%[Note that the worst offenders here are the ones with a big discrepency in how deep into the declining phase we are! In other words, really the issue here is with determining D]

There are three entries for which the presence of incomplete condensation and differentiation are both ambiguous. These are Ross 640 and both entries for SDSS\,J0956+5912, but note that these are practically identical.

For Ross 640, incomplete condensation is invoked when using the SN grid to $3.0\sigma$. This appears to be driven by Ca, which sinks relatively slowly for SN (especially relative to Fe) such that the pollutant is inferred to be rich in Ca, implying thermal processing. This occurs when using white dwarf parameters from \textit{Gaia}, which we consider marginally preferable in this case as the photometry avoids saturation. When using parameters from \citet{Blouin2018,Caron2023}, differentiation is additionally invoked (to $1.4\sigma$) for the SN grid only, for the reasons previously outlined.

For the two entries corresponding to white dwarf SDSS\,J0956+5912 which use data from \citet{Hollands2022}, incomplete condensation and differentiation are both invoked to $4.3\sigma$, only for the SN timescales. This is a combination of Ca sinking slower and siderophiles sinking faster for the KN grid.

\begin{figure}
    \centering
    \includegraphics[width=\columnwidth, keepaspectratio]{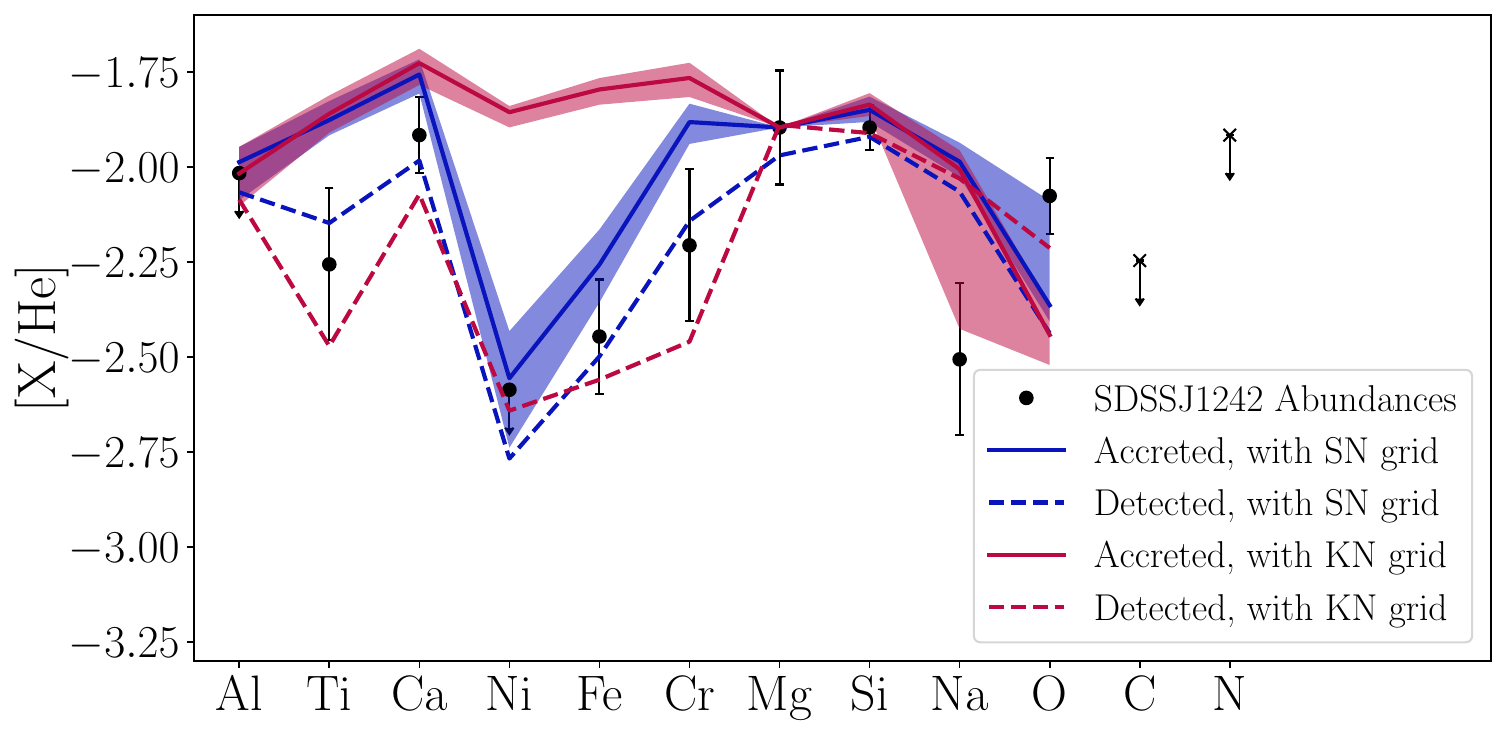}
    \caption{The geological interpretation of the metals accreted by SDSS\,J124231.07+522626.6 changes when switching between the SN (blue) and KN (red) timescale grids, corresponding to different physical assumptions and treatments of the white dwarf atmospheric physics. The plot features are similar to Figure~\ref{fig:WD0310-688}. Our Bayesian framework finds (strong) evidence for accretion of mantle-rich material, and therefore core-mantle differentiation, only when using the SN grid. The data points for C and N are both upper bounds, and were both excluded when fitting the data because our incomplete condensation model does not accurately handle sub-Solar values for these elements. All four compositions shown here predict negligible quantities of C and N, so are consistent with the upper bounds.}
    \label{fig:SDSSJ1242}
\end{figure}

\subsection{Overview and discussion}
\label{sec:bvk_overview}

Figure~\ref{fig:DMP_DA_BVK} shows the regions of parameter space in which discrepancies between the STELUM and Koester timescales are most likely to be relevant for H-dominated white dwarfs. The most affected region is a narrow band at around \SI{12000}{K} due to differences in the mixing-length parameter. However, this part of parameter space is relatively sparsely populated. The white dwarfs we identified as having ambiguous geological interpretations have lower $T_{\rm eff}$. This is because at lower temperatures, sinking timescales become longer, and declining phase solutions are viable. However, this is dependent on our prior distribution of accretion event duration. If the accretion event duration is assumed to always be at least $10^5$\;yr, it is likely that none of these systems would be affected by the choice of grid used here.

Figure~\ref{fig:DMP_DB_BVK} shows the equivalent for He-dominated white dwarfs. Here, the population of He-dominated white dwarfs (grey dots) is most concentrated in the most severely affected region of parameter space. The large difference seen at low temperature is likely due to the improved treatment of the boundary condition in the Koester code. However, this plot shows that the difference between the Koester and STELUM timescales can be relevant for almost any He-dominated white dwarf. Such systems are far more likely to be in the declining phase due to their long sinking timescales, exacerbating this issue. Therefore, the input physics appears to be the most important source of uncertainty for He-rich white dwarfs. In fact, the comparison presented here does not even tell the full story, as some modelling aspects are treated identically in the Koester and STELUM frameworks. For instance, both codes use the He equation of state of \citet{Saumon1995}, which is highly uncertain in the regime of cool white dwarf envelopes and directly affects the depth of the convection zone \citep{Koester2015}. Both codes also handle diffusion using the screened Coulomb potential approximation of \citet{Paquette1986a}, which breaks down in cool He-rich white dwarfs \citep{Heinonen2020}. More theoretical work on sinking timescales is clearly needed. In the meantime, our tests highlight that the Koester timescales tend to shift inferred pollutant compositions towards core-rich material and away from mantle-rich material, while the STELUM timescales have the opposite effect.

%[TODO: the most affected systems seem to be the ones which are deep into declining phase (quote c for Ross 640 and 0956?)] - update: the numbers do not support this!

%We do not necessarily favour a particular choice out of the STELUM or Koester timescales. Rather, we aim to highlight the potential impacts of this choice.

\section{Systems with robust interpretations}
\label{sec:robust}

\begin{figure*}
    \centering
    \includegraphics[width=\textwidth, keepaspectratio]{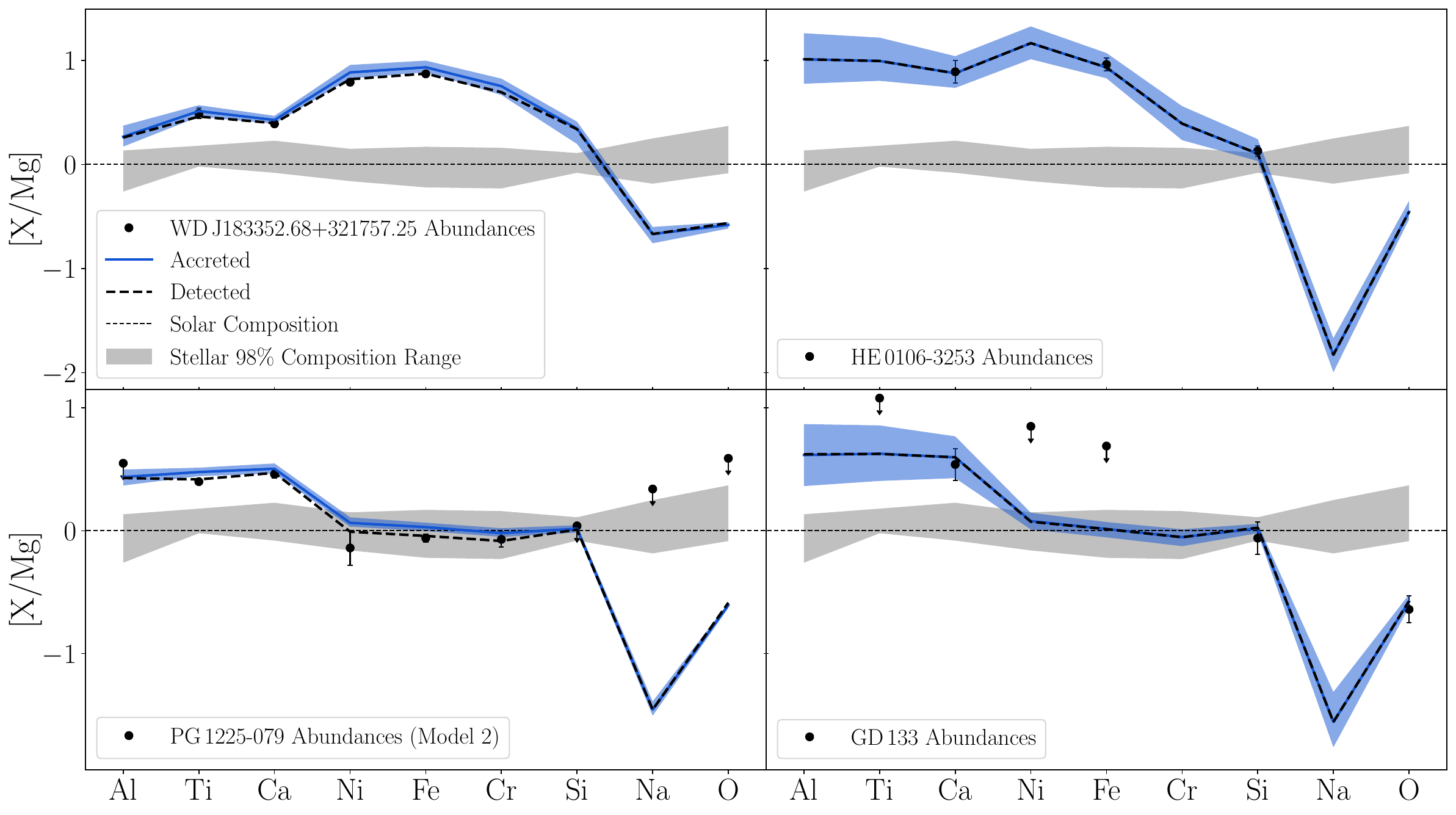}
    \caption{Representative model fits to the abundance data for four systems whose geological interpretation is robust against changes in white dwarf parameter values, sinking timescale grids, and treatment of thermohaline mixing. The blue line shows the inferred (median) composition of the accreted material, along with a $1\sigma$ confidence interval. The dashed black line shows the resulting detected composition, after accounting for differential sinking, excluding a confidence interval for visual clarity. The dashed black line may be compared against the data. For WDJ\,183352+321757 and PG\,1225$-$079, the example fit shown here uses the KN grid, and for PG\,1225$-$079 we adopt abundance values for the `Model 2' entry. For HE\,0106$-$3253 and GD\,133 we use the S3D grid, and include thermohaline mixing. Because thermohaline mixing is inferred to be active, and because it disables differential sinking, the accreted and detected compositions are close to identical in these cases. In all cases we use white dwarf parameters from \textit{Gaia}.}
    \label{fig:robust_systems}
\end{figure*}

So far, we have focussed our attention on systems for which the interpretation of accreted material depends on the treatment of the white dwarf itself. It is also important to point out that there are are systems whose geological interpretation does not change, and so appears to be qualitatively robust. This can occur when certain metals are so strongly enhanced or depleted (relative to uncertainties) that a certain geological interpretation is inevitable, regardless of the sinking timescales used. We describe the four most statistically significant examples of such systems, drawn from the samples investigated in this work. These are WDJ\,183352+321757, HE\,0106$-$3253, PG\,1225$-$079 and GD\,133. To confirm that our conclusions for these systems are robust, we have rerun our code using all relevant combinations of timescale grids\footnote{Specifically, the KN, KF, and S3D/SV grids.}, white dwarf parameters, and inclusion/exclusion of thermohaline mixing. We illustrate an example of our model fits to the data for each of these systems in Figure~\ref{fig:robust_systems}.

Our metric predicts that WDJ\,183352+321757 should be highly sensitive to the choice of STELUM or Koester timescale grids. In practice, its atmosphere is so strongly enriched in Fe and Ni that a core--mantle differentiation is always reached regardless, with very high confidence ($>$$30\sigma$). We also consistently find strong evidence for incomplete condensation ($>$$10\sigma$). The overall interpretation, which is qualitatively consistent in all cases, is that the accreted material originated in a parent body which formed at high temperature (with a median temperature between \SI{1570}{K} and \SI{1580}{K} in all cases), and which differentiated into a core and mantle in Earth-like proportions - the molar core fraction is close to 0.16 in all cases, close to an Earth-like value of roughly 0.17 \citep{McDonough2003}. The accreted material itself was predominantly core-like, with a molar core fraction ranging between 0.59 and 0.60. We also find that accretion has likely reached the declining phase, with probability $>99.6\%$.
%PEWDD:
%kn 11.1/31.3/0.6/0.16/1570/99.9
%ko 11.2/31.3/0.60/0.16/1580/99.8
%vo 10.7/31.2/0.59/0.16/1570/99.8

%GR:
%kn 10.8/31.3/0.6/0.16/1570/99.9
%ko 10.9/31.3/0.6/0.16/1570/99.9
%vo 11.4/31.3/0.59/0.16/1570/99.6

Similarly, we find that for HE\,0106$-$3253, our results show strong evidence for both incomplete condensation and core--mantle differentiation, due to its relatively high Ca and Fe abundances respectively. The confidence level for incomplete condensation ranges from $7.8\sigma-9.4\sigma$, with the equivalent range for differentiation being $7.8\sigma-9.6\sigma$. In all cases, we find that this white dwarf has accreted core-rich material (with molar core fraction ranging from 0.49 to 0.64). We also infer that the parent body of the accreted material had a relatively low (molar) core fraction, with values ranging from 0.04 to 0.07. This is comparable to Mars, for which the equivalent value is approximately 0.08 \citep{Yoshizaki2020}.
%For reference the equivalent values for the Earth and Moon are approximately 0.17 and 0.005 \citep{McDonough2003,Sossi2025}.
%PEWDD:
%knn 9.3/9.6/0.64/0.05
%knt 7.8/7.8/0.49/0.07
%kon 9.4/9.5/0.64/0.05
%kot 7.8/7.8/0.49/0.07
%3pn 9.0/9.0/0.6/0.04
%3pt 7.8/7.8/0.47/0.07

%GR:
%knn 9.3/8.0/0.65/0.05
%knt 7.8/7.5/0.49/0.07
%kon 9.2/9.2/0.65/0.05
%kot 7.8/7.8/0.49/0.07
%3pn 8.4/8.4/0.55/0.03
%3pt 7.8/6.9/0.49/0.07

\citet{Klein2011} and \citet{Xu2013} found that PG\,1225$-$079 had likely accreted refractory-enriched (i.e., volatile depleted) material. We find that this conclusion is robust. In all cases, and for all sets of abundances and parameters from both of these papers, incomplete condensation is confidently inferred (to $>8\sigma$). This is driven by the relatively high abundances of the refractory metals Ti and Ca. The inferred condensation temperature is between \SI{1390}{K} and \SI{1500}{K} in all cases. We do not include the Sc, V and Mn detections or C, Be, S, Zn and Sr upper bounds in our modelling, but these do not obviously contradict our conclusion. We also consistently find a high ($>96\%$) probability that accretion has reached the declining phase.
%M1
%PEWDD:
%kn 16.6/99/1390
%ko 16.5/99/1390
%vo 16.0/99/1400
%GR:
%kn 16.7/99/1390
%ko 16.5/ 99/1390
%vo 16.1/99/1400

%M2
%PEWDD:
%kn 16.4/99/1400
%ko 17.0/97.7/1400
%vo 17.0/99/1400
%GR:
%kn 16.9/98.7/1400
%ko 17.0/ 98.1/1400
%vo 16.8/97.1/1400

%M3
%PEWDD:
%kn 16.6/97.6/1390
%ko 17.0/98.6/1480
%vo 16.9/99.6/1460
%GR:
%kn 16.7/99/1500
%ko 16.5/99.3/1400
%vo 17.2/99.3/1400

%Xu2013
% 8.9/99.0/1400
% 8.9/98.4/1400
% 9.2/99.0/1400

GD\,133 exhibits a depletion of O, as well as a slight enrichment in the refractory element Ca, relative to Solar. Our framework robustly interprets this as evidence of incomplete condensation for every combination of parameters, with significance varying between 5.4 and 6.1$\sigma$. The associated temperature is consistently inferred to be high, with a median value between \SI{1410}{K} and \SI{1520}{K}.

%As a representative example, when using the S3D timescale grid, white dwarf parameters from PEWDD, and including thermohaline mixing, the temperature is $1420^{+30}_{-40}$\,K.
%PEWDD:
%knn 5.4/1440
%knt 5.8/1520
%kon 5.8/1460
%kot 5.9/1420
%3pn 6.1/1450
%3pt 5.5 1420 +30 -40
%GR:
%knn 5.4/1440
%knt 5.8/1510
%kon 5.9/1460
%kot 5.7/1410
%3pn 6.0/1440
%3pt 5.8/1420

We emphasise that these results, and in particular the extremely high confidence levels assigned to them in some cases, are dependent on the geological forward models within our Bayesian framework. Beyond a certain confidence level, the specific values are not very meaningful because the `weak link' is the model itself. Regardless, the key point we make here is that these results appear to be robust against the systematics under consideration in this paper. We also note that our samples were intentionally drawn from those regions of parameter space which are least conducive to robust results. Outside of these regions, we expect that there should be a greater proportion of systems with robust interpretations, all else being equal.

\section{Summary and Conclusions}
\label{sec:conclusions}

We have shown that the assumptions made about white dwarf physics can affect the geological conclusions drawn from material accreted by white dwarfs. This is because those assumptions affect the relative sinking timescales of different metals, which propagate into the inferred composition of the accreted material. We introduced a metric to quantify the discrepancy between two grids of sinking timescales as a function of $T_{\rm eff}$ and $\log(g)$. Our key findings are as follows.

%Different treatments of white dwarf atmospheric physics, including the extent of convective overshoot, affect the predicted (relative) metal sinking timescales. We have demonstrated that this change can propagate into a discrepancy in the geological interpretation of accreted exoplanetary material. The most relevant physical modelling assumptions depending on the dominant element in the white dwarf's atmosphere, as well as its temperature.

\textit{Convective overshoot:} For H-dominated white dwarfs, there is a sensitive regime between approximately \SI{12000}{K} and \SI{18000}{K}. In this zone, including a full 3D treatment of convective overshoot causes Fe and O to sink on (relatively) longer timescales compared to the no overshoot case. Including overshoot will, in general, decrease the inferred oxygen and iron content of accreted material. In borderline cases, neglecting overshoot can introduce spurious evidence of core-like material and/or an oxygen excess. We recommend using sinking timescales from the full 3D treatment of convective overshoot where possible, especially in this temperature range.

For He-dominated white dwarfs, the impact of convective overshoot is relatively minor, although this needs to be fully confirmed with 3D overshoot modelling. We show that, for most He-dominated white dwarfs (approximately speaking, those below about \SI{16000}{K}), ignoring convective overshoot is preferable to using overshoot fixed to 1 pressure scale height, as benchmarked against a variable treatment of overshoot.

%For H-dominated atmospheres, boundary conditions are important for a narrow temperature band at around \SI{12000}{K}. They are also potentially important at temperatures below about \SI{10000}{K}, but this is contingent on declining phase solutions being plausible for such systems.

\textit{Thermohaline mixing:} The primary effect of thermohaline mixing is to increase the inferred accretion rate at H-dominated white dwarfs. The strength of this effect is primarily temperature-dependent, becoming stronger for hotter white dwarfs. For sufficiently high temperatures, thermohaline mixing becomes strong enough to disable differential sinking, potentially affecting geological interpretation. This is also dependent on the inferred accretion rate. The impact of thermohaline mixing on geology is to cause the inferred composition of accreted material to tend towards more mantle-rich/core-poor solutions, and also towards more volatile-rich solutions. While thermohaline mixing is expected to be important for H-dominated white dwarfs, the larger accretion rates implied by thermohaline mixing may be harder to explain dynamically. We have quantified this expectation in terms of Bayesian evidence, finding a mean improvement in $\ln(\mathcal{Z})$ of 4.8 when neglecting thermohaline mixing. We also find that the increased accretion rate inferred for G\,29-38 is harder to reconcile with X-ray measurements.

\textit{Other physical assumptions:} For He-dominated white dwarfs, differences in input physics (as captured by differences between STELUM and Koester sinking timescales) are potentially important across the full range of parameter space we tested, especially below about \SI{10000}{K} where the Koester code uses detailed boundary conditions. However, the sensitive regime is not clearly defined because the long sinking timescales associated with He-dominated atmospheres mean that any such white dwarf could plausibly be in the declining phase of accretion, which exacerbates the effects of timescale discrepancies. The clearest impact is that the Koester timescales generally tend towards core-rich/mantle-poor solutions compared to the STELUM  timescales. Since both of these grids (or close approximations) are publicly available, we recommend that both options be considered, and any discrepancies in the resulting interpretation reported.

%For He-dominated white dwarfs, there is less clear dependence on temperature. This is because any He-dominated white dwarf can be in the declining phase of accretion, which makes it vulnerable to changes in sinking timescales. The most important effect is [mixing length?], especially for cooler white dwarfs. [Add comments on how to handle this practically! Can we focus on certain elements, or determine D using Ca/Al?]

The systematic effects described above can be mitigated against by modelling systems multiple times, covering a range of possible assumptions. We performed this exercise and identified four systems which, in the context of our model, yield robust geological interpretations at a high level of confidence. Two systems, PG\,1225$-$079 and GD\,133, have accreted material which experienced incomplete condensation at high temperature (roughly \SI{1400}{K}). Two other systems, WD\,J183352+321757 and  HE\,0106$-$3253, witnessed both incomplete condensation and core--mantle differentiation, followed by accretion of a core-rich fragment of the resulting material.

In the longer term, models of white dwarf pollution should introduce uncertainties on (relative) sinking timescales, and propagate these through the forward model. These uncertainties should capture the discrepancies which can plausibly be introduced by a change in white dwarf modelling assumptions, and any correlation between them. In many cases, these uncertainties will be small (relative to uncertainties on the abundances themselves). However, in cases where they are non-negligible, this strategy should ensure that any geological conclusions are robust.

Geological modelling can itself be used to investigate whether certain treatments of element transport processes are favoured. This would require a compositional model of the reservoir(s) of material which pollute white dwarfs, and a model for the typical timing of accretion events. The distribution of detected metal abundances across a large population of white dwarfs could be predicted using different sets of sinking timescales, and tested against data from a representative real sample.

%And maybe also say that future work could work out whether geology points towards a particular treatment of overshoot by modelling a population under both assumptions

%Other qs to consider: For the He ones, is there still a Al-Ti-Ca system equivalent where, regardless of the grid you're using, you can kind of tell how deep into the declining phase you are? I guess in principle the answer must be yes? But perhaps the ratio varies across parameter space so the advice will be avoid anything with high Ti/Ca above 15000K or anything with low Al/Ti below 5000K or something like that

%Basically it would be good to know which WDs we should focus on and which ones not to focus on. For the He-dominated ones, maybe this comes down to which ones you can tell are in declining phase a priori?

\section*{Acknowledgements}

AMB was supported by a Leverhulme Trust Grant (ID RPG-2020-366). This research received funding from the European Research Council under the European Union’s Horizon 2020 research and innovation programme number 101002408 (MOS100PC). TC was supported by NASA through the NASA Hubble Fellowship grant HST-HF2-51527.001-A awarded by the Space Telescope Science Institute, which is operated by the Association of Universities for Research in Astronomy, Inc., for NASA, under contract NAS5-26555. The authors would like to acknowledge the University of Warwick Research Technology Platform SCRTP for assistance in the research described in this paper. We thank Amy Bonsor and Paula Izquierdo for useful discussions. We also thank the anonymous referee for their helpful feedback, which improved the quality of the manuscript.

%%%%%%%%%%%%%%%%%%%%%%%%%%%%%%%%%%%%%%%%%%%%%%%%%%
\section*{Data Availability}
\label{sec:data}

The geological inference code used in this work is publicly available at \url{https://github.com/andrewmbuchan4/PyllutedWD_Public}. White dwarf data are publicly available at \url{https://github.com/jamietwilliams/PEWDD}. The S3D and SV sinking timescale grids may be found at \url{https://zenodo.org/records/17313701}. We plan to implement the SV grid on the MWDD in the near future to incorporate the effect of convective overshoot. Therefore, our comments on the SN and SV grids apply to the MWDD versions before and after this change, respectively. Other sinking timescales calculated in this work will be shared upon request.


\bibliographystyle{mnras}
\bibliography{references.bib}

\begin{thebibliography}{}
\makeatletter
\relax
\def\mn@urlcharsother{\let\do\@makeother \do\$\do\&\do\#\do\^\do\_\do\%\do\~}
\def\mn@doi{\begingroup\mn@urlcharsother \@ifnextchar [ {\mn@doi@} {\mn@doi@[]}}
\def\mn@doi@[#1]#2{\def\@tempa{#1}\ifx\@tempa\@empty \href {http://dx.doi.org/#2} {doi:#2}\else \href {http://dx.doi.org/#2} {#1}\fi \endgroup}
\def\mn@eprint#1#2{\mn@eprint@#1:#2::\@nil}
\def\mn@eprint@arXiv#1{\href {http://arxiv.org/abs/#1} {{\tt arXiv:#1}}}
\def\mn@eprint@dblp#1{\href {http://dblp.uni-trier.de/rec/bibtex/#1.xml} {dblp:#1}}
\def\mn@eprint@#1:#2:#3:#4\@nil{\def\@tempa {#1}\def\@tempb {#2}\def\@tempc {#3}\ifx \@tempc \@empty \let \@tempc \@tempb \let \@tempb \@tempa \fi \ifx \@tempb \@empty \def\@tempb {arXiv}\fi \@ifundefined {mn@eprint@\@tempb}{\@tempb:\@tempc}{\expandafter \expandafter \csname mn@eprint@\@tempb\endcsname \expandafter{\@tempc}}}

\bibitem[\protect\citeauthoryear{Badenas-Agusti, Viaña, Vanderburg, Blouin, Dufour, Xu  \& Sha}{Badenas-Agusti et~al.}{2024}]{BadenasAgusti2024}
Badenas-Agusti M.,  Viaña J.,  Vanderburg A.,  Blouin S.,  Dufour P.,  Xu S.,   Sha L.,  2024, \mn@doi [Monthly Notices of the Royal Astronomical Society] {10.1093/mnras/stae421}, 529, 1688

\bibitem[\protect\citeauthoryear{Bauer \& Bildsten}{Bauer \& Bildsten}{2018}]{Bauer2018}
Bauer E.~B.,  Bildsten L.,  2018, \mn@doi [The Astrophysical Journal Letters] {10.3847/2041-8213/aac492}, 859, L19

\bibitem[\protect\citeauthoryear{Bauer \& Bildsten}{Bauer \& Bildsten}{2019}]{Bauer2019}
Bauer E.~B.,  Bildsten L.,  2019, \mn@doi [The Astrophysical Journal] {10.3847/1538-4357/ab0028}, 872, 96

\bibitem[\protect\citeauthoryear{{B{\'e}dard}}{{B{\'e}dard}}{2024}]{Bedard2024}
{B{\'e}dard} A.,  2024, \mn@doi [\apss] {10.1007/s10509-024-04307-5}, \href {https://ui.adsabs.harvard.edu/abs/2024Ap&SS.369...43B} {369, 43}

\bibitem[\protect\citeauthoryear{{B{\'e}dard}, {Brassard}, {Bergeron}  \& {Blouin}}{{B{\'e}dard} et~al.}{2022a}]{Bedard2022_STELUM}
{B{\'e}dard} A.,  {Brassard} P.,  {Bergeron} P.,   {Blouin} S.,  2022a, \mn@doi [\apj] {10.3847/1538-4357/ac4497}, \href {https://ui.adsabs.harvard.edu/abs/2022ApJ...927..128B} {927, 128}

\bibitem[\protect\citeauthoryear{{B{\'e}dard}, {Bergeron}  \& {Brassard}}{{B{\'e}dard} et~al.}{2022b}]{Bedard2022_DQ}
{B{\'e}dard} A.,  {Bergeron} P.,   {Brassard} P.,  2022b, \mn@doi [\apj] {10.3847/1538-4357/ac609d}, \href {https://ui.adsabs.harvard.edu/abs/2022ApJ...930....8B} {930, 8}

\bibitem[\protect\citeauthoryear{{B{\'e}dard}, {Bergeron}  \& {Brassard}}{{B{\'e}dard} et~al.}{2023}]{Bedard2023}
{B{\'e}dard} A.,  {Bergeron} P.,   {Brassard} P.,  2023, \mn@doi [\apj] {10.3847/1538-4357/acbb62}, \href {https://ui.adsabs.harvard.edu/abs/2023ApJ...946...24B} {946, 24}

\bibitem[\protect\citeauthoryear{Blouin, Dufour  \& Allard}{Blouin et~al.}{2018}]{Blouin2018}
Blouin S.,  Dufour P.,   Allard N.~F.,  2018, \mn@doi [The Astrophysical Journal] {10.3847/1538-4357/aad4a9}, 863, 184

\bibitem[\protect\citeauthoryear{{B{\"o}hm-Vitense}}{{B{\"o}hm-Vitense}}{1958}]{BohmVitense1958}
{B{\"o}hm-Vitense} E.,  1958, \zap, \href {https://ui.adsabs.harvard.edu/abs/1958ZA.....46..108B} {46, 108}

\bibitem[\protect\citeauthoryear{Buchan}{Buchan}{2023}]{BuchanThesis}
Buchan A.,  2023, PhD thesis, Apollo - University of Cambridge Repository, \mn@doi{10.17863/CAM.104327}, \url {https://www.repository.cam.ac.uk/handle/1810/361540}

\bibitem[\protect\citeauthoryear{Buchan, Bonsor, Shorttle, Wade, Harrison, Noack  \& Koester}{Buchan et~al.}{2022}]{Buchan2021}
Buchan A.~M.,  Bonsor A.,  Shorttle O.,  Wade J.,  Harrison J.,  Noack L.,   Koester D.,  2022, \mn@doi [Monthly Notices of the Royal Astronomical Society] {10.1093/mnras/stab3624}, 510, 3512

\bibitem[\protect\citeauthoryear{Buchan, Bonsor, Rogers, Brouwers, Shorttle  \& Tremblay}{Buchan et~al.}{2024}]{Buchan2024}
Buchan A.~M.,  Bonsor A.,  Rogers L.~K.,  Brouwers M.~G.,  Shorttle O.,   Tremblay P.-E.,  2024, \mn@doi [Monthly Notices of the Royal Astronomical Society] {10.1093/mnras/stae1608}, 532, 2705

\bibitem[\protect\citeauthoryear{{Caron}, {Bergeron}, {Blouin}  \& {Leggett}}{{Caron} et~al.}{2023}]{Caron2023}
{Caron} A.,  {Bergeron} P.,  {Blouin} S.,   {Leggett} S.~K.,  2023, \mn@doi [\mnras] {10.1093/mnras/stac3733}, \href {https://ui.adsabs.harvard.edu/abs/2023MNRAS.519.4529C} {519, 4529}

\bibitem[\protect\citeauthoryear{Cresswell, Fraser, Bauer, Anders  \& Brown}{Cresswell et~al.}{2025}]{Cresswell2025}
Cresswell I.~G.,  Fraser A.~E.,  Bauer E.~B.,  Anders E.~H.,   Brown B.~P.,  2025, \mn@doi [The Astrophysical Journal Letters] {10.3847/2041-8213/addbd5}, 986, L10

\bibitem[\protect\citeauthoryear{{Cukanovaite}, {Tremblay}, {Freytag}, {Ludwig}, {Fontaine}, {Brassard}, {Toloza}  \& {Koester}}{{Cukanovaite} et~al.}{2019}]{Cukanovaite2019}
{Cukanovaite} E.,  {Tremblay} P.~E.,  {Freytag} B.,  {Ludwig} H.~G.,  {Fontaine} G.,  {Brassard} P.,  {Toloza} O.,   {Koester} D.,  2019, \mn@doi [\mnras] {10.1093/mnras/stz2656}, \href {https://ui.adsabs.harvard.edu/abs/2019MNRAS.490.1010C} {490, 1010}

\bibitem[\protect\citeauthoryear{Cukanovaite, Tremblay, Bergeron, Freytag, Ludwig  \& Steffen}{Cukanovaite et~al.}{2021}]{Cukanovaite2021}
Cukanovaite E.,  Tremblay P.-E.,  Bergeron P.,  Freytag B.,  Ludwig H.-G.,   Steffen M.,  2021, \mn@doi [Monthly Notices of the Royal Astronomical Society] {10.1093/mnras/staa3684}, 501, 5274

\bibitem[\protect\citeauthoryear{Cunningham, Tremblay, Freytag, Ludwig  \& Koester}{Cunningham et~al.}{2019}]{Cunningham2019}
Cunningham T.,  Tremblay P.-E.,  Freytag B.,  Ludwig H.-G.,   Koester D.,  2019, \mn@doi [Monthly Notices of the Royal Astronomical Society] {10.1093/mnras/stz1759}, 488, 2503

\bibitem[\protect\citeauthoryear{Cunningham et~al.,}{Cunningham et~al.}{2021}]{Cunningham2021}
Cunningham T.,  et~al., 2021, \mn@doi [Monthly Notices of the Royal Astronomical Society] {10.1093/mnras/stab553}, 503, 1646

\bibitem[\protect\citeauthoryear{{Cunningham}, {Wheatley}, {Tremblay}, {G{\"a}nsicke}, {King}, {Toloza}  \& {Veras}}{{Cunningham} et~al.}{2022}]{Cunningham2022}
{Cunningham} T.,  {Wheatley} P.~J.,  {Tremblay} P.-E.,  {G{\"a}nsicke} B.~T.,  {King} G.~W.,  {Toloza} O.,   {Veras} D.,  2022, \mn@doi [\nat] {10.1038/s41586-021-04300-w}, \href {https://ui.adsabs.harvard.edu/abs/2022Natur.602..219C} {602, 219}

\bibitem[\protect\citeauthoryear{Cunningham et~al.,}{Cunningham et~al.}{2025}]{Cunningham2025}
Cunningham T.,  et~al., 2025, \mn@doi [Monthly Notices of the Royal Astronomical Society] {10.1093/mnras/staf428}, 539, 2021

\bibitem[\protect\citeauthoryear{Deal, Deheuvels, Vauclair, Vauclair  \& Wachlin}{Deal et~al.}{2013}]{Deal2013}
Deal M.,  Deheuvels S.,  Vauclair G.,  Vauclair S.,   Wachlin F.~C.,  2013, \mn@doi [A&A] {10.1051/0004-6361/201322206}, 557, L12

\bibitem[\protect\citeauthoryear{Dohnanyi}{Dohnanyi}{1969}]{Dohnanyi1969}
Dohnanyi J.~S.,  1969, \mn@doi [\jgr] {10.1029/JB074i010p02531}, \href {https://ui.adsabs.harvard.edu/abs/1969JGR....74.2531D} {74, 2531}

\bibitem[\protect\citeauthoryear{Doyle et~al.,}{Doyle et~al.}{2023}]{Doyle2023}
Doyle A.~E.,  et~al., 2023, \mn@doi [The Astrophysical Journal] {10.3847/1538-4357/acbd44}, 950, 93

\bibitem[\protect\citeauthoryear{{Dufour}, {Blouin}, {Coutu}, {Fortin-Archambault}, {Thibeault}, {Bergeron}  \& {Fontaine}}{{Dufour} et~al.}{2017}]{MWDD}
{Dufour} P.,  {Blouin} S.,  {Coutu} S.,  {Fortin-Archambault} M.,  {Thibeault} C.,  {Bergeron} P.,   {Fontaine} G.,  2017, in {Tremblay} P.~E.,  {Gaensicke} B.,   {Marsh} T.,  eds,  Astronomical Society of the Pacific Conference Series Vol. 509, 20th European White Dwarf Workshop. p.~3 (\mn@eprint {arXiv} {1610.00986})

\bibitem[\protect\citeauthoryear{{Dupuis}, {Fontaine}, {Pelletier}  \& {Wesemael}}{{Dupuis} et~al.}{1992}]{Dupuis1992}
{Dupuis} J.,  {Fontaine} G.,  {Pelletier} C.,   {Wesemael} F.,  1992, \mn@doi [\apjs] {10.1086/191728}, \href {https://ui.adsabs.harvard.edu/abs/1992ApJS...82..505D} {82, 505}

\bibitem[\protect\citeauthoryear{{Elms} et~al.,}{{Elms} et~al.}{2024}]{Elms2024}
{Elms} A.~K.,  et~al., 2024, \mn@doi [\mnras] {10.1093/mnras/stae2265}, \href {https://ui.adsabs.harvard.edu/abs/2024MNRAS.534.2758E} {534, 2758}

\bibitem[\protect\citeauthoryear{Farihi, Jura  \& Zuckerman}{Farihi et~al.}{2009}]{Farihi2009}
Farihi J.,  Jura M.,   Zuckerman B.,  2009, \mn@doi [The Astrophysical Journal] {10.1088/0004-637x/694/2/805}, 694, 805

\bibitem[\protect\citeauthoryear{Farihi, Brinkworth, Gänsicke, Marsh, Girven, Hoard, Klein  \& Koester}{Farihi et~al.}{2011}]{Farihi2011}
Farihi J.,  Brinkworth C.~S.,  Gänsicke B.~T.,  Marsh T.~R.,  Girven J.,  Hoard D.~W.,  Klein B.,   Koester D.,  2011, \mn@doi [The Astrophysical Journal] {10.1088/2041-8205/728/1/l8}, 728, L8

\bibitem[\protect\citeauthoryear{Farihi, G{\"a}nsicke  \& Koester}{Farihi et~al.}{2013}]{Farihi2013}
Farihi J.,  G{\"a}nsicke B.~T.,   Koester D.,  2013, \mn@doi [Science] {10.1126/science.1239447}, 342, 218

\bibitem[\protect\citeauthoryear{{Feynman}, {Metropolis}  \& {Teller}}{{Feynman} et~al.}{1949}]{Feynman1949}
{Feynman} R.~P.,  {Metropolis} N.,   {Teller} E.,  1949, \mn@doi [Physical Review] {10.1103/PhysRev.75.1561}, \href {https://ui.adsabs.harvard.edu/abs/1949PhRv...75.1561F} {75, 1561}

\bibitem[\protect\citeauthoryear{{Fontaine}, {Brassard}  \& {Bergeron}}{{Fontaine} et~al.}{2001}]{Fontaine2001}
{Fontaine} G.,  {Brassard} P.,   {Bergeron} P.,  2001, \mn@doi [\pasp] {10.1086/319535}, \href {https://ui.adsabs.harvard.edu/abs/2001PASP..113..409F} {113, 409}

\bibitem[\protect\citeauthoryear{{Fontaine}, {Brassard}, {Dufour}  \& {Tremblay}}{{Fontaine} et~al.}{2015a}]{Fontaine2015a}
{Fontaine} G.,  {Brassard} P.,  {Dufour} P.,   {Tremblay} P.~E.,  2015a, in {Dufour} P.,  {Bergeron} P.,   {Fontaine} G.,  eds,  Astronomical Society of the Pacific Conference Series Vol. 493, 19th European Workshop on White Dwarfs. p.~113

\bibitem[\protect\citeauthoryear{{Fontaine}, {Dufour}, {Chayer}, {Dupuis}  \& {Brassard}}{{Fontaine} et~al.}{2015b}]{Fontaine2015b}
{Fontaine} G.,  {Dufour} P.,  {Chayer} P.,  {Dupuis} J.,   {Brassard} P.,  2015b, in {Dufour} P.,  {Bergeron} P.,   {Fontaine} G.,  eds,  Astronomical Society of the Pacific Conference Series Vol. 493, 19th European Workshop on White Dwarfs. p.~117

\bibitem[\protect\citeauthoryear{{Freytag}, {Ludwig}  \& {Steffen}}{{Freytag} et~al.}{1996}]{Freytag1996}
{Freytag} B.,  {Ludwig} H.~G.,   {Steffen} M.,  1996, \aap, \href {https://ui.adsabs.harvard.edu/abs/1996A&A...313..497F} {313, 497}

\bibitem[\protect\citeauthoryear{{Gaia Collaboration} et~al.,}{{Gaia Collaboration} et~al.}{2023}]{Gaia2023}
{Gaia Collaboration} et~al., 2023, \mn@doi [A&A] {10.1051/0004-6361/202243940}, 674, A1

\bibitem[\protect\citeauthoryear{Gentile~Fusillo, Gänsicke, Farihi, Koester, Schreiber  \& Pala}{Gentile~Fusillo et~al.}{2017}]{GentileFusillo2017}
Gentile~Fusillo N.~P.,  Gänsicke B.~T.,  Farihi J.,  Koester D.,  Schreiber M.~R.,   Pala A.~F.,  2017, \mn@doi [Monthly Notices of the Royal Astronomical Society] {10.1093/mnras/stx468}, 468, 971

\bibitem[\protect\citeauthoryear{Gentile Fusillo et~al.,}{Gentile Fusillo et~al.}{2021}]{GentileFusillo2021}
Gentile Fusillo N.~P.,  et~al., 2021, \mn@doi [Monthly Notices of the Royal Astronomical Society] {10.1093/mnras/stab2672}, 508, 3877

\bibitem[\protect\citeauthoryear{Gänsicke, Koester, Farihi, Girven, Parsons  \& Breedt}{Gänsicke et~al.}{2012}]{Gaensicke2012}
Gänsicke B.~T.,  Koester D.,  Farihi J.,  Girven J.,  Parsons S.~G.,   Breedt E.,  2012, \mn@doi [Monthly Notices of the Royal Astronomical Society] {10.1111/j.1365-2966.2012.21201.x}, 424, 333

\bibitem[\protect\citeauthoryear{Harrison, Bonsor  \& Madhusudhan}{Harrison et~al.}{2018}]{Harrison2018}
Harrison J.~H.,  Bonsor A.,   Madhusudhan N.,  2018, \mn@doi [Monthly Notices of the Royal Astronomical Society] {10.1093/mnras/sty1700}, 479, 3814

\bibitem[\protect\citeauthoryear{Harrison, Bonsor, Kama, Buchan, Blouin  \& Koester}{Harrison et~al.}{2021}]{Harrison2021}
Harrison J. H.~D.,  Bonsor A.,  Kama M.,  Buchan A.~M.,  Blouin S.,   Koester D.,  2021, \mn@doi [Monthly Notices of the Royal Astronomical Society] {10.1093/mnras/stab736}, 504, 2853–2867

\bibitem[\protect\citeauthoryear{{Heinonen}, {Saumon}, {Daligault}, {Starrett}, {Baalrud}  \& {Fontaine}}{{Heinonen} et~al.}{2020}]{Heinonen2020}
{Heinonen} R.~A.,  {Saumon} D.,  {Daligault} J.,  {Starrett} C.~E.,  {Baalrud} S.~D.,   {Fontaine} G.,  2020, \mn@doi [\apj] {10.3847/1538-4357/ab91ad}, \href {https://ui.adsabs.harvard.edu/abs/2020ApJ...896....2H} {896, 2}

\bibitem[\protect\citeauthoryear{Hollands, Tremblay, Gänsicke  \& Koester}{Hollands et~al.}{2021}]{Hollands2022}
Hollands M.~A.,  Tremblay P.-E.,  Gänsicke B.~T.,   Koester D.,  2021, \mn@doi [Monthly Notices of the Royal Astronomical Society] {10.1093/mnras/stab3696}, 511, 71

\bibitem[\protect\citeauthoryear{Hoskin et~al.,}{Hoskin et~al.}{2020}]{Hoskin2020}
Hoskin M.~J.,  et~al., 2020, \mn@doi [Monthly Notices of the Royal Astronomical Society] {10.1093/mnras/staa2717}, 499, 171

\bibitem[\protect\citeauthoryear{{Hummer} \& {Mihalas}}{{Hummer} \& {Mihalas}}{1988}]{Hummer1988}
{Hummer} D.~G.,  {Mihalas} D.,  1988, \mn@doi [\apj] {10.1086/166600}, \href {https://ui.adsabs.harvard.edu/abs/1988ApJ...331..794H} {331, 794}

\bibitem[\protect\citeauthoryear{{Iglesias} \& {Rogers}}{{Iglesias} \& {Rogers}}{1996}]{Iglesias1996}
{Iglesias} C.~A.,  {Rogers} F.~J.,  1996, \mn@doi [\apj] {10.1086/177381}, \href {https://ui.adsabs.harvard.edu/abs/1996ApJ...464..943I} {464, 943}

\bibitem[\protect\citeauthoryear{Izquierdo, Toloza, Gänsicke, Rodríguez-Gil, Farihi, Koester, Guo  \& Redfield}{Izquierdo et~al.}{2020}]{Izquierdo2020}
Izquierdo P.,  Toloza O.,  Gänsicke B.~T.,  Rodríguez-Gil P.,  Farihi J.,  Koester D.,  Guo J.,   Redfield S.,  2020, \mn@doi [Monthly Notices of the Royal Astronomical Society] {10.1093/mnras/staa3987}, 501, 4276

\bibitem[\protect\citeauthoryear{Jura}{Jura}{2003}]{Jura2003}
Jura M.,  2003, \mn@doi [The Astrophysical Journal] {10.1086/374036}, 584, L91–L94

\bibitem[\protect\citeauthoryear{Jura, Muno, Farihi  \& Zuckerman}{Jura et~al.}{2009}]{Jura2009}
Jura M.,  Muno M.~P.,  Farihi J.,   Zuckerman B.,  2009, \mn@doi [The Astrophysical Journal] {10.1088/0004-637X/699/2/1473}, 699, 1473

\bibitem[\protect\citeauthoryear{Kawka \& Vennes}{Kawka \& Vennes}{2012}]{Kawka2012}
Kawka A.,  Vennes S.,  2012, \mn@doi [A&A] {10.1051/0004-6361/201118210}, 538, A13

\bibitem[\protect\citeauthoryear{Kawka, Vennes, Ferrario  \& Paunzen}{Kawka et~al.}{2019}]{Kawka2019}
Kawka A.,  Vennes S.,  Ferrario L.,   Paunzen E.,  2019, \mn@doi [Monthly Notices of the Royal Astronomical Society] {10.1093/mnras/sty3048}, 482, 5201

\bibitem[\protect\citeauthoryear{Kenyon \& Bromley}{Kenyon \& Bromley}{2017}]{KenyonBromley2017}
Kenyon S.~J.,  Bromley B.~C.,  2017, \mn@doi [The Astrophysical Journal] {10.3847/1538-4357/aa7b85}, 844, 116

\bibitem[\protect\citeauthoryear{Klein, Jura, Koester  \& Zuckerman}{Klein et~al.}{2011}]{Klein2011}
Klein B.,  Jura M.,  Koester D.,   Zuckerman B.,  2011, \mn@doi [The Astrophysical Journal] {10.1088/0004-637x/741/1/64}, 741, 64

\bibitem[\protect\citeauthoryear{Klein, Doyle, Zuckerman, Dufour, Blouin, Melis, Weinberger  \& Young}{Klein et~al.}{2021}]{Klein2021}
Klein B.~L.,  Doyle A.~E.,  Zuckerman B.,  Dufour P.,  Blouin S.,  Melis C.,  Weinberger A.~J.,   Young E.~D.,  2021, \mn@doi [The Astrophysical Journal] {10.3847/1538-4357/abe40b}, 914, 61

\bibitem[\protect\citeauthoryear{Koester}{Koester}{2009}]{Koester2009}
Koester D.,  2009, \mn@doi [Astronomy \& Astrophysics] {10.1051/0004-6361/200811468}, 498, 517–525

\bibitem[\protect\citeauthoryear{{Koester}}{{Koester}}{2010}]{Koester2010}
{Koester} D.,  2010, \memsai, \href {https://ui.adsabs.harvard.edu/abs/2010MmSAI..81..921K} {81, 921}

\bibitem[\protect\citeauthoryear{{Koester} \& {Kepler}}{{Koester} \& {Kepler}}{2015}]{Koester2015}
{Koester} D.,  {Kepler} S.~O.,  2015, \mn@doi [\aap] {10.1051/0004-6361/201527169}, \href {https://ui.adsabs.harvard.edu/abs/2015A&A...583A..86K} {583, A86}

\bibitem[\protect\citeauthoryear{{Koester} \& {Wolff}}{{Koester} \& {Wolff}}{2000}]{Koester2000}
{Koester} D.,  {Wolff} B.,  2000, \aap, \href {https://ui.adsabs.harvard.edu/abs/2000A&A...357..587K} {357, 587}

\bibitem[\protect\citeauthoryear{{Koester}, {G{\"a}nsicke}  \& {Farihi}}{{Koester} et~al.}{2014}]{Koester2014}
{Koester} D.,  {G{\"a}nsicke} B.~T.,   {Farihi} J.,  2014, \mn@doi [\aap] {10.1051/0004-6361/201423691}, \href {https://ui.adsabs.harvard.edu/abs/2014A&A...566A..34K} {566, A34}

\bibitem[\protect\citeauthoryear{Koester, Kepler  \& Irwin}{Koester et~al.}{2020}]{Koester2020}
Koester D.,  Kepler S.~O.,   Irwin A.~W.,  2020, \mn@doi [A\&A] {10.1051/0004-6361/202037530}, 635, A103

\bibitem[\protect\citeauthoryear{{Kupka}, {Zaussinger}  \& {Montgomery}}{{Kupka} et~al.}{2018}]{Kupka2018}
{Kupka} F.,  {Zaussinger} F.,   {Montgomery} M.~H.,  2018, \mn@doi [\mnras] {10.1093/mnras/stx3119}, \href {https://ui.adsabs.harvard.edu/abs/2018MNRAS.474.4660K} {474, 4660}

\bibitem[\protect\citeauthoryear{Limbach et~al.,}{Limbach et~al.}{2024}]{Limbach2024}
Limbach M.~A.,  et~al., 2024, \mn@doi [The Astrophysical Journal Letters] {10.3847/2041-8213/ad74ed}, 973, L11

\bibitem[\protect\citeauthoryear{McDonough}{McDonough}{2003}]{McDonough2003}
McDonough W.~F.,  2003, \mn@doi [Treatise on Geochemistry] {10.1016/B0-08-043751-6/02015-6}, 2-9, 547

\bibitem[\protect\citeauthoryear{Melis \& Dufour}{Melis \& Dufour}{2017}]{Melis2017}
Melis C.,  Dufour P.,  2017, \mn@doi [The Astrophysical Journal] {10.3847/1538-4357/834/1/1}, 834, 1

\bibitem[\protect\citeauthoryear{O’Brien et~al.,}{O’Brien et~al.}{2023}]{OBrien2023}
O’Brien M.~W.,  et~al., 2023, \mn@doi [Monthly Notices of the Royal Astronomical Society] {10.1093/mnras/stac3303}, 518, 3055

\bibitem[\protect\citeauthoryear{O’Brien, Tremblay, Klein, Melis, Koester, Buchan, Veras  \& Doyle}{O’Brien et~al.}{2025}]{OBrien2025}
O’Brien M.~W.,  Tremblay P.-E.,  Klein B.~L.,  Melis C.,  Koester D.,  Buchan A.~M.,  Veras D.,   Doyle A.~E.,  2025, \mn@doi [Monthly Notices of the Royal Astronomical Society] {10.1093/mnras/staf398}, 539, 171

\bibitem[\protect\citeauthoryear{{Paquette}, {Pelletier}, {Fontaine}  \& {Michaud}}{{Paquette} et~al.}{1986a}]{Paquette1986a}
{Paquette} C.,  {Pelletier} C.,  {Fontaine} G.,   {Michaud} G.,  1986a, \mn@doi [\apjs] {10.1086/191111}, \href {https://ui.adsabs.harvard.edu/abs/1986ApJS...61..177P} {61, 177}

\bibitem[\protect\citeauthoryear{{Paquette}, {Pelletier}, {Fontaine}  \& {Michaud}}{{Paquette} et~al.}{1986b}]{Paquette1986b}
{Paquette} C.,  {Pelletier} C.,  {Fontaine} G.,   {Michaud} G.,  1986b, \mn@doi [\apjs] {10.1086/191112}, \href {https://ui.adsabs.harvard.edu/abs/1986ApJS...61..197P} {61, 197}

\bibitem[\protect\citeauthoryear{{Putirka} \& {Xu}}{{Putirka} \& {Xu}}{2021}]{Putirka2021}
{Putirka} K.~D.,  {Xu} S.,  2021, \mn@doi [Nature Communications] {10.1038/s41467-021-26403-8}, \href {https://ui.adsabs.harvard.edu/abs/2021NatCo..12.6168P} {12, 6168}

\bibitem[\protect\citeauthoryear{Raddi, Gänsicke, Koester, Farihi, Hermes, Scaringi, Breedt  \& Girven}{Raddi et~al.}{2015}]{Raddi2015}
Raddi R.,  Gänsicke B.~T.,  Koester D.,  Farihi J.,  Hermes J.~J.,  Scaringi S.,  Breedt E.,   Girven J.,  2015, \mn@doi [Monthly Notices of the Royal Astronomical Society] {10.1093/mnras/stv701}, 450, 2083

\bibitem[\protect\citeauthoryear{Rogers et~al.,}{Rogers et~al.}{2024a}]{Rogers2024}
Rogers L.~K.,  et~al., 2024a, \mn@doi [Monthly Notices of the Royal Astronomical Society] {10.1093/mnras/stad3557}, 527, 6038

\bibitem[\protect\citeauthoryear{Rogers et~al.,}{Rogers et~al.}{2024b}]{Rogers2024b}
Rogers L.~K.,  et~al., 2024b, \mn@doi [Monthly Notices of the Royal Astronomical Society] {10.1093/mnras/stae1520}, 532, 3866

\bibitem[\protect\citeauthoryear{{Saumon}, {Chabrier}  \& {van Horn}}{{Saumon} et~al.}{1995}]{Saumon1995}
{Saumon} D.,  {Chabrier} G.,   {van Horn} H.~M.,  1995, \mn@doi [\apjs] {10.1086/192204}, \href {https://ui.adsabs.harvard.edu/abs/1995ApJS...99..713S} {99, 713}

\bibitem[\protect\citeauthoryear{{Stanton} \& {Murillo}}{{Stanton} \& {Murillo}}{2016}]{Stanton2016}
{Stanton} L.~G.,  {Murillo} M.~S.,  2016, \mn@doi [\pre] {10.1103/PhysRevE.93.043203}, \href {https://ui.adsabs.harvard.edu/abs/2016PhRvE..93d3203S} {93, 043203}

\bibitem[\protect\citeauthoryear{Steele, Debes, Xu, Yeh  \& Dufour}{Steele et~al.}{2021}]{Steele2021}
Steele A.,  Debes J.,  Xu S.,  Yeh S.,   Dufour P.,  2021, \mn@doi [The Astrophysical Journal] {10.3847/1538-4357/abc262}, 911, 25

\bibitem[\protect\citeauthoryear{Swan, Farihi, Koester, Hollands, Parsons, Cauley, Redfield  \& Gänsicke}{Swan et~al.}{2019}]{Swan2019}
Swan A.,  Farihi J.,  Koester D.,  Hollands M.,  Parsons S.,  Cauley P.~W.,  Redfield S.,   Gänsicke B.~T.,  2019, \mn@doi [Monthly Notices of the Royal Astronomical Society] {10.1093/mnras/stz2337}, 490, 202

\bibitem[\protect\citeauthoryear{Swan, Farihi, Melis, Dufour, Desch, Koester  \& Guo}{Swan et~al.}{2023}]{Swan2023a}
Swan A.,  Farihi J.,  Melis C.,  Dufour P.,  Desch S.~J.,  Koester D.,   Guo J.,  2023, \mn@doi [Monthly Notices of the Royal Astronomical Society] {10.1093/mnras/stad2867}, 526, 3815

\bibitem[\protect\citeauthoryear{{Tassoul}, {Fontaine}  \& {Winget}}{{Tassoul} et~al.}{1990}]{Tassoul1990}
{Tassoul} M.,  {Fontaine} G.,   {Winget} D.~E.,  1990, \mn@doi [\apjs] {10.1086/191420}, \href {https://ui.adsabs.harvard.edu/abs/1990ApJS...72..335T} {72, 335}

\bibitem[\protect\citeauthoryear{Tremblay, Ludwig, Steffen  \& Freytag}{Tremblay et~al.}{2013}]{Tremblay2013}
Tremblay P.-E.,  Ludwig H.-G.,  Steffen M.,   Freytag B.,  2013, \mn@doi [A&A] {10.1051/0004-6361/201322318}, 559, A104

\bibitem[\protect\citeauthoryear{Tremblay et~al.,}{Tremblay et~al.}{2020}]{Tremblay2020}
Tremblay P.-E.,  et~al., 2020, \mn@doi [Monthly Notices of the Royal Astronomical Society] {10.1093/mnras/staa1892}, 497, 130

\bibitem[\protect\citeauthoryear{Vennes \& Kawka}{Vennes \& Kawka}{2013}]{VennesKawka2013}
Vennes S.,  Kawka A.,  2013, \mn@doi [The Astrophysical Journal] {10.1088/0004-637X/779/1/70}, 779, 70

\bibitem[\protect\citeauthoryear{Wachlin, Vauclair, Vauclair  \& Althaus}{Wachlin et~al.}{2017}]{Wachlin2017}
Wachlin F.~C.,  Vauclair G.,  Vauclair S.,   Althaus L.~G.,  2017, \mn@doi [A&A] {10.1051/0004-6361/201630094}, 601, A13

\bibitem[\protect\citeauthoryear{Wang, Lineweaver  \& Ireland}{Wang et~al.}{2019}]{WANG2019}
Wang H.~S.,  Lineweaver C.~H.,   Ireland T.~R.,  2019, \mn@doi [Icarus] {https://doi.org/10.1016/j.icarus.2019.03.018}, 328, 287

\bibitem[\protect\citeauthoryear{Williams, Gänsicke, Swan, O’Brien, Izquierdo, Cutolo  \& Cunningham}{Williams et~al.}{2024}]{Williams2024}
Williams J.~T.,  Gänsicke B.~T.,  Swan A.,  O’Brien M.~W.,  Izquierdo P.,  Cutolo A.-M.,   Cunningham T.,  2024, \mn@doi [A&A] {10.1051/0004-6361/202450509}, 691, A352

\bibitem[\protect\citeauthoryear{Wilson, Gaensicke, Koester, Toloza, Pala, Breedt  \& Parsons}{Wilson et~al.}{2015}]{Wilson2015}
Wilson D.~J.,  Gaensicke B.~T.,  Koester D.,  Toloza O.,  Pala A.~F.,  Breedt E.,   Parsons S.~G.,  2015, \mn@doi [Monthly Notices of the Royal Astronomical Society] {https://doi.org/10.1093/mnras/stv1201}, 451, 3237–3248

\bibitem[\protect\citeauthoryear{Wyatt, Farihi, Pringle  \& Bonsor}{Wyatt et~al.}{2014}]{Wyatt2014}
Wyatt M.~C.,  Farihi J.,  Pringle J.~E.,   Bonsor A.,  2014, \mn@doi [Monthly Notices of the Royal Astronomical Society] {10.1093/mnras/stu183}, 439, 3371

\bibitem[\protect\citeauthoryear{Xu, Jura, Klein, Koester  \& Zuckerman}{Xu et~al.}{2013}]{Xu2013}
Xu S.,  Jura M.,  Klein B.,  Koester D.,   Zuckerman B.,  2013, \mn@doi [The Astrophysical Journal] {10.1088/0004-637x/766/2/132}, 766, 132

\bibitem[\protect\citeauthoryear{Xu, Jura, Koester, Klein  \& Zuckerman}{Xu et~al.}{2014}]{Xu2014}
Xu S.,  Jura M.,  Koester D.,  Klein B.,   Zuckerman B.,  2014, \mn@doi [The Astrophysical Journal] {10.1088/0004-637x/783/2/79}, 783, 79

\bibitem[\protect\citeauthoryear{Xu, Zuckerman, Dufour, Young, Klein  \& Jura}{Xu et~al.}{2017}]{Xu2017}
Xu S.,  Zuckerman B.,  Dufour P.,  Young E.~D.,  Klein B.,   Jura M.,  2017, \mn@doi [The Astrophysical Journal] {10.3847/2041-8213/836/1/l7}, 836, L7

\bibitem[\protect\citeauthoryear{Xu, Dufour, Klein, Melis, Monson, Zuckerman, Young  \& Jura}{Xu et~al.}{2019}]{Xu2019}
Xu S.,  Dufour P.,  Klein B.,  Melis C.,  Monson N.~N.,  Zuckerman B.,  Young E.~D.,   Jura M.~A.,  2019, \mn@doi [The Astronomical Journal] {10.3847/1538-3881/ab4cee}, 158, 242

\bibitem[\protect\citeauthoryear{Yoshizaki \& McDonough}{Yoshizaki \& McDonough}{2020}]{Yoshizaki2020}
Yoshizaki T.,  McDonough W.~F.,  2020, \mn@doi [Geochimica et Cosmochimica Acta] {https://doi.org/10.1016/j.gca.2020.01.011}, 273, 137

\bibitem[\protect\citeauthoryear{Zuckerman, Koester, Melis, Hansen  \& Jura}{Zuckerman et~al.}{2007}]{Zuckerman2007}
Zuckerman B.,  Koester D.,  Melis C.,  Hansen B.~M.,   Jura M.,  2007, \mn@doi [The Astrophysical Journal] {10.1086/522223}, 671, 872

\bibitem[\protect\citeauthoryear{Zuckerman, Koester, Dufour, Melis, Klein  \& Jura}{Zuckerman et~al.}{2011}]{Zuckerman2011}
Zuckerman B.,  Koester D.,  Dufour P.,  Melis C.,  Klein B.,   Jura M.,  2011, \mn@doi [The Astrophysical Journal] {10.1088/0004-637x/739/2/101}, 739, 101

\makeatother
\end{thebibliography}

%\appendix

\supplementarysection

\section*{Appendix}

\subsection*{Sample selection metric}

We wish to quantify the potential impact of switching to a new set of timescales on the interpretation of the metal abundances of a white dwarf. We derive a simple expression, which can be applied in a pre-processing phase of modelling, to estimate this impact in a quantitative way.

We start by considering the relative abundances of two metals, \textit{i} and \textit{j}, as they evolve over time. We assume that material accretes at a constant rate starting at time $t=0$, for a time of length $t_{\rm event}$ before switching off. The (number) abundance ratio of \textit{i} to \textit{j}, which we label as $R_{ij}$, evolves as
\begin{equation}
    R_{ij} = R^0_{ij}\frac{\tau_i}{\tau_j}\exp{\left(\frac{t}{\tau_j}-\frac{t}{\tau_i}\right)}\frac{\left[\exp{\left(\frac{t_l}{\tau_i}\right)-1}\right]}{\left[\exp{\left(\frac{t_l}{\tau_j}\right)-1}\right]},
\label{eq:Rij_full}
\end{equation}
{\noindent}where $\tau_X$ is the sinking timescale of element $X$ and $t_l = \min(t, t_{\rm event})$. As time increases, $R_{ij}$ deviates further from its initial value in the pollutant, $R^0_{ij}$. The greatest potential discrepency that can be introduced by a change in sinking timescales occurs at late times, so we consider the late time behaviour (i.e., $t \ge t_{\rm event}$). Taking log (base 10) of both sides (since abundances are typically given on a log scale), the late time form of equation~\ref{eq:Rij_full} becomes
\begin{equation}
    \log_{10}(R_{ij}) \approx \log_{10}\left(R^0_{ij}\frac{\tau_i}{\tau_j}\right) + \frac{t}{\ln(10)}{\left(\frac{1}{\tau_j}-\frac{1}{\tau_i}\right)}~.
\label{eq:Rij_simple}
\end{equation}
{\noindent}Substituting $t = \beta\tau_i$, this can be simplified to:
\begin{equation}
    \log_{10}(R_{ij}) \approx \log_{10}\left(R^0_{ij}\frac{\tau_i}{\tau_j}\right) + \frac{\beta}{\ln(10)}{\left(\frac{\tau_i}{\tau_j}-1\right)}
\label{eq:Rij}
\end{equation}
{\noindent}where $\beta$ represents how deep we are into the declining phase, and assuming that all relevant diffusion timescales are of a similar order of magnitude, it does not matter much which one we use as a reference (i.e., we approximate $\beta = t/\tau$ without specifying which element). Note that when $\beta=0$, we recover the result that in the steady state phase of accretion, the ratio of $i$ and $j$ is modified according to the ratio of their sinking timescales. Equation~\ref{eq:Rij} can be thought of as a steady state modification (first term) plus a modification associated with the declining phase (second term). Now consider the change in this ratio that would result from swapping to a new set of timescales, with the sets labelled as $a$ and $b$. The difference, $\Delta$, is
\begin{equation}
    \Delta = \log_{10}\left(\frac{\tau_i^a\tau_j^b}{\tau_j^a\tau_i^b}\right) + \frac{\beta}{\ln(10)}\left(\frac{\tau_i^a}{\tau_j^a}-\frac{\tau_i^b}{\tau_j^b}\right).
\label{eq:Delta}
\end{equation}

Broadly speaking, this change is significant when its magnitude exceeds the observational error. We therefore define our metric for \textit{i,j} as
\begin{equation}
    M_{a,b}^{i,j} = \frac{\vert\log_{10}\left(\frac{\tau_i^a\tau_j^b}{\tau_j^a\tau_i^b}\right) + \frac{\beta}{\ln(10)}\left(\frac{\tau_i^a}{\tau_j^a}-\frac{\tau_i^b}{\tau_j^b}\right)\vert}{\sqrt{\sigma_i^2+\sigma_j^2}}
\label{eq:Mij}
\end{equation}
{\noindent}where $\sigma_X$ is the 1 sigma uncertainty associated with $X$. The geological interpretation can, in principle, be affected by any pair \textit{i,j}. We therefore iterate over all possible pairs, giving our final metric:
\begin{equation}
    M_{a,b} = \max_{i,j}(M_{a,b}^{i,j}),
\label{eq:M}
\end{equation} 
{\noindent}for all $i,j\in E$ where $E$ is the set of all modellable elements.

In order to use this metric, we must estimate $\beta$, the number of sinking timescales we are into the declining phase. We firstly consider that for white dwarfs with short sinking timescales (e.g., warm H-dominated white dwarfs), the probability of observation in the declining phase is very low, so $\beta \approx 0$. We calculate an \textit{a priori} probability of observation in declining phase, $p$ as:
\begin{equation}
    p = \frac{n\tau}{t_{\rm event} + n\tau}
\label{eq:apriori_p},
\end{equation} where $n$ is the maximum number of sinking timescales into the declining phase for which pollution could plausibly be detected and $\tau$ is any relevant sinking timescale (assuming they are all of a similar order of magnitude). We assume $n$ = 5 and $t_{\rm event} = $\SI{100}{kyr}. In practice, the purpose of $p$ is to act as a switch, being close to 0 for white dwarfs with short sinking timescales and close to 1 for white dwarfs with long sinking timescales.

We next consider that for white dwarfs deep in the declining phase, metal abundances become strongly correlated with their sinking timescales. We calculate the Pearson correlation coefficient, $c$, between the (log) of metal abundances and the corresponding sinking timescales, for white dwarfs with at least 4 modellable elements. We scale $\beta$ with $c$ if $c > 0$ (otherwise we ignore it). Finally, we scale $\beta$ by $n$ in order to ensure that the maximum value of $\beta$ is $n$. This gives
\begin{equation}
    \beta = pn\max(0,c), 
\label{eq:D}
\end{equation} in which $p$ and $c$ are both functions of the set of timescales being considered, so we calculate them for both grids $a$ and $b$ and adopt their mean values.

Our metric satisfies $M_{a,b}^{i,j} = M_{a,b}^{j,i} = M_{b,a}^{i,j}$ and $M_{a,a}^{i,j} = M_{a,b}^{i,i} = 0$. For warm H-dominated white dwarfs with $\tau \ll t_{\rm event}$, $p$ (and therefore $\beta$) is negligible and the numerator of Equation~\ref{eq:Mij} reduces to the more familiar steady state scaling. For white dwarfs deep into the declining phase, $M$ increases, but only if there is a difference in the timescale ratios. The terms in the numerator of Equation~\ref{eq:Mij} are always the same sign so the declining phase can only increase the discrepancy (this also motivates ignoring negative values of $c$).

Our metric, especially the estimation of $\beta$, is only an approximation, and should be thought of as a heuristic indicator of which white dwarfs are most likely to suffer from a change in the adopted grid of sinking timescales. A complete evaluation would need to account for the values of the abundances themselves. This could be achieved by running our full Bayesian framework, but would defeat the purpose of using a simple metric to narrow down the number of target white dwarfs.

\subsubsection*{Metric performance}

%However, the discrepancy between two grids of sinking timescales, calculated with different assumptions, is not uniform across $T_{\rm eff}$/$\log(g)$ space. We have introduced a metric to quantify this variation.

%Can we wrap this up with some statements like x\% of DAs are potnentially affected - pointing towards running the whole of PEWDD, but leaving that for future work is OK! Not necessarly by us!

%\subsection{Application of discrepancy metric}
Our discrepancy metric can be thought of as predicting the strength of the effect of switching from one set of sinking timescales to another, measured in units of sigma. Crudely, one might expect that whether a geological result is robust depends on whether the sigma significance of the result is greater than the discrepancy metric. We investigate whether this is actually the case by plotting all geological results, from all 5 tests (excluding duplicate entries), by their sigma significance and the associated discrepancy metric.

\begin{figure}
    \centering
    \includegraphics[width=\columnwidth, keepaspectratio]{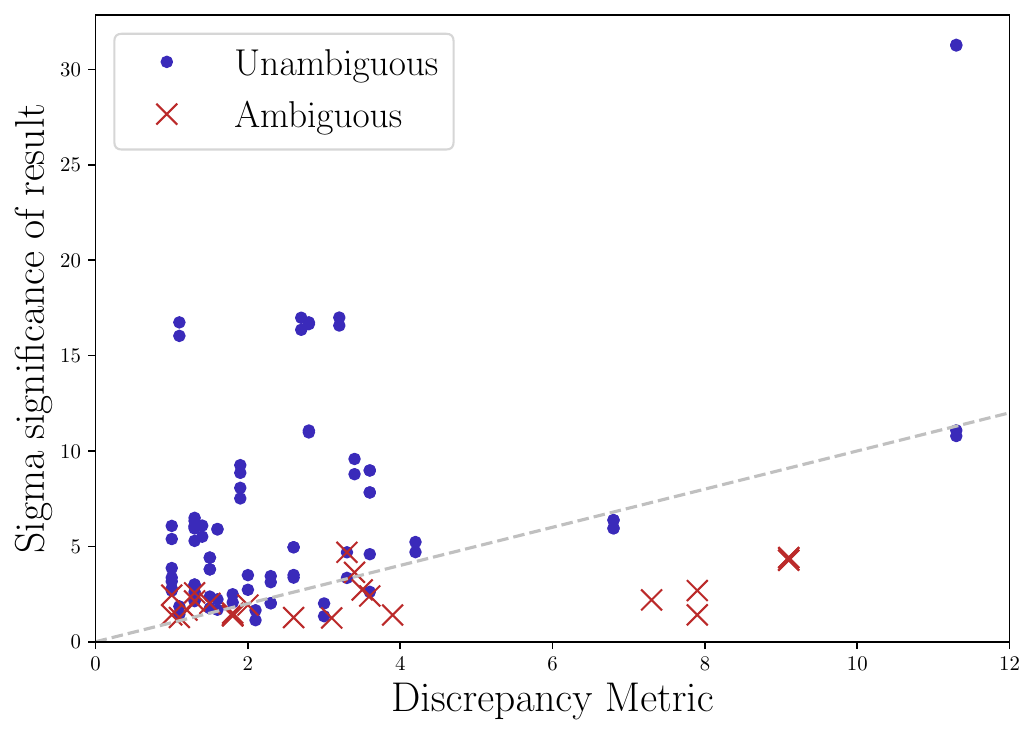}
    \caption{Performance of our discrepancy metric as a predictor for whether a result is ambiguous, i.e., can be lost when switching to a different timescale grid. If the metric was perfect, there would be no ambiguous results above the dashed line, which follows $y=x$. Some ambiguous results are above the dashed line: these are results which the metric falsely predicts to be robust. These false predictions are mostly limited to low metric values (below 1.8). For higher metric values, the performance is much stronger, with a prediction accuracy of 95\%.}
    \label{fig:metric_plot}
\end{figure}

The results are shown in Figure~\ref{fig:metric_plot}. We label ambiguous results with red crosses. These are results which are not recovered when switching to the other timescale grid in question. Unambiguous results are shown with blue circles. If the discrepancy metric were perfect, there would be no ambiguous results for which the sigma significance is greater than the metric. This is not the case: there are results which the metric falsely predicts to be robust, mostly for metric values below 1.8. We find that, above this threshold, the performance is much stronger, with a prediction accuracy of 95\%. In practical terms, suppose that a model finds evidence of a geological process with sigma significance $\sigma$, and one wishes to know whether it is robust to a change to another set of sinking timescales without rerunning the model. If the metric value associated with that change in timescale grid is $M$, then if $\sigma>M>1.8$, one could be 95\% confident that the result is robust. If $\sigma<M$, Figure~\ref{fig:metric_plot} shows that the result is likely not robust. If $\sigma>M$, but $M<1.8$, Figure~\ref{fig:metric_plot} shows that the result might still be robust as long as $\sigma $ is sufficiently large.

\subsection*{Timescale ratios}

\begin{figure*}
    \centering
    \includegraphics[width=0.95\textwidth, keepaspectratio]{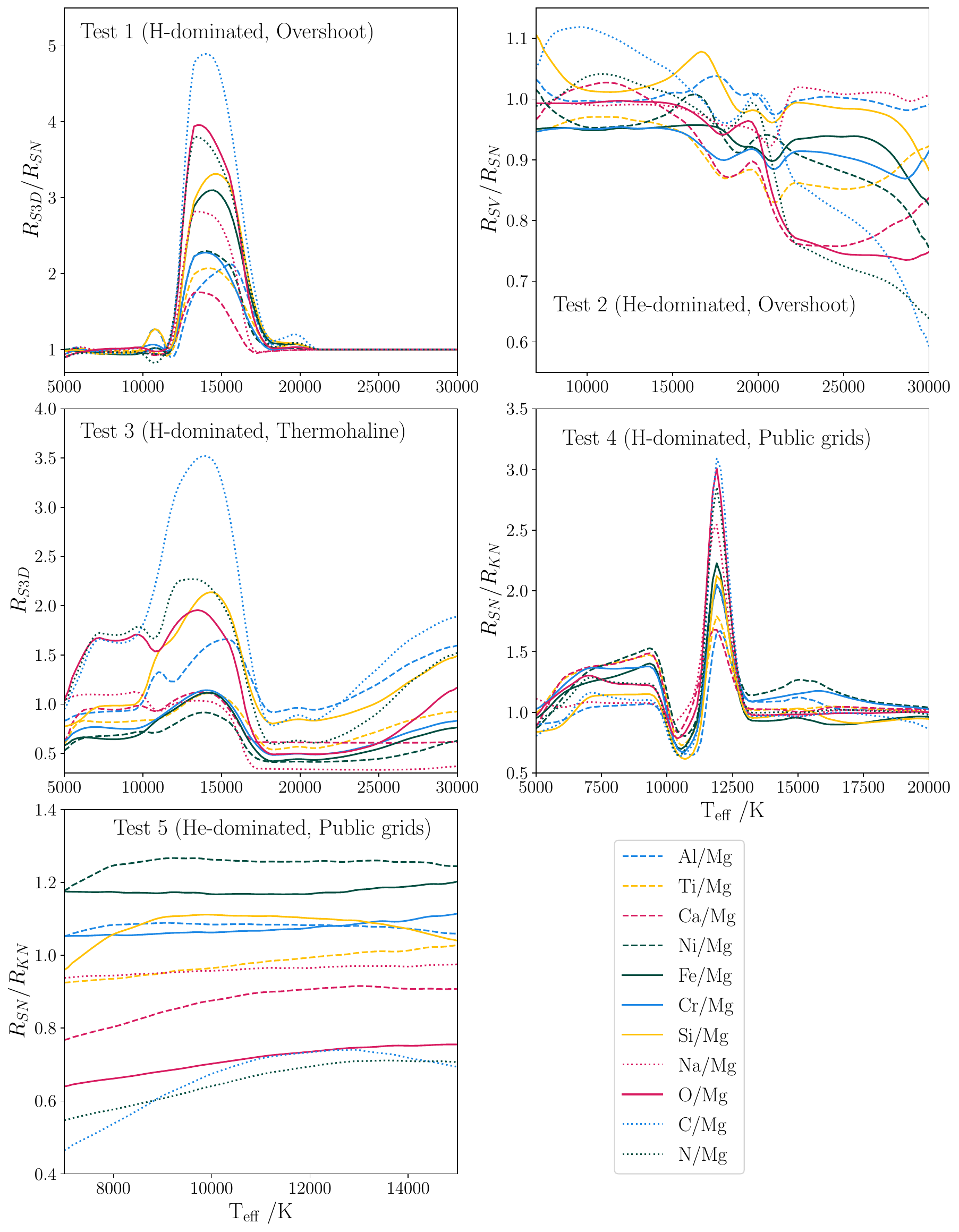}
    \caption{The dependence of key sinking timescale ratios on effective temperature, $T_{\rm eff}$. Each panel shows the ratio of $R$ as calculated using two different grids, labelled on the vertical axis. $R$ denotes any of 11 sinking timescale ratios, indicated in the legend. For the panel labelled `Test 3', the vertical axis is simply the value of $R$ when using the S3D grid. For the panel labelled `Test 5', the KN grid takes Ca/He as an additional variable. We set this to a (low) value of -15, to minimise the effect of this variable. We have smoothed the lines using a Savitzky-Golay filter to aid visual clarity.}
    \label{fig:timescale_crosssection}
\end{figure*}

The behaviour described in Sections~\ref{sec:overshoot}-\ref{sec:public_grids} is driven by changes in sinking timescale ratios under different assumptions. Figure~\ref{fig:timescale_crosssection} illustrates this underlying behaviour for each of the 5 test cases for all modellable elements. Here, we fix the value of $\log(g)$ to 8 since the primary dependence is on $T_{\rm eff}$. The Fe/Mg line of the top left panel is a cross-section through Figure~\ref{fig:da_overshoot_timescales} at $\log(g)=8$. The other panels can be loosely thought of as cross-sections through Figures~\ref{fig:DMP_DB_OVERSHOOT}, \ref{fig:thermohaline_paramspace}, \ref{fig:DMP_DA_BVK} and \ref{fig:DMP_DB_BVK}. This is not an exact equivalence, since those figures take into account accretion phase and which elements are detected.

% Don't change these lines
\bsp	% typesetting comment
\label{lastpage}
\end{document}